\documentclass[a4paper,11pt]{article}
\usepackage{graphicx}

\pdfoutput=1 

\usepackage{jcappub} 

\usepackage[T1]{fontenc} 

\usepackage[utf8]{inputenc}
\usepackage{amsmath,mathtools}

\usepackage{tensor}
\usepackage{amssymb}
\usepackage{empheq}
\usepackage{physics}
\usepackage{cleveref}
\usepackage{bigints}
\usepackage{verbatim}

\usepackage{braket}
\usepackage{cancel}
\usepackage{booktabs}
\usepackage{color}
\usepackage{hyperref}
\usepackage{subcaption}

\newcommand{\EqCorrection}[1]{#1}

\newcommand{\roughly}[1]{\mathrel{\raise.3ex\hbox{$#1$\kern-0.85em
\lower1ex\hbox{$\sim$}}}}

\newcommand{\tphi}{{\phi}}

\newcommand{\MPL}{M_{\rm Pl}}

\newcommand{\nn}{\nonumber}

\begin{document}

\title{Cuscuton Fifth Force: Implications of Non-Dynamical Scalars for Structure Growth}

\author[a,b]{Adam Smith,}
\author[c]{Ilaria Andrei,}
\author[a,d,e, f]{Maria Mylova,}
\author[e,g,h]{and Niayesh Afshordi}

\affiliation[a]{School of Mathematical and Physical Sciences, University of Sheffield
}
\affiliation[b]{Department of Physics and Astronomy, University of Manchester
}
\affiliation[c]{Institute of Physics, University of Tartu
}
\affiliation[d]{Kavli Institute for the Physics and Mathematics of the Universe (WPI)
}
\affiliation[e]{Perimeter Institute for Theoretical Physics
}
\affiliation[f]{Cosmology, Gravity and AstroparticlePhysics Group at Center for Theoretical Physics of the Universe,
Institute for Basic Science (IBS)
}
\affiliation[g]{Department of Physics and Astronomy, University of Waterloo
}
\affiliation[h]{Waterloo Centre for Astrophysics, University of Waterloo
}
\emailAdd{asmith69@sheffield.ac.uk}
\emailAdd{ilaria.andrei@ut.ee}
\emailAdd{mariamylova2026@ibs.re.kr}
\emailAdd{nafshordi@pitp.ca}
\date{\today}
\abstract{
 The cuscuton is a scalar field that obeys a constraint rather than an evolution equation, so it modifies gravity without adding a
propagating scalar degree of freedom. We study the cosmology of effective
cuscuton theory when matter is minimally coupled in the Jordan frame, and therefore conformally coupled
to the cuscuton in the Einstein-frame. We derive the background and linear
perturbation equations and show that the algebraic cuscuton response mediates a
fifth force even though the scalar itself does not propagate. The same
conformal factor also changes Einstein-frame particle masses and the friction of
matter perturbations, tying the strength and sign of the force to the
background evolution rather than to an independent coupling. We study a power-law family of models which
contain both tracking solutions and sign-switching solutions; in the latter the effective cuscuton
density runs from negative during matter domination to a positive,
cosmological-constant-like value today. Using a modified version
of \texttt{CLASS}, including the induced particle-mass dependence of
recombination, we illustrate the corresponding changes to CMB temperature and
linear matter power spectra for benchmark models, without fitting to data. In the strongest benchmarks the fifth force changes the effective gravitational coupling on sub-horizon scales by about one percent, smaller than the accompanying changes to the matter power spectrum and the acoustic
scale. For the sign-switching solutions this change is negative over most of cosmic history, so a
non-propagating field can weaken rather than strengthen gravitational
attraction. 
}
\maketitle
\section{Introduction}\label{sec:introduction}

The remarkable economy of $\Lambda$CDM has made it the standard language of
precision cosmology. Its economy, however, should not be confused with a
microscopic explanation. The nature and smallness of the cosmological constant
remain obscure, and recent distance measurements have sharpened the case for
testing dynamics beyond a strictly constant dark energy density
\cite{Planck:2018vyg,Brout:2022vxf,DESI:2024mwx,DESI:2025zgx,DESI:2025fii}. A scalar field is
the most economical way to introduce such dynamics, and the space of
second-order scalar-tensor theories built from one is by now completely
classified \cite{Horndeski:1974wa,Deffayet:2009mn,Kobayashi:2011nu}. An ordinary light
scalar, however, also introduces a new propagating degree of freedom,  sensitivity to radiative corrections, and when coupled to matter, a fifth force that is already
strongly constrained away from cosmological scales
\cite{Will:2014kxa,Berti:2015itd,Brax:2022vlf}, and which must therefore be
screened, or coupled weakly enough to evade those bounds
\cite{Vainshtein:1972sx,Damour:1990tw,Khoury:2003rn,Burrage:2017qrf,Joyce:2014kja}.

The cuscuton offers a different possibility. It arises as the incompressible
limit of a scalar field theory whose phase space is locally degenerate, so that
its apparently infinite sound speed carries no microscopic information
\cite{Afshordi:2006ad}. On cosmological backgrounds the field supplies no
independent initial conditions. Its equation of motion is not an evolution
equation but a constraint, and that constraint modifies how the metric responds
to matter \cite{Afshordi:2007yx}. The cuscuton has consequently become a useful
laboratory for modified gravity, cosmological perturbations, limiting-curvature
constructions, and theories carrying only the two tensorial degrees of freedom
of general relativity
\cite{Afshordi:2009tt,Bhattacharyya:2016mah,Boruah:2017tvg,Iyonaga:2018vnu,Iyonaga:2020bmm,Quintin:2019orx,Lin:2017oow}.
Comparable constraints do not always leave the scalar non-propagating, for example in in mimetic gravity
\cite{Chamseddine:2013kea} which instead leaves the scalar propagating
as pressureless dust. 
The construction below is a natural
extension of the effective field theory (EFT) of dark energy
\cite{Gubitosi:2012hu} to non-propagating degrees of freedom.
Recently, there has been a lot of progress in cuscuton-like theories. These are modified gravity theories which only propagate two tensorial degrees of freedom\cite{DeFelice:2015hla, Chagoya:2016inc, Lin:2017oow, Mukohyama:2019unx, Iyonaga:2018vnu, Aoki:2018brq, Gao:2019twq, DeFelice:2020eju,  Tasinato:2020fni, Aoki:2020lig, Aoki:2021zuy, Hu:2021yaq, DeFelice:2022uxv, Yao:2020tur, Yao:2023qjd}.

Here we start from the effective cuscuton construction of
Ref.~\cite{Mylova:2023ddj}, which is formulated geometrically around the
preferred spacetime foliation that the cuscuton defines. At the order retained below some  scale $\Lambda$, associated with the microscopic description of the cuscuton, the
theory propagates only the two tensor polarizations, even though scalar
derivatives appear once the frames are transformed. This leaves us with only one physical
choice to make for the metric to which matter couples minimally;
take it to
be the Jordan-frame metric \cite{Jordan:1955,Dicke:1961gz,Flanagan:2004bz}. 

In the Einstein-frame this produces a universal
conformal interaction, time-dependent particle masses, and  exchange of energy and momentum
between the cuscuton and non-relativistic matter.
This behaviour is 
familiar to coupled quintessence and dilaton models
\cite{Wetterich:1994bg,Amendola:1999er,Damour:1990tw,Gasperini:2001pc,Brax:2010gi},
whose cosmological perturbations have been followed through recombination in
closely related multi-field settings \cite{Smith:2024ayu,Smith:2024ibv},
and constrained by current data on interacting dark sectors \cite{Giare:2024smz}. Radiation
stays uncoupled at the classical fluid level, since its stress tensor is
traceless, while recombination does not escape, because the electron mass and the
Thomson cross-section both inherit conformal factors
\cite{Ali_Ha_moud_2011,Schoneberg:2024ynd,Toda:2025kcq,Smith:2025icl,Smith:2025uaq}. 

This setup makes a simple question unexpectedly subtle: can a field with no
propagating scalar mode mediate a fifth force? The answer is yes. The cuscuton
perturbation is fixed algebraically by the matter perturbation and metric
potentials, but this constrained response still appears in the matter Euler
equation. In the quasi-static limit it can be packaged as an effective Newton
constant of the kind used to parameterise linear growth in modified gravity
\cite{Amendola:2007rr,Bellini:2014fua}. Unlike in a conventional scalar-tensor theory, however, its strength
cannot be increased independently of the cosmological solution. The same
background evolution that controls the force also controls particle mass drift,
the matter dilution law, Hubble friction, and, near the recombination epoch, atomic
physics. These correlated effects typically become important before the direct
fifth force correction dominates a cosmological observable.

To expose this relation, we study a power-law family for the two independent
functional combinations that enter the background and perturbation equations.
It contains two qualitatively distinct limits. The first is a tracking case in
which the cuscuton follows the dominant density before approaching a late-time
vacuum. The second is a sign-switching case in which the effective cuscuton
energy density is negative during matter domination and becomes positive at
late times. The latter provides a dynamical counterpart of phenomenological
sign-switching dark energy models \cite{Akarsu:2021fol,Akarsu:2025gwi}, and of the
negative density  fits for dark energy that motivate them \cite{Visinelli:2019qqu}, with the
additional ingredient that the conformal interaction changes the relation
between early- and late-time particle masses.

We solve the background and linearized system with a modified version of
\texttt{CLASS} \cite{Lesgourgues:2011re,Diego_Blas_2011}, consistently passing the
Einstein-frame electron mass to the HyRec recombination module
\cite{Ali_Ha_moud_2011,Lee:2020obi}. The spectra
presented here are benchmark predictions rather than a likelihood analysis.
They show that the CMB and matter power spectrum change by up to tens of percent
while the direct fifth force stays at the percent level. For the displayed
benchmarks, the quasi-static correction to Newton's constant reaches
approximately $10^{-2}$, in the tracking case at $\lambda=-10^{2}$ and in the
sign-switching case at $\zeta=10^{-2}$, and falls by one and two orders of magnitude, respectively,
across the plotted parameter ranges. In the sign-switching examples the
correction is negative for $a\lesssim0.8$, so the constrained cuscuton response
weakens rather than strengthens the attraction between matter perturbations
over most of the evolution. This
cosmological result cannot be extrapolated directly to the Solar System because a
localized solution must first be shown to preserve the timelike cuscuton
foliation.

We deliberately restrict the present analysis to illustrative benchmarks.
A parameter scan and likelihood analysis, including a quantitative test of the
sign-switching case, are deferred to a follow-up study. Accordingly, we make
no claim here that the displayed models are observationally viable, excluded,
or preferred over $\Lambda$CDM.

The remainder of this paper is organised as follows. In
\cref{sec:cuscuton} we introduce the effective cuscuton theory, transform the
action to the Einstein-frame and derive the scalar, matter and gravitational
field equations in the ADM decomposition. In
\cref{sec:cuscuton_cosmology} we specialise these equations to a homogeneous
cosmological background and derive the linear perturbation equations in
Newtonian gauge, including the algebraic constraints that determine the
background and perturbed cuscuton configurations. In
\cref{sec:quasistatic} we take the quasi-static limit of the perturbation
system and derive the modified matter-growth equation, separating the effects
of the conformal coupling into a modified friction term and an effective
gravitational coupling. In \cref{sec:cuscuton_dark_energy} we apply the theory
to a power-law family of effective potentials and cuscuton coefficients. We
study both tracking and sign-switching regimes of the parameter space, implement their background
and perturbation evolution in a modified version of \texttt{CLASS}, and
present the resulting particle mass evolution, CMB temperature anisotropies,
matter power spectra and fifth force corrections. We summarise our results and
discuss directions for future work in \cref{sec:conclusions}.
\Cref{sec:perts_in_other_gauges} gives the perturbation equations in unitary and
synchronous gauges, \cref{app:numerics} records the numerical implementation and
the consistency checks it was validated against, and \cref{app:useful_formulae}
gives the transformation identities used in the conformal mapping.
\section{Effective cuscuton theory with conformally coupled matter}\label{sec:cuscuton}

The effective cuscuton theory is an EFT of modified gravity
constructed so as to preserve only the two tensorial degrees of freedom of
general relativity. It introduces an additional scalar field $\phi$ that labels
a preferred foliation of spacetime and enters through constraint equations
that modify the response of the geometry and matter sector.

The theory is most naturally formulated in unitary gauge, where the constant-$\phi$ hypersurfaces define the preferred time slicing. The action is therefore organised in terms of geometric quantities intrinsic and extrinsic to these
hypersurfaces. For this foliation to be well defined, the scalar gradient must remain timelike,
\begin{equation}
    X\equiv-g^{\mu\nu}\partial_\mu\phi\,\partial_\nu\phi>0,
\end{equation}
so that the constant-$\phi$ hypersurfaces are spacelike.

Since the theory is treated as an effective description rather than as a
complete microscopic model, the frame in which matter is minimally coupled is
an additional physical choice \cite{Brans:1961sx,Dicke:1961gz,Flanagan:2004bz,Fujii:2003pa}. Matter could be taken to couple minimally to
the Einstein-frame metric, in which case no explicit conformal interaction
would appear in that frame. In this work we instead entertain the possibility
that matter is minimally coupled in the Jordan frame. After transforming to
the Einstein-frame, this choice generates a universal conformal coupling
between the cuscuton and the matter sector. The resulting Einstein-frame
particle masses depend on $\phi$, and the associated exchange of
energy and momentum modifies both the cosmological background evolution and the
growth of matter perturbations.

We now introduce the Jordan-frame effective action relevant for this choice of
matter coupling.
\subsection{Action and field equations}

Following the EFT construction of cuscuton gravity developed
in \cite{Mylova:2023ddj}, we introduce a dimensionless scalar field $\tphi$ and
a mass scale $\Lambda$ set by the UV completion. As in that construction,
$\Lambda$ may sit at or below the Planck mass, with the Einstein--Hilbert term
carrying its own normalization \cite{Mylova:2023ddj,Adshead:2017srh}. The
coefficient of the Jordan-frame Ricci scalar, $\Lambda^2 f(\tphi)$, is then an
effective Planck mass in the sense of \cite{Bellini:2014fua}, and its constant
part is not observable and a constant rescaling of $f$ can be absorbed into
$\Lambda$ and into the normalization of $V$ and $c_1$ without changing any
equation below. What power counting constrains is the variation of $f$, which
must remain of order unity for the expansion in $\Lambda$ to hold. In every
benchmark of \cref{sec:cuscuton_dark_energy} the ratio $f/f_0$, with $f_0$ its
present-day value, stays between $0.2$ and $2.0$ over the full evolution. We set $\Lambda=\MPL$ throughout and
define Newton's constant in the usual way $G_N\equiv(8\pi\MPL^2)^{-1}$.

We begin with the Jordan-frame action
\begin{align}\label{eq:action}
S_J=\Lambda^4\!\int \! d^4x\,\sqrt{-g_J}\,\Bigg\{
\frac{f(\tphi)}{2\Lambda^2}\,R_J
+\frac{c_1(\tphi)}{\Lambda}\sqrt{X_J}
-V(\tphi)
+\frac{c_2(\tphi)}{\Lambda^2}\sqrt{X_J}\,K_J
\Bigg\} + S_{\rm sm}( g^J_{\mu\nu}, \Psi_i),
\end{align}
where $f(\tphi)$ is an arbitrary function determining the effective
gravitational coupling. The quantity $K_J$ denotes the Jordan-frame trace of
the extrinsic curvature of the preferred constant-$\tphi$ hypersurfaces, while
$c_1(\tphi)$ and $c_2(\tphi)$ parameterise higher-order operators in the
effective cuscuton description. The Standard Model sector,
$S_{\rm sm}(g^J_{\mu\nu},\Psi_i)$, is minimally coupled to the Jordan-frame
metric, which is therefore the physical frame in which standard matter follows
geodesics.

All Jordan-frame quantities in \cref{eq:action} are built from $g^J_{\mu\nu}$
and its inverse $g_J^{\mu\nu}$ in the same way that their Einstein-frame
counterparts below are built from $g_{\mu\nu}$,
\begin{equation}
X_J \equiv -\,g_J^{\mu\nu}\,\partial_\mu \tphi\,\partial_\nu \tphi,
\qquad
n^J_\mu \equiv \frac{\partial_\mu \tphi}{\sqrt{X_J}},
\qquad
K_J = \nabla^J_\mu n_J^\mu,
\label{eq:jordan_frame_defs}
\end{equation}
where $\nabla^J_\mu$ is the covariant derivative compatible with
$g^J_{\mu\nu}$.

The purpose of this paper is best illustrated by performing a Weyl
transformation to the Einstein-frame, in which the gravitational sector assumes
canonical Einstein--Hilbert form. This is achieved via
\begin{equation}\label{eq:conf_trans}
g_{\mu\nu}=f(\tphi)\,g^{J}_{\mu\nu},
\qquad
\sqrt{-g}=f^{2}(\tphi)\sqrt{-g_J}.
\end{equation}

Using the standard Weyl transformation properties of the Ricci scalar and extrinsic curvature \cite{Wald:1984rg,Fujii:2003pa} (see Appendix~\ref{app:useful_formulae}), the action can be written in the Einstein-frame as
\begin{align}
S = \int d^4x\,\sqrt{-g}\,\bigg\{
\frac{\Lambda^2}{2} R
+3\Lambda^2\bigg( \frac{\Box f}{2f} - \frac{3}{4}\frac{\partial_\mu f\partial^\mu f}{f^2}\bigg)-\Lambda^4 \frac{ V(\tphi)}{f^2}
+\Lambda^3\, \frac{c_1(\tphi)}{f^{3/2}}\, \sqrt{X}
\nn\\
+\Lambda^2 \frac{c_2(\tphi)}{f}\,
\Big( \sqrt{X}\,K - \frac{3}{2}\frac{\sqrt{X}}{f}\, n^\mu\partial_\mu f \Big)
\bigg\}
+S_{\rm sm}(f^{-1}(\tphi) g_{\mu\nu}, \Psi_i),
\label{eq:EF_intermediate}
\end{align}
where $n^\mu$ are the Einstein-frame unit normal vectors from which the Einstein-frame extrinsic curvature, $K$, is built as
\begin{equation}
K=\nabla_\mu n^\mu .
\end{equation}
Throughout this work the foliation is defined by the cuscuton field itself, and we therefore define the unit normal vector to constant--$\tphi$ hypersurfaces as
\begin{equation}
n_\mu \equiv \frac{\partial_\mu \tphi}{\sqrt{X}},
\qquad
X \equiv -\,g^{\mu\nu}\partial_\mu \tphi\,\partial_\nu \tphi,
\label{eq:normal_def}
\end{equation}
which satisfies $n_\mu n^\mu=-1$. The orientation is fixed once a monotonic
background branch for $\tphi$ has been selected.

The contribution proportional to $\Box f$ introduces derivative couplings that
mix scalar and geometric quantities. After integration by parts, it combines
with the derivative term generated by the transformed $c_2(\phi)$ operator. In the
minimal effective cuscuton construction the degeneracy relation fixes
$c_2(\tphi)$ in terms of the variation of the gravitational coupling
$f(\tphi)$~\cite{Mylova:2023ddj}. In our sign conventions for the Weyl
transformation and extrinsic curvature, this relation is
\begin{equation}\label{eq:c_2_cancellation}
c_2(\tphi) = -\, f_{,\tphi} .
\end{equation}

This choice removes redundant higher-derivative operators and selects the minimal cuscuton effective theory. Substituting this relation \cref{eq:c_2_cancellation} into \cref{eq:EF_intermediate} yields the simplified Einstein-frame action (see Appendix~\ref{app:useful_formulae})
\begin{align}\label{eq:einstein_frame_action}
S = \Lambda^2\int d^4x\,\sqrt{-g}\,\bigg\{
\frac{1}{2} R
+ \frac{3}{4}\frac{\partial_\mu f\partial^\mu f}{f^2}
-\, \frac{f_{,\tphi}}{f}\,
\sqrt{X}\,K
-\Lambda^2 \frac{ V(\tphi)}{f^2}
&+\Lambda\, \frac{c_1(\tphi)}{f^{3/2}}\, \sqrt{X}
 \bigg\}\nonumber
\\&\qquad+S_{\rm sm}( \tphi, g_{\mu\nu}, \Psi_i).
\end{align}
The term quadratic in derivatives of $f$ might naively be interpreted as a
kinetic term. It instead arises from the mixing of the cuscuton operators with
gravity under the Weyl transformation. At the level of the equations of motion
the apparent dynamics cancel and the cuscuton remains non-propagating. We make
this explicit in \cref{sec:background_cosmology}, where the background equations
reduce to an algebraic constraint for $\bar f$, and in
\cref{sec:pert_newtonian}, where the linear perturbation $\delta f$ is likewise
constrained.

\subsection{Einstein-frame ADM decomposition}

Given that the cuscuton action mixes quantities intrinsic to the constant--$\tphi$ hypersurfaces with four-dimensional spacetime curvature, it is convenient to perform a $3+1$ Arnowitt--Deser--Misner (ADM) decomposition of spacetime using these hypersurfaces as the foliation. In this decomposition, using cosmic time $t$, the spacetime metric is written as
\begin{equation}
ds^{2} = -N^{2} dt^{2}
+ \gamma_{ij}\left(dx^{i}+N^{i}dt\right)\left(dx^{j}+N^{j}dt\right),
\label{eq:ADM_metric}
\end{equation}
where $\gamma_{ij}$ is the induced spatial metric, $N$ is the lapse function, and $N^{i}$ is the shift vector.

 With this convention, in unitary gauge where $\tphi=\tphi(t)$ and the
monotonic branch has $\dot\tphi>0$, the normal is past directed and takes the
explicit ADM form
\begin{equation}
n_\mu = (N,0,0,0),
\qquad
n^\mu = \left(-\frac{1}{N},\,\frac{N^{i}}{N}\right).
\label{eq:normal_ADM}
\end{equation}

The induced metric on the spatial hypersurfaces is given by the projector
\begin{equation}
\gamma_{\mu\nu} = g_{\mu\nu} + n_\mu n_\nu ,
\label{eq:projector}
\end{equation}
and the extrinsic curvature, which measures the embedding of the hypersurfaces into the full spacetime, is defined as
\begin{equation}
K_{\mu\nu}
\equiv
\gamma_\mu^{\;\alpha}\gamma_\nu^{\;\beta}
\nabla_\alpha n_\beta .
\label{eq:extrinsic_def}
\end{equation}
In ADM variables this reduces to
\begin{equation}
K_{ij}
=
-\frac{1}{2N}
\left(
\dot{\gamma}_{ij}
- D_i N_j
- D_j N_i
\right),
\qquad
K \equiv \gamma^{ij}K_{ij},
\label{eq:Kij_ADM}
\end{equation}
where $D_i$ denotes the covariant derivative compatible with $\gamma_{ij}$ and a dot represents $\partial_t$.

With these definitions the Gauss--Codazzi relations give the standard $3+1$
decomposition of the four-dimensional Ricci scalar
\cite{Arnowitt:1962hi,Gourgoulhon:2007ue},
\begin{equation}
R
=
R^{(3)}
+ K_{ij}K^{ij}
- K^2
+ 2\nabla_\mu\!\left(n^\mu K - a^\mu\right),
\label{eq:R_ADM}
\end{equation}
where $R^{(3)}$ is the Ricci scalar of the spatial metric $\gamma_{ij}$, and
\begin{equation}
a^\mu \equiv n^\nu\nabla_\nu n^\mu,
\label{eq:acceleration_def}
\end{equation}
is the four-acceleration of the congruence of curves normal to the hypersurfaces.

Substituting the ADM decomposition outlined above into the Einstein-frame action \cref{eq:einstein_frame_action} and discarding boundary terms, the action can be written in terms of the spatial hypersurface quantities as
\begin{align}
S = \int d^4x\,\sqrt{-g}\,\bigg\{
\frac{\Lambda^2}{2}\Big(
R^{(3)} + K_{ij}K^{ij} &- K^2\Big)-\,\Lambda^2 \frac{f_{,\tphi}}{f}\,
\sqrt{X}\,K
\nn\\
&+\Lambda^2 \frac{3}{4}\frac{\partial_\mu f\,\partial^\mu f}{f^2}
-\Lambda^4 \frac{V(\tphi)}{f^2}
+\Lambda^3 \frac{c_1(\tphi)}{f^{3/2}} \sqrt{X}
\bigg\}
+S_{\rm sm} .
\label{eq:ADM_precanonical}
\end{align}
This is the decomposed action from which the Einstein equations can be consistently derived.


\subsection{Equations of motion}\label{sec:eom}

We now collect the full system of equations of motion required below. The cuscuton equation
is obtained by varying the covariant Einstein-frame action
\cref{eq:einstein_frame_action}, while the matter exchange equations follow from
the conformally coupled matter action, $S_{\rm sm}$. The gravitational equations are most
conveniently obtained from the ADM action \cref{eq:ADM_precanonical} by varying the
lapse, shift and induced spatial metric.

\subsubsection*{Cuscuton Klein--Gordon equation}

The Cuscuton's Klein--Gordon equation follows from variation of the Einstein-frame covariant action
\cref{eq:einstein_frame_action} with respect to the original field $\tphi$, or equivalently with
respect to $f(\tphi)$. Since the operator
$\sqrt{X}\,K$ contains second derivatives through $K=\nabla_\mu n^\mu$, the
corresponding Euler--Lagrange equation takes the generalised covariant form
\begin{equation}
\frac{\partial \mathcal{L}}{\partial \tphi}
-\nabla_\mu\!\left(
\frac{\partial \mathcal{L}}
{\partial(\nabla_\mu \tphi)}
\right)
+\nabla_\mu\nabla_\nu
\!\left(
\frac{\partial \mathcal{L}}
{\partial(\nabla_\mu\nabla_\nu \tphi)}
\right)
=0 ,
\end{equation}
where $\mathcal{L}$ is the Lagrangian density of \cref{eq:einstein_frame_action}.
Using the identity
\begin{equation}
\nabla_\mu\nabla_\nu \gamma^{\mu\nu}
=
n^\mu\nabla_\mu K
+
K^2
+
\nabla_\mu a^\mu ,
\end{equation}
with $n^\mu$ and $a^\mu$ defined in \cref{eq:normal_def} and \cref{eq:acceleration_def} respectively,
and rewriting the field equation in terms of
$f(\tphi)$, the scalar equation of motion becomes
\begin{align}
\left(
\frac{\Box f}{f}
-
\frac{\nabla_\mu f\nabla^\mu f}{f^2}
\right)
&+
\frac{2}{3}\Lambda^2
\left(
\frac{V_{,f}}
{f}
-
2\frac{V}{f^2}
\right)\nonumber
\\&-
\frac{2}{3}
\left[
\Lambda
\frac{c_1(\tphi)}
{\sqrt f\,f_{,\tphi}}
K
-
\left(
n^\mu\nabla_\mu K+K^2-\nabla_\mu a^\mu
\right)
\right]
=
-\frac{1}{3\Lambda^2}T^\mu_{\;\mu\,({\rm sm})}
 .
\label{eq:f_eom_final}
\end{align}

Here we have introduced the total Einstein-frame energy-momentum tensor of
the fluid sector,
\begin{equation}
T^{\mu\nu}_{({\rm sm})}
\equiv
\frac{2}{\sqrt{-g}}
\frac{\delta S_{\rm sm}}{\delta g_{\mu\nu}}
=
\sum_I T^{\mu\nu}_{(I)},
\end{equation}
where the sum runs over all fluid species, including radiation. For a cosmology containing pressureless matter and radiation the trace reduces to
\begin{equation}
T^\mu{}_{\mu\,({\rm sm})}=-\rho_m ,
\end{equation}
where $\rho_m$
 is the matter density, since the radiation contribution to the trace vanishes in the relativistic limit.

\subsubsection*{Matter sector}\label{sec:matter_sector}

The matter sector is treated as a collection of cosmological perfect fluids.
The energy-momentum tensor for each species $I$ is written as
\begin{equation}
T^{\mu\nu}_{(I)}
=
(\rho_I+p_I)\,u^\mu_{(I)}u^\nu_{(I)}
+
p_I g^{\mu\nu},
\end{equation}
where $u^\mu_{(I)}$ is the fluid four-velocity, $\rho_I$ is the energy density,
and $p_I$ is the pressure.

In the Einstein-frame, the conformal coupling of matter to the scalar field
induces an exchange of energy-momentum between the fluid sector and the
cuscuton. An additional effect that is important to keep track of in non-minimally coupled cosmologies is the exchange of energy-momentum between different coupled species. The Bianchi identity, together with the cuscuton equation
of motion, gives the exchange equation for each fluid species,
\begin{equation}
\nabla_\mu T^{\mu\nu}_{(I)}
=
-\frac{1}{2}
T^\alpha{}_{\alpha\,(I)}
\frac{\nabla^\nu f}{f}
+
J^\nu_{(I)},
\label{eq:matter_exchange_f}
\end{equation}
where $J^\nu_{(I)}$ denotes the energy-momentum transferred to species $I$
through scattering processes. 
For the cosmological epochs considered here, Thomson scattering of photons
off free electrons provides the dominant momentum exchange between the photon
and baryon fluids. For these interactions, conservation
of the total energy-momentum of  baryons and photons implies
\begin{equation}
J^\nu_{(b)}
=
-J^\nu_{(\gamma)}.
\label{eq:thomson_exchange}
\end{equation}
Projecting \cref{eq:matter_exchange_f} along the four-velocity of species $I$
gives the continuity equation
\begin{equation}\label{eq:continuity_eq_cov}
u_{\nu\,(I)}\nabla_\mu T^{\mu\nu}_{(I)}
=
-\frac{1}{2}
T^\alpha{}_{\alpha\,(I)}
u^\mu_{(I)}
\frac{\nabla_\mu f}{f}
+
u_{\nu\,(I)}J^\nu_{(I)},
\end{equation}
which determines the evolution of the fluid energy density. Projecting
orthogonally to $u^\mu_{(I)}$ gives the Euler equation,
\begin{equation}\label{eq:euler_eq_cov}
(\rho_I+p_I)a^\mu_{(I)}
+
h^\mu{}_\nu\nabla^\nu p_I
=
-\frac{1}{2}
T^\alpha{}_{\alpha\,(I)}
h^\mu{}_\nu
\frac{\nabla^\nu f}{f}
+
h^\mu{}_\nu J^\nu_{(I)},
\end{equation}
where
\begin{equation}
a^\mu_{(I)}
\equiv
u^\nu_{(I)}\nabla_\nu u^\mu_{(I)},
\qquad
h^\mu{}_\nu
\equiv
g^\mu{}_\nu
+
u^\mu_{(I)}u_{\nu\,(I)}
\end{equation}
denote, respectively, the fluid four-acceleration and the projector onto the
local rest frame of species $I$.
\footnote{At the homogeneous Friedmann--Robertson--Walker(FRW) background level the fluid four-velocity is
aligned oppositely to our past-directed normal to the constant-$\phi$
hypersurfaces, so \EqCorrection{$u^\mu=-n^\mu$},
$h^\mu{}_\nu=\gamma^\mu{}_\nu$, and the corresponding
four-accelerations vanish. These quantities differ once perturbations induce
peculiar velocities in the fluids, so we keep them separately defined
throughout this work.}

The exchange equations of baryons and photons from \cref{eq:matter_exchange_f}  therefore become
\begin{equation}
\label{eq:einstein_frame_matter_coupling}
\nabla_\mu T^{\mu\nu}_{(b)}
=
\frac{1}{2}\rho_b\frac{\nabla^\nu f}{f}
+
J^\nu_{(b)},
\qquad
\nabla_\mu T^{\mu\nu}_{(\gamma)}
=
J^\nu_{(\gamma)}
=
-J^\nu_{(b)}.
\end{equation}
For a general non-relativistic species, conservation of particle number,
together with $\rho_I=m_I n_I$, where $n_I$ is the number density of the fluid
and $m_I$ is its mass, allows the continuity equation
\cref{eq:continuity_eq_cov} to be written
as
\begin{equation}
u^\mu_{(I)}\nabla_\mu\ln m_I
=
-\frac{1}{2}
u^\mu_{(I)}\nabla_\mu\ln f,
\end{equation}
where we have used the fact that Thomson scattering produces no net
background energy transfer. The Einstein-frame particle masses therefore
scale as
\begin{equation}
m_I(\phi)
=
m_{I0}
\sqrt{\frac{f_0}{f(\phi)}}
\propto
f^{-1/2}.
\label{eq:mass_scaling_here}
\end{equation}
This mass evolution is particularly important for Thomson scattering because
the electron mass enters directly into the baryon-photon cross section.
Assuming that the fine-structure constant, $\alpha$, remains fixed, as is
consistent with current bounds on its variation
\cite{Uzan:2010pm,Planck:2014ylh},
\begin{equation}
\sigma_T
=
\frac{8\pi\alpha^2}{3m_e^2}
\propto f.
\end{equation}
The Thomson energy-momentum transfer terms
$J^\nu_{(b)}=-J^\nu_{(\gamma)}$ therefore acquire an explicit dependence on
the evolving conformal factor through $\sigma_T$.
Keeping these collision terms is therefore essential for consistently
following the momentum-transfer rate of baryons and photons, the time and width of
photon decoupling, and the resulting baryon and photon velocity
perturbations. We include this dependence using the HyRec recombination solver
\cite{Ali_Ha_moud_2011,Lee:2020obi} within \texttt{CLASS}, modifying the recombination code
to update the electron mass consistently with the background evolution of
$f(\phi)$.

The frame dictionary makes clear that this varying Einstein-frame mass is not
an additional observable. With $a_E=\sqrt f\,a_J$ and
$dt_E=\sqrt f\,dt_J$, conformal time is invariant and
\begin{equation}
 a_E=\sqrt f\,a_J,
 \qquad
 \mathcal H_E=\mathcal H_J+\frac12\frac{f'}{f},
 \qquad
 1+z=\frac{a_{J0}}{a_J}
 =\frac{a_{E0}}{a_E}\sqrt{\frac{f}{f_0}},
 \label{eq:frame_background_dictionary}
\end{equation}
where a prime denotes the derivative with respect to conformal
time. Every remaining weight follows from \cref{eq:conf_trans}: proper lengths
scale as $\ell_E=\sqrt f\,\ell_J$, so a quantity of mass dimension $d$ obeys
$Q_E=f^{-d/2}Q_J$. The matter and atomic quantities used by the recombination
calculation therefore obey
\begin{equation}
\begin{gathered}
 T_E=f^{-1/2}T_J,\qquad
 m_{e,E}=f^{-1/2}m_{e,J},\qquad
 \sigma_{T,E}=f\,\sigma_{T,J},\\[4pt]
 n_{e,E}=f^{-3/2}n_{e,J},\qquad
 \rho_E=f^{-2}\rho_J.
\end{gathered}
 \label{eq:frame_matter_dictionary}
\end{equation}
Consequently the differential optical depth is unchanged,
\begin{equation}
 a_E n_{e,E}\sigma_{T,E}=a_J n_{e,J}\sigma_{T,J},
 \label{eq:optical_depth_frame_invariance}
\end{equation}
and the ratio of an atomic binding energy to the photon temperature is also
frame invariant. The separation of a prediction into an 'expansion-history'
part and a 'mass-dependent recombination' part is therefore frame dependent;
only their combined effect on dimensionless observables is physical.

\subsubsection*{Einstein equations}

The gravitational field equations are obtained by variation of the action
with respect to the ADM variables $(N,N^i,\gamma_{ij})$. 
The resulting equations correspond to the Hamiltonian constraint,
momentum constraint, and evolution equations for the spatial metric.

The Hamiltonian and momentum constraints below are written in the preferred foliation adapted to the cuscuton, corresponding to unitary gauge, $\tphi=\tphi(t)$. For the spatial metric equation, however, we retain the form prior to imposing the unitary-gauge condition $D_i f=0$. This form is useful below when the equations are expressed in a generic perturbative gauge; its unitary-gauge specialization is obtained simply by setting $D_i f=0$.

The Hamiltonian constraint comes from variation with respect to the lapse
function $N$. In the equations below the cuscuton kinetic factor is written in
terms of $f$ alone, using
$
\partial_\mu f\,\partial^\mu f = -X f_{,\tphi}^{\,2}.
$
Variation of the lapse then yields
\begin{align}
\frac{1}{2\Lambda^2}
\left(
K_{ij}K^{ij}-K^2-R^{(3)}
\right)
=
&-\frac{3}{4\Lambda^2}
\frac{\partial_\mu f\partial^\mu f}{f^2}
-\frac{V}{f^2}
+
\frac{K}{\Lambda^2 f\sqrt{-\partial_\mu f\partial^\mu f}}
(\mathcal{L}_n f)^2
-\frac{1}{\Lambda^4}
T_{\mu\nu}n^\mu n^\nu,
\label{eq:ham_constraint}
\end{align}
where $\mathcal{L}_n f\equiv n^\mu\nabla_\mu f$ denotes the Lie derivative
of $f$ along the unit normal.
The momentum constraint, obtained by variation with respect to the shift vector $N^i$ is
\begin{align}
\frac{1}{\Lambda^{2}}\,D_{j}
\left(K^{j}_{\;i}-\delta^{j}_{\;i}K\right)
-
\frac{1}{\Lambda^2}\,
D_i\!\left(
\frac{\sqrt{-\partial_\mu f\partial^\mu f}}{f}
\right)
=
\frac{1}{\Lambda^{4}}T_{\mu\nu}n^\mu \gamma^\nu_{\;i}.
\label{eq:momentum_constraint}
\end{align}
The additional scalar contributions appearing in
Eqs.~(\ref{eq:ham_constraint}) and (\ref{eq:momentum_constraint})
originate from the $\sqrt{X_\phi}K$ interaction and represent the
modification of the gravitational constraints by the cuscuton.

The $i$--$j$ equation, obtained from variation of
\cref{eq:ADM_precanonical} with respect to $\gamma_{ij}$, is given by
\begin{align}
\frac{2}{N\sqrt{\gamma}\Lambda^2}(\partial_t-\mathcal{L}_{\vec N})\Pi^{ij}
=
-R^{(3)\,ij}+\frac12\gamma^{ij}R^{(3)}-C_N^{ij}+C_{K^2}^{ij}+C_{\rm mix}^{ij}
+C_{f}^{ij}+\,\frac{T_{(m)}^{ij}}{\Lambda^2},
\label{eq:ij_EL_compact}
\end{align}
where $\Pi^{ij}$ is the canonical momentum conjugate to the spatial metric
\begin{equation}
    \Pi^{ij}
\equiv \frac{\partial \mathcal{L}}{\partial \dot\gamma_{ij}}
=
-\frac{\Lambda^2}{2}\sqrt{\gamma}
\left(
K^{ij}-\gamma^{ij}K
-\frac{\sqrt{-\partial_\mu f\partial^\mu f}}{f}\,\gamma^{ij}
\right),
\label{eq:Pi_def}
\end{equation}

\noindent
with the individual contributions
\begin{align}
C_{N}^{ij}
&\equiv
-\left[
\frac{1}{N}\big(D^iD^jN-\gamma^{ij}D^2N\big)
\right],
\label{eq:C_N_def}
\\[4pt]
C_{K^2}^{ij}
&\equiv
\left[
\frac12\gamma^{ij}\big(K_{ab}K^{ab}-K^2\big)
-2\big(K^{i}{}_{k}K^{jk}-K K^{ij}\big)
\right],
\label{eq:C_K2_def}
\\[4pt]
C_{\rm mix}^{ij}
&\equiv
\left(
\frac{\sqrt{-\partial_\mu f\partial^\mu f}}{f}(2 \,K^{ij} -\,K\,\gamma^{ij}) -\frac{K}{f\sqrt{-\partial_\mu f\partial^\mu f}}\,D^i f\,D^j f
\right),
\label{eq:C_mix_def}
\\[4pt]
C_{f}^{ij} &\equiv \left[ \frac{3}{4f^2}\gamma^{ij} \big(\partial_\mu f\,\partial^\mu f\big) -\frac{3}{2f^2}\partial^{i}f\,\partial^{j}f \right] -\Lambda^2\,\frac{V}{f^2}\,\gamma^{ij} \nonumber\\ &\qquad\qquad\qquad +\Lambda\frac{c_1(\tphi)}{f^{3/2}f_{,\tphi}} \left( \gamma^{ij}\sqrt{-\partial_\mu f\partial^\mu f} +\frac{1}{\sqrt{-\partial_\mu f\partial^\mu f}}\,D^i f\,D^j f \right). \label{eq:C_f_def} 
\end{align}
With these equations we have all of the ingredients required to study the
cosmology at both the background and linear-perturbation levels, as we will do in
the next section.

\section{Cosmological background and perturbations}\label{sec:cuscuton_cosmology}

Having derived the covariant field equations, we now specialise the theory to a spatially homogeneous and isotropic cosmological background and to linear
scalar perturbations about it. From the equations presented in the preceding section,
it is apparent that, once the functional dependence of
\begin{equation}
\label{eq:free_functions}
    \frac{V(\phi)}{f^2(\phi)},
    \qq{and}
    \frac{c_1(\phi)}
    {\sqrt{f(\phi)}\,f_{,\phi}(\phi)},
\end{equation}
on $f(\phi)$ is fixed, one never has to work directly with the intrinsic
cuscuton field $\phi$. Provided that $f(\phi)$ is monotonic, the function $f(\phi)$ can itself be interpreted as the cuscuton
variable. In the interest of considering the theory as generally as possible
while remaining agnostic about the detailed form of too many functions, we
therefore work with $f(\phi)$ as the cuscuton field from here on.

\subsection{Background cosmology}\label{sec:background_cosmology}

At the background level, we work with a spatially flat FRW line element,
\begin{equation}
ds^2=-dt^2+a^2(t)\,d\mathbf x^2,
\end{equation}
and we denote background quantities with an overbar.

The first and second Friedmann equations, derived from
\cref{eq:ham_constraint} and the trace of \cref{eq:ij_EL_compact}, are
\begin{align}
H^2
=
\frac{1}{3\Lambda^2}\left(
-\frac{3\Lambda^2}{4}
\frac{\dot f^2}{f^2}
+\Lambda^4\frac{V}{f^2}
+3\Lambda^2 H\frac{\dot f}{f}
+\bar\rho
\right),
\label{eq:Friedmann_raw}
\\
3H^2 + 2\dot{H}
=
\left(
\frac{\ddot f}{f}
-\frac{1}{4}\frac{\dot f^2}{f^2}
+\Lambda^2\frac{V}{f^2}
-\Lambda\frac{c_1(\tphi)}{f^{3/2}f_{,\tphi}}\dot f
\right)
-\frac{\bar p}{\Lambda^2},
\label{eq:second_Friedmann_f}
\end{align}
where $\bar\rho \equiv \sum_I \rho_I= \bar{\rho}_m + \bar{\rho}_r$ and $\bar{p}\equiv \sum_I \bar{p}_I = \bar{p}_m + \bar{p}_r$ are the total energy density and pressure exerted by matter species and radiation.
The effective energy density and pressure of the cuscuton at the background
level are therefore

\begin{equation}\label{eq:eff_energy}
    \rho_f
    \equiv
    -\frac{3\Lambda^2}{4}\frac{\dot f^2}{f^2}
    +\Lambda^4\frac{V}{f^2}
    +3\Lambda^2 H\frac{\dot f}{f},
\end{equation}
and
\begin{equation}\label{eq:eff_pressure}
    p_f
    \equiv
    -\Lambda^2\frac{\ddot f}{f}
    +\frac{1}{4}\Lambda^2\frac{\dot f^2}{f^2}
    -\Lambda^4\frac{V}{f^2}
    +\Lambda^3\frac{c_1(\tphi)}{f^{3/2}f_{,\tphi}}\dot f .
\end{equation}
It is
often convenient to rewrite \cref{eq:Friedmann_raw} in the manifestly
positive form
\begin{equation}
\frac{\Lambda^2}{2}
\left(
\sqrt{6}\,H
-
\sqrt{\frac{3}{2}}\frac{\dot f}{f}
\right)^2
=
\Lambda^4\frac{V}{f^2}
+
\bar\rho,
\label{eq:un1_discuss}
\end{equation}
which makes the tying between the cuscuton and the geometry extremely apparent. The cuscuton acts as a modification to the Einstein-frame Hubble rate, while simultaneously sourcing it from the RHS of this equation through its potential energy.

The matter continuity equations, derived from \cref{eq:continuity_eq_cov},
\begin{equation}
\dot{\bar\rho}_m+3H\bar\rho_m
=
-\frac{1}{2}\bar\rho_m\frac{\dot f}{f},
\qquad
\dot{\bar\rho}_r+4H\bar\rho_r=0,
\label{eq:cont_discuss}
\end{equation}
show explicitly how the conformal coupling induces an energy exchange between
non-relativistic matter and the scalar sector, while radiation remains
uncoupled. The right-hand side of \cref{eq:cont_discuss} is the
Einstein-frame manifestation of the fact that particle masses scale as
$m\propto A(f)=f^{-1/2}$.

The scalar equation of motion,
\begin{equation}
\begin{split}
\sqrt{\frac{3}{2}}
\left(
\frac{\ddot f}{f}
-
\frac{\dot f^2}{f^2}
+
3H\frac{\dot f}{f}
\right)
&-
\sqrt{\frac{2}{3}}\Lambda^2
\left(
\frac{V_{,\tphi}}{f f_{,\tphi}}
-
2\frac{V}{f^2}
\right)
\\
&-
\sqrt{6}\Lambda
\frac{c_1(\tphi)}{\sqrt f\,f_{,\tphi}}H
-
\sqrt{6}\big(\dot H+3H^2\big)
=
-\sqrt{\tfrac{2}{3}}\frac{1}{2\Lambda^2}\,\bar\rho_m,
\end{split}
\label{eq:cuscback_discuss}
\end{equation}
contains the same structure as a canonical independent dynamical scalar field,
but once combined with the Friedmann
equation, \cref{eq:cuscback_discuss} does not introduce an independent
propagating degree of freedom for the background. Instead it enforces a
constraint that algebraically ties $\bar f$ to the matter content through the
functions $V(f)$ and $c_1(f)$. 

This can be obtained by differentiating \cref{eq:Friedmann_raw} and combining it with \cref{eq:cuscback_discuss} into the purely
algebraic relation
\begin{equation}
\bar\rho_m+\bar\rho_r
+
\Lambda^4
\left(
\frac{V}{f^2}
+
\frac{V_{,\tphi}}{f f_{,\tphi}}
\right)
+
\frac{\sqrt{6}}{2}\Lambda^3
\frac{c_1(\tphi)}{\sqrt f\,f_{,\tphi}}
\left(
\sqrt{6}H
-
\sqrt{\frac{3}{2}}\frac{\dot f}{f}
\right)
=0,
\label{eq:un2_discuss}
\end{equation}
so that the combination $\sqrt{6}H-\sqrt{3/2}\,\dot f/f$ is the
only place where the time derivative of the scalar appears. Eliminating this
combination using \cref{eq:un1_discuss} yields the Einstein-frame cuscuton constraint equation
in the presence of matter and radiation,
\begin{equation}
\Lambda^4\frac{V}{f^2}+\bar\rho
=
\frac{\Lambda^4 f f_{,\tphi}^{\,2}}{3c_1^2(\tphi)}
\left(
\frac{\bar\rho}{\Lambda^4}
+
\frac{V}{f^2}
+
\frac{V_{,\tphi}}{f f_{,\tphi}}
\right)^2.
\label{eq:un3_discuss}
\end{equation}
 Given $\bar\rho_m(a)$ and $\bar\rho_r(a)$ from
\cref{eq:cont_discuss}, \cref{eq:un3_discuss} determines $\bar f(a)$ algebraically, after
which $H(a)$ follows from \cref{eq:un1_discuss}. In this way the background
cosmology can be evolved without ever treating $f$ as an independently
propagating field.\footnote{Since \cref{eq:un3_discuss} is obtained by squaring
\cref{eq:un2_discuss}, it may contain an additional algebraic solution.
Below we retain only solutions that also satisfy the original unsquared
relation \cref{eq:un2_discuss}.}

\subsection{Perturbations}
\label{sec:pert_newtonian}

Although the preferred foliation makes unitary gauge natural, we present the
perturbation equations in Newtonian gauge to display the metric potentials and
the fifth force mediated by the constrained scalar. The corresponding unitary-
and synchronous-gauge systems are collected in
\cref{sec:perts_in_other_gauges}.

Working in Newtonian gauge and expanding to linear order about a
spatially-flat FRW background,
\begin{equation}
ds^2 = -\left(1+2\Phi\right)dt^2
+a^2(t)\left(1-2\Psi\right)d\mathbf{x}^2,
\end{equation}
the
cuscuton is decomposed as a standard scalar,
\begin{equation}
f(t,\mathbf{x})=\bar f(t)+\delta f(t,\mathbf{x}).
\end{equation}
We treat the matter sector as a set of perfect fluids with density contrasts $\delta_I \equiv \frac{\delta\rho_I}{\bar\rho_I}$ and transform to Fourier space using the convention
\begin{equation}
    Y(\mathbf{x},t)
    = \int\!\frac{d^3k}{(2\pi)^3}\,Y(\mathbf{k},t)\,e^{i\mathbf{k}\cdot\mathbf{x}},
\end{equation}
with $k^2 \equiv \delta^{ij} k_i k_j$. For each species we introduce a
scalar velocity potential $v_I$ through $u^{(I)}_i = a\,\partial_i v_I$, and we
define the cosmic time (Ma--Bertschinger \cite{Ma:1995ey}) velocity divergence
\begin{equation}
\theta_I \equiv -\frac{k^2}{a}\,v_I
\qquad\Rightarrow\qquad
T^{0}{}_{i\,(I)} = -a^2(\bar\rho_I+\bar p_I)\,\frac{i k_i}{k^2}\,\theta_I  .
\end{equation}
The $0$--$0$ and $0$--$i$ Einstein equations, from \cref{eq:ham_constraint} and \cref{eq:momentum_constraint}, in Newtonian gauge may be written as
\begin{align}
\frac{k^{2}}{a^{2}}\,\Psi
&\;+\; \dot{\Psi}\,
\left(
3H-\frac{3}{2}\frac{\dot{\bar f}}{\bar f}
\right)\;+\;
\frac{\delta f}{2\bar f}
\left[
\frac{k^2}{a^2}
+\Lambda^2
\left(
\frac{V_{,f}}
{\bar f}
-
2\frac{V}{\bar f^2}
\right)
\right]
\nn\\
&\;+\;
\left(
\frac{\dot{\delta f}}{\bar f}
-
\frac{\dot{\bar f}}{\bar f^2}\delta f
\right)
\left(
\frac{3}{2}H
-
\frac{3}{4}\frac{\dot{\bar f}}{\bar f}
\right)
+\Lambda^2\frac{V}{\bar f^2}\Phi
\;=\;-\; \sum_{I}
\frac{\bar\rho_I}{2\Lambda^{2}}\Bigl(\delta_{I} + 2\Phi \Bigr),
\label{eq:Einstein00_Newtonian}
\end{align}
and
\begin{align}
&\frac{k^2}{a^2}
\Bigg[
\dot\Psi
+\left(
H-\frac{1}{2}\frac{\dot{\bar f}}{\bar f}
\right)\Phi
+\frac{1}{2}
\left(
\frac{\dot{\delta f}}{\bar f}
-
\frac{\dot{\bar f}}{\bar f^2}\delta f
\right)
\nn\\
&\qquad\qquad
-\frac{\delta f}{4\bar f}
\left(
6H
-
3\frac{\dot{\bar f}}{\bar f}
+
2\Lambda
\frac{c_1(\bar{\tphi})}
{\sqrt{\bar f}\,f_{,\tphi}}
\right)
\Bigg]
=
-\frac{k^2}{2a\Lambda^2}
\sum_I(\bar\rho_I+\bar p_I)v_I .
\label{eq:Einstein0i_Newtonian}
\end{align}
From \cref{eq:ij_EL_compact}, the trace of the $i$--$j$ equation gives 
\begin{align}
&\ddot{\Psi}
+3H\dot{\Psi}
+\left(H
-\frac12\frac{\dot f}{\bar f}\right)\dot{\Phi}
+\frac12\frac{d^2}{dt^2}
\left(\frac{\delta f}{\bar f}\right)
+\frac{d}{dt}
\left(\frac{\delta f}{\bar f}\right)
\left(
-\frac{\Lambda}{2}
\frac{c_1(\bar{\tphi})}
{\sqrt{\bar f}\,f_{,\tphi}}
+\frac34\frac{\dot f}{f}
\right)
\nn\\
&\quad
+(\Psi-\Phi)
\left[
\frac12\Lambda^2
\frac{V}{\bar f^2}
-\frac{\bar\rho_r}{6\Lambda^2}
+\frac{k^2}{3a^2}
-3H^2
-2\dot H
-\frac34\Lambda
\frac{c_1(\bar{\tphi})}
{\sqrt{\bar f}\,f_{,\tphi}}
\frac{\dot f}{f}
+\frac{d^2}{dt^2}\ln\bar f
+\frac34
\left(\frac{\dot f}{f}\right)^2
\right]
\nn\\
&\quad
+\frac{\Psi+\Phi}{12}
\left(
6\Lambda^2
\frac{V}{\bar f^2}
-\frac{2\bar\rho_r}{\Lambda^2}
-3\Lambda
\frac{c_1(\bar{\tphi})}
{\sqrt{\bar f}\,f_{,\tphi}}
\frac{\dot f}{f}
\right)
\nn\\
&\quad
+\frac{\delta f}{\bar f}
\left[
-\Lambda H
\frac{c_1(\bar{\tphi})}
{\sqrt{\bar f}\,f_{,\tphi}}
+\frac34H\frac{\dot f}{f}
-\frac34\frac{d^2}{dt^2}\ln\bar f
\right.
-3H^3\frac{f}{\dot f}
+\ddot H\frac{f}{\dot f}
\nn\\
&\quad\left.
+\frac{f}{2\dot f}
\left(
\Lambda
\frac{c_1(\bar{\tphi})}
{\sqrt{\bar f}\,f_{,\tphi}}
\frac{d^2}{dt^2}\ln\bar f
-\frac{d^3}{dt^3}\ln\bar f
\right)
+H\frac{f}{\dot f}
\left(
\Lambda^2
\frac{V}{\bar f^2}
-\frac{\bar\rho_r}{\Lambda^2}
+\dot H
+\frac{d^2}{dt^2}\ln\bar f
\right)
\right]
=
\frac{\delta p_r}{2\Lambda^2}.
\label{eq:grouped_constraint_like_eq_logf}
\end{align}
and here the traceless part of the $i$--$j$ Einstein equation gives
\begin{equation}
    \Psi-\Phi=\frac{1}{2}\pi,
\end{equation}
where the total scalar anisotropic-stress perturbation of the fluid
sector is defined by
\begin{equation}
    \pi \equiv
    -\frac{3a^2}{\Lambda^2 k^2}
    \left(
        \frac{k_i k^j}{k^2}
        -\frac{1}{3}\delta_i{}^j
    \right)
    \delta T^i{}_j .
\end{equation}
The cuscuton does not contribute anisotropic stress at linear order. We neglect
the fluid anisotropic stress only in the late-time, quasi-static analytic
reduction below, so that $\pi=0$ and hence
\begin{equation}
    \Psi-\Phi=0 .
    \label{eq:slip_Newtonian}
\end{equation}
Anisotropic stress is retained in the full numerical runs shown in \cref{sec:tracking-case}.

For pressureless matter, the conformal coupling
\cref{eq:conf_trans} induces the modified continuity equation \cref{eq:continuity_eq_cov}, which at linear order gives
\begin{align}
\dot{\delta}_m+\theta_m-3\dot{\Psi}
&=
-\frac{1}{2}
\frac{d}{dt}\left(\frac{\delta f}{\bar f}\right),
\label{eq:matter_cont_Newtonian}
\end{align}
and the corresponding baryon and cold dark-matter Euler equations are
\begin{align}
\dot{\theta}_b+2H\theta_b-\frac{k^2}{a^2}\Phi
&=
\frac{1}{2}
\left(
\frac{\dot{\bar f}}{\bar f}\theta_b
-
\frac{k^2}{a^2}\frac{\delta f}{\bar f}
\right)
+
\frac{4\bar\rho_\gamma}{3\bar\rho_b}\,
n_e\sigma_T(f)
\left(\theta_\gamma-\theta_b\right),
\label{eq:baryon_euler}
\\
\dot{\theta}_c+2H\theta_c-\frac{k^2}{a^2}\Phi
&=
\frac{1}{2}
\left(
\frac{\dot{\bar f}}{\bar f}\theta_c
-
\frac{k^2}{a^2}\frac{\delta f}{\bar f}
\right),
\label{eq:cdm_euler}
\end{align}
as given by \cref{eq:euler_eq_cov}. The final term in the baryon Euler equation
parameterises the baryon-photon Thomson scattering during tight coupling, where electrons and protons are too energetic to combine, causing them to efficiently exchange momentum with the photons that scatter off their charges. As
discussed in \cref{sec:matter_sector}, it acquires an explicit dependence on
$f$ through the cuscuton-induced electron-mass scaling.

The linearised cuscuton equation, \cref{eq:f_eom_final}, in Newtonian gauge can be written as
\begin{equation}
\begin{aligned}
&
\frac{d^2}{dt^2}\left(\frac{\delta f}{\bar f}\right)
+
3H
\frac{d}{dt}\left(\frac{\delta f}{\bar f}\right)
-
\left[
2\Lambda H
\frac{\bar c_1}
{\sqrt{\bar f}\,\bar f_{,\tphi}}
+
\frac{4}{3}\Lambda^2
\left(
\frac{\bar{V}_{,\tphi}}
{\bar f\,\bar f_{,\tphi}}
-
2\frac{\bar{V}}{\bar f^2}
\right)+\frac{2}{3}\frac{k^2}{a^2}
\right]\Phi
\\
&
+
\left(
2\Lambda
\frac{\bar c_1}
{\sqrt{\bar f}\,\bar f_{,\tphi}}
+
12H
-
3\frac{\dot{\bar f}}{\bar f}
\right)\dot{\Psi}
+
\Bigg[
\frac{k^2}{a^2}
-
2\Lambda H
\bar f
\frac{d}{df}
\left(
\frac{c_1}
{\sqrt{f}\,f_{,\tphi}}
\right)
-
\frac{2}{3}\Lambda^2
\bar f
\frac{d}{df}
\left(
\frac{V_{,\tphi}}
{f\,f_{,\tphi}}
-
2\frac{V}{f^2}
\right)
\\
&\qquad\qquad\qquad
-
\frac{2k^2\bar f}{3a^2\dot{\bar f}}
\left(
4H + \Lambda \frac{c_1}{\sqrt{\bar{f}} \bar{f},_\phi}
\right)
\Bigg]
\frac{\delta f}{\bar f}
+
2\ddot{\Psi}+
\left(
2H-\frac{\dot{\bar f}}{\bar f}
\right)\dot{\Phi}
=
-\frac{\bar\rho_m}{3\Lambda^2}
\left(\delta_m+2\Phi\right).
\end{aligned}
\label{eq:delta_f_Newtonian}
\end{equation}
The first-order cuscuton constraint follows by subtracting
\cref{eq:delta_f_Newtonian} from
\cref{eq:grouped_constraint_like_eq_logf} and substituting for the remaining
$\dot{\delta f}$ terms using \cref{eq:Einstein00_Newtonian}. This gives


\begin{align}
0={}&\sqrt{\frac{3}{2}}\frac{\delta f}{\bar f}\Bigg\{\Bigg(H-\frac12\frac{\dot{\bar f}}{\bar f}\Bigg)\Bigg[\frac{\dot{\bar\rho}}{\Lambda^2}
+\frac32\Lambda\frac{\bar c_1}{\sqrt{\bar f}\,\bar f_{,\phi}}\Big(2\dot H-\frac{d^2}{dt^2}\ln\bar f\Big)\Bigg]
-2\frac{k^2}{a^2}H\left(2H-\frac{\dot{\bar f}}{\bar f}+\frac{\Lambda}{2}\frac{\bar c_1}{\sqrt{\bar f}\,\bar f_{,\phi}}\right)
\nonumber\\
&\qquad\quad-\Lambda\frac{\bar c_1}{\sqrt{\bar f}\,\bar f_{,\phi}}\frac{\dot{\bar f}}{\bar f}\Bigg[\frac{\bar\rho}{\Lambda^2}
-\frac38\Big(2H-\frac{\dot{\bar f}}{\bar f}\Big)\Big(6H-3\frac{\dot{\bar f}}{\bar f}+2\Lambda\frac{\bar c_1}{\sqrt{\bar f}\,\bar f_{,\phi}}\Big)\Bigg]\Bigg\}
\nonumber\\
&-\sqrt{\frac{3}{2}}\frac{\dot{\bar f}}{\bar f}\Big(H-\frac12\frac{\dot{\bar f}}{\bar f}+\frac{\Lambda}{2}\frac{\bar c_1}{\sqrt{\bar f}\,\bar f_{,\phi}}\Big)\Big(\frac{\delta\rho}{\Lambda^2}+2\frac{k^2}{a^2}\Psi\Big)
-\sqrt6\,\frac{\dot{\bar f}}{\bar f}\Big(H-\frac12\frac{\dot{\bar f}}{\bar f}\Big)\frac{k^2}{a^2}\Psi ,
\end{align}

where we are once again using compressed notation for the total fluid density perturbation $\delta \rho \equiv \sum_I \delta \rho_I$.
This equation is the linear counterpart of the background constraint
\cref{eq:un3_discuss}: $\delta f$, and hence $\delta\phi$, is determined
algebraically by the remaining perturbations and the background expansion.

The numerical implementation used below evolves the synchronous-gauge system
only. The Newtonian-gauge equations in this section are retained for analytic
interpretation and for defining the quasi-static potentials; we do not claim an
independent numerical Newtonian-gauge validation.

\section{Fifth forces in the quasi-static limit}\label{sec:quasistatic}

The conformal coupling between the cuscuton and the matter sector modifies
the growth of density perturbations in two distinct ways. It changes the
friction term in the matter evolution equation through the time dependence
of the Einstein-frame particle masses, and it introduces an additional force
through the spatial perturbation $\delta f$. In this section we isolate these
effects in the quasi-static limit and express their combined contribution as
an effective gravitational coupling governing the growth of pressureless
matter perturbations.

\subsection{The quasi-static approximation}

The quasi-static approximation, whose regime of validity in modified gravity
cosmologies has been examined in detail in \cite{Sawicki:2015zya}, applies to
Fourier modes whose physical wavelength is much smaller than the Hubble radius,
\begin{equation}
    \frac{k^2}{a^2}\gg H^2 .
    \label{eq:qs_ordering}
\end{equation}
On these scales, the characteristic time variation of a cosmological
perturbation is assumed to be set by the background expansion rate,
\begin{equation}
    |\dot X|\lesssim H|X|,
\end{equation}
where $X$ denotes any of the scalar perturbations entering the gravitational
and cuscuton field equations. Spatial gradient terms therefore dominate over
terms containing only time derivatives.
The time derivatives in the
matter continuity and Euler equations are however retained, since they govern
the growth of $\delta_m$ itself.
We apply this approximation during matter domination, when
$\bar\rho_r\ll\bar\rho_m$ and the radiation perturbation may be neglected.
 We also neglect the fluid anisotropic
stress, so that $\Phi=\Psi$.

Under these assumptions, the scalar perturbation equation (\ref{eq:delta_f_Newtonian}) reduces to
\begin{equation}
\frac{k^2}{a^2}\frac{\delta f}{\bar f}\left[
1
-\frac{2}{3}\frac{\bar{f}}{\dot{\bar f}}\left(4H + \Lambda\frac{\bar c_1}{\sqrt{\bar{f}}\,\bar{f}_{,\tphi}}\right)
\right] - \frac{2}{3}\frac{k^2}{a^2}\Phi
=
-\frac{\bar\rho_m}{3\Lambda^2}\delta_m .
\label{scalarqs}
\end{equation}
All time derivatives of $\delta f/\bar f$ and of the metric potentials drop out at this order: the
only surviving terms are the spatial gradient terms and the matter source, so the cuscuton perturbation is
fixed algebraically by $\Phi$ and $\delta_m$ alone. The corresponding quasi-static part of the $0$--$0$
Einstein equation (\ref{eq:Einstein00_Newtonian}) is
\begin{equation}
\frac{k^{2}}{a^{2}}
\left(
\Psi+\frac{1}{2}\frac{\delta f}{\bar f}
\right)
=
-\frac{\bar\rho_m}{2\Lambda^{2}}\,\delta_m .
\label{einsteinqs}
\end{equation}
The trace of the spatial Einstein equation (\ref{eq:grouped_constraint_like_eq_logf}) contains no
$k^2\,\delta f$ terms, so its leading quasi-static content is only $k^2(\Psi-\Phi)=0$, which is already
implied by \cref{eq:slip_Newtonian}; it supplies no further relation. Using $\Phi=\Psi$ to eliminate the
metric potential between \cref{scalarqs} and \cref{einsteinqs} gives the algebraic constraint
\begin{equation}
\frac{k^2}{a^2}\frac{\delta f}{\bar f}
=
\frac{\bar\rho_m}{\Lambda^2}\,
\frac{
\displaystyle \frac{\dot{\bar f}}{\bar f}
}{
\displaystyle
4H+\Lambda\frac{\bar c_1}{\sqrt{\bar f}\,\bar f_{,\tphi}}-2\frac{\dot{\bar f}}{\bar f}
}\,\delta_m ,
\qquad
\frac{k^2}{a^2}\Phi
=
-\frac{\bar\rho_m}{2\Lambda^2}\,
\frac{
\displaystyle 4H+\Lambda\frac{\bar c_1}{\sqrt{\bar f}\,\bar f_{,\tphi}}-\frac{\dot{\bar f}}{\bar f}
}{
\displaystyle
4H+\Lambda\frac{\bar c_1}{\sqrt{\bar f}\,\bar f_{,\tphi}}-2\frac{\dot{\bar f}}{\bar f}
}\,\delta_m .
\label{eq:deltaf_constraint_qs}
\end{equation}
The cuscuton perturbation therefore remains non-dynamical in the quasi-static regime: $\delta f$ is fixed
algebraically by the matter-density perturbation and the background evolution, with no dependence on
$\dot\delta_m$. The $0$--$i$ Einstein equation (\ref{eq:Einstein0i_Newtonian}) is then satisfied
identically at leading quasi-static order once the continuity equation below is used, so the reduced
system contains exactly the two modes of $\delta_m$.

The pressureless-matter continuity and Euler equations reduce to
\begin{align}
    \dot{\delta}_m+\theta_m &\simeq 0 ,
    \label{eq:cont_qs}
    \\[4pt]
    \dot{\theta}_m+2H\theta_m-\frac{k^2}{a^2}\Phi
    &\simeq
    \frac{1}{2}
    \left(
    \frac{\dot{\bar f}}{\bar f}\theta_m
    -
    \frac{k^2}{a^2}\frac{\delta f}{\bar f}
    \right).
    \label{eq:euler_qs}
\end{align}
Here the conformal source in the full continuity equation is subleading because
the algebraic solution gives $\delta f/\bar f=\mathcal O[(aH/k)^2\delta_m]$;
it is therefore consistently discarded at the leading quasi-static order,
while the mass-drift contribution to the Euler friction is retained.
Differentiating \cref{eq:cont_qs} and using \cref{eq:euler_qs} gives
\begin{equation}
\ddot{\delta}_m
+
\left(
2H-\frac{1}{2}\frac{\dot{\bar f}}{\bar f}
\right)\dot{\delta}_m
=
-\frac{k^2}{a^2}\Phi
+
\frac{1}{2}\frac{k^2}{a^2}\frac{\delta f}{\bar f}.
\label{eq:growth_force_qs}
\end{equation}
The first term on the right-hand side is the usual force sourced by the
Newtonian potential, while the second arises from the conformal coupling
to the cuscuton perturbation.
Using $\Phi=\Psi$, we eliminate the metric potential from
\cref{eq:growth_force_qs} with the quasi-static Einstein equation
\cref{einsteinqs}, so that the right-hand side becomes
$\bar\rho_m\delta_m/(2\Lambda^2)+(k^2/a^2)\,\delta f/\bar f$. The remaining cuscuton perturbation is
then eliminated using the algebraic constraint \cref{eq:deltaf_constraint_qs}. This yields
the closed growth equation
\begin{equation}
\ddot{\delta}_m
+
\left(
2H-\frac{1}{2}\frac{\dot{\bar f}}{\bar f}
\right)\dot{\delta}_m
=
\frac{\bar\rho_m\delta_m}{2\Lambda^2}
\left(
1
+
\frac{
\displaystyle 2\frac{\dot{\bar f}}{\bar f}
}{
\displaystyle
4H
+\Lambda
\frac{\bar c_1}{\sqrt{\bar f}\,\bar f_{,\tphi}}
-2\frac{\dot{\bar f}}{\bar f}
}
\right).
\label{eq:growth_qs}
\end{equation}
Because the cuscuton perturbation and metric potential equations \cref{eq:deltaf_constraint_qs} involve $\delta_m$ but not $\dot\delta_m$, the direct fifth force contribution associated with $\delta f$ enters only through the gravitational source on the right-hand side of \cref{eq:growth_qs}; it does not contribute to the friction term. The friction term is nevertheless modified separately by the background evolution of $f$. In particular, the time dependence of the Einstein-frame particle masses changes the coefficient of $\dot{\delta}_m$, motivating the definition
\begin{equation}\label{eq:eff_hubble_friction}
    H_{\rm eff}
    \equiv
    H-\frac{1}{4}\frac{\dot{\bar f}}{\bar f}.
\end{equation}
The conformal coupling therefore affects
the growth of structure both through this modified friction term and through the gravitational source on the right-hand side of
\cref{eq:growth_qs}. These two effects have the same microscopic origin
but enter the matter-growth equation in physically distinct ways.
Using the normalization $\Lambda=\MPL$ stated in
\cref{sec:cuscuton}, we define the effective gravitational coupling by
\begin{equation}
\ddot{\delta}_m
+
\left(
2H-\frac{1}{2}\frac{\dot{\bar f}}{\bar f}
\right)\dot{\delta}_m
=
4\pi G_{\rm eff}\bar\rho_m\delta_m .
\label{eq:growth_geff}
\end{equation}
It follows that
\begin{equation}
G_{\rm eff}
=
G_N
\left(
1
+
\frac{
\displaystyle 2\frac{\dot{\bar f}}{\bar f}
}{
\displaystyle
4H
+\Lambda
\frac{\bar c_1}{\sqrt{\bar f}\,\bar f_{,\tphi}}
-2\frac{\dot{\bar f}}{\bar f}
}
\right),
\label{eq:eff_newt}
\end{equation}
where $G_N\equiv(8\pi\MPL^2)^{-1}$ is Newton's constant. This is the standard
way of packaging a scale-independent modification of linear growth
\cite{Amendola:2007rr,Bellini:2014fua}, and in scalar-tensor theories with a
propagating scalar the analogous coefficient is built from the conformal
coupling strength rather than, as here, from the constraint
\cite{Amendola:1999er,Joyce:2014kja}. We report the correction below as the
ratio $(G_{\rm eff}-G_N)/G_N$, which does not depend on the constant
normalization of $f$ discussed at the start of \cref{sec:cuscuton}.

Although the cuscuton carries no propagating scalar degree of freedom, its
constraint equation makes $\delta f$ respond to the matter-density
perturbation. The resulting term in the Euler equation acts as a fifth
force and modifies the gravitational coupling relevant for linear
structure growth. At leading quasi-static order, $G_{\rm eff}$ depends only on background quantities and is
scale independent. The correction can either enhance or suppress the
matter source depending on the background evolution and the sign of the
denominator in \cref{eq:eff_newt}. In the limit
$\dot{\bar f} = 0$, both the additional friction and the
fifth force correction vanish, and the standard general relativity prediction
$G_{\rm eff}= G_N$ is recovered.

The denominator in \cref{eq:eff_newt} also identifies a potential singular
response of the constraint system. The benchmark solutions used below do not
cross this locus over the plotted range.

\section{The cuscuton as dark energy}\label{sec:cuscuton_dark_energy}

Because the cuscuton is tied to the preferred cosmological foliation, it
naturally evolves on Hubble time scales. At late times this is the behavior
required of a dark energy component. The absence of a propagating scalar removes
the usual light-particle interpretation and its associated scalar initial conditions
\cite{Afshordi:2006ad,Afshordi:2007yx}. However, the radiative stability of the full
effective action, and the size of the operators neglected beyond the order
retained here, are separate EFT questions of the kind addressed by
power counting in derivatively coupled cosmologies
\cite{Burgess:2009ea,Adshead:2017srh,Babic:2019ify}. As shown above, the
cuscuton adjusts to the ambient matter density rather than independently rolling
down a potential, allowing even simple functional choices for its potential and couplings to differ sharply
from standard quintessence \cite{Ratra:1987rm,Ferreira:1997hj,Caldwell:1997ii,Copeland:2006wr}.

In principle, the Einstein-frame action \cref{eq:einstein_frame_action} contains three
free functions of the cuscuton, $c_1(\phi)$, $V(\phi)$, and $f(\phi)$.
However, the background and perturbation equations derived above show that these
functions enter only through the two combinations defined in \cref{eq:free_functions}.
The separate choice of the variable $\phi$ is therefore redundant and specifying the two
combinations above as functions of $f$ fixes the physical content of the model, without
requiring an independent choice of $f(\phi)$.

We therefore consider the power-law ansatz
\begin{equation}
\frac{V(\phi)}{f^2(\phi)}
=
V_0\,f^{\lambda},
\qquad
\frac{c_1(\phi)}
{\sqrt{f(\phi)}\,f_{,\phi}(\phi)}
=
C_0\,f^{\frac{1}{\zeta}},
\label{eq:power_law_ansatz}
\end{equation}
where $V_0$, $\lambda$, $C_0$ and $\zeta$ are constants. With this choice the cuscuton constraint
\cref{eq:un3_discuss} can be solved directly for $f$, after which the background expansion,
matter coupling, and perturbation equations are fully determined. In the
tracking case $C_0$ fixes the present-day cuscuton density while $V_0$ is a
redundant normalization. In the sign-switching case these roles are reversed:
$V_0$ fixes the late-time density and only the magnitude of $C_0$ is redundant,
since a rescaling of $f$ absorbs $|C_0|$ but not its sign. We therefore set
$V_0=1$ in the tracking case and $C_0=-1$ in the sign-switching case. Once the present-day dark energy density is matched,
$\lambda$ and $\zeta$ are the respective shape parameters of the two limiting
families.

\begin{table}[t]
\centering
\begin{tabular}{llll}
\toprule
Case & Defining limit & Shape parameter & Density normalization \\
\midrule
Tracking & $\zeta\to\infty$ & $\lambda$ & $C_0$ ($V_0=1$) \\
Sign-switching & $\lambda=0$ & $\zeta$ & $V_0$ ($C_0=-1$) \\
\bottomrule
\end{tabular}
\caption{Parameter bookkeeping for the two benchmark cases. The coefficient in
parentheses is fixed to the value shown in the numerical examples.}
\label{tab:branch_parameters}
\end{table}

The following subsections focus on two benchmark limits of \cref{eq:power_law_ansatz}. The
first is the constant-$C_0$ tracking regime, corresponding to
$\zeta\to\infty$, in which the cuscuton potential contribution follows the ambient
cosmological density before approaching a late-time vacuum. The second is the
$\lambda=0$ sign-switching regime, in which the effective potential is flat while the characteristic Einstein-frame cuscuton kinetic
coefficient varies with $f$. This second case naturally realizes an effective dark energy
density that can be negative at high density and positive at late times.

For each case, we compare the intrinsic cuscuton equation of state
\begin{equation}
    w_f \equiv \frac{p_f}{\rho_f},
\end{equation}
with the effective equation of state inferred after accounting for the Einstein-frame
reshuffling of energy between non-relativistic matter and the cuscuton sector, given by 
\begin{equation}
w_{\rm eff}(a)
\;=\;
\frac{p_f}{\rho_f+\left(\sqrt{\dfrac{f(\phi_0)}{f(\phi)}}-1\right)\dfrac{\rho_{m0}}{a^{3}}}
\;=\;
\frac{w_f}{1+\left(\sqrt{\dfrac{f(\phi_0)}{f(\phi)}}-1\right)\dfrac{\rho_{m0}}{a^{3}\rho_f}},
\label{eq:weff_def_here}
\end{equation}
where $\rho_{m0}$ is the present-day matter density and $\phi_0$ is the present
value of the field. This definition packages the background into the
constant-mass matter plus effective dark energy form commonly used in
observational analyses \cite{DES:2025upx,Brout:2022vxf,DESI:2025fii,DESI:2025zgx}.
We also track the associated particle mass evolution $m(a)/m_0$ relative to the
present-day value $m_0$, and the quasi-static
fifth force correction $(G_{\rm eff}-G_N)/G_N$ defined in \cref{eq:eff_newt}.

The numerical examples shown below were computed using a modified version of
\texttt{CLASS}\footnote{The base versions of the code are \texttt{CLASS} v3.3.4 with HyRec2020} \cite{Lesgourgues:2011re,Diego_Blas_2011,Lee:2020obi} in which the Einstein-frame particle masses (e.g. proton and electron masses) are
evolved consistently with the background cuscuton.  \Cref{tab:benchmark_parameters} summarises the parameter choices taken for each model variant. The cosmological parameters shared by this model and $\Lambda$CDM are fixed to the
Planck-constrained values for $\Lambda$CDM \cite{Planck:2018vyg}. We fix $C_0$
so that the present cuscuton fraction
$\Omega_f\equiv\rho_f/(3\Lambda^2H_0^2)$ matches the Planck $\Lambda$CDM
dark energy fraction in the tracking case, and fix $V_0$ analogously in the
sign-switching case. We evolve the linear perturbations together with the
background by integrating the synchronous-gauge system in
\Cref{sec:synchronous_gauge}. Gauge-invariant CMB spectra are obtained directly
from this evolution, while the Newtonian metric potentials shown below are
constructed from the corresponding metric gauge transformation. No
Newtonian-gauge evolution system was implemented. The matter power spectrum is
built from the synchronous-gauge density contrast.

\begin{table}[t]
\centering
\renewcommand{\arraystretch}{1.25}
\caption{Parameters used for the benchmark cosmologies. Standard cosmological
parameters are fixed to the Planck 2018 TT,TE,EE+lowE+lensing best-fit
$\Lambda$CDM values \cite{Planck:2018vyg}.}
\label{tab:benchmark_parameters}

\begin{tabular}{lcc}
\hline
Parameter & \multicolumn{2}{c}{Value} \\

\hline

$H_0\,[\mathrm{km\,s^{-1}\,Mpc^{-1}}]$
    & \multicolumn{2}{c}{$67.36$} \\

$\omega_b$
    & \multicolumn{2}{c}{$0.02237$} \\

$\omega_{\rm cdm}$
    & \multicolumn{2}{c}{$0.1200$} \\

$n_s$
    & \multicolumn{2}{c}{$0.9649$} \\

$\ln(10^{10}A_s)$
    & \multicolumn{2}{c}{$3.044$} \\

$\tau_{\rm reio}$
    & \multicolumn{2}{c}{$0.0544$} \\

$\Omega_k$
    & \multicolumn{2}{c}{$0$} \\

$\sum m_\nu\,[\mathrm{eV}]$
    & \multicolumn{2}{c}{$0$} \\

\hline
    & Tracking & Sign-switching \\
\hline

$\lambda$
    & $-10^{3} \leq \lambda \leq -10^{2}$
    & $0$ \\

$\zeta$
    & $\infty$
    & $10^{-4} \leq \zeta \leq 10^{-2}$ \\

$V_0$
    & $1$
    & Fixed by $\Omega_{f0}=0.68613$ \\

$C_0$
    & Fixed by $\Omega_{f0}=0.68613$
    & $-1$ \\

\hline
\end{tabular}
\end{table}

\subsection{\texorpdfstring{Tracking cuscutons ($\zeta \to \infty$)}{Tracking cuscutons}}
\label{sec:tracking-case}

We first consider the constant kinetic coefficient case of the power-law ansatz.
The behaviour it produces is the cuscuton analogue of the tracker and scaling
solutions familiar from quintessence
\cite{Ferreira:1997hj,Zlatev:1998tr,Steinhardt:1999nw}, with the difference that
the cuscuton is fixed by a constraint rather than rolling. The ansatz is
\begin{equation}
    \frac{V(\tphi)}{f^2(\tphi)}
    =
    V_0 f^\lambda(\tphi),
    \qquad
    \frac{c_1(\tphi)}
    {\sqrt{f(\tphi)}\,f_{,\tphi}(\tphi)}
    =
    C_0 .
\end{equation}
With the convention used in \cref{eq:power_law_ansatz}, this corresponds to the
$\zeta\to\infty$ limit. The background cuscuton constraint
\cref{eq:un3_discuss} then reduces to a quadratic equation in $f^\lambda$,
\begin{equation}\label{eq:quad_equation}
    a\,f^{2\lambda} + b\,f^\lambda + c = 0,
\end{equation}
with coefficients
\begin{align}
a
=
\frac{V_0^{\,2}}{C_0^{\,2}}
\left(
\frac{\lambda}{\sqrt{3}}+\sqrt{3}
\right)^2,
\;\;
b
=
\frac{V_0}{C_0^{\,2}}
\left[
\frac{2\bar\rho}{\Lambda^4}
\left(
1+\frac{\lambda}{3}
\right)
-C_0^{\,2}
\right],
\qq{and}
c
=
\frac{1}{C_0^{\,2}}
\left(
\frac{1}{3}\frac{\bar\rho^2}{\Lambda^8}
-
C_0^{\,2}\frac{\bar\rho}{\Lambda^4}
\right).
\end{align}
When the cuscuton's effective energy density is dominated by potential energy, $\bar\rho\ll \Lambda^4 V_0 f^\lambda$, the nontrivial solution to \cref{eq:quad_equation} is
\begin{equation}
f_{\rm vac}
=
\left[
\frac{C_0^{\,2}}
{V_0
\left(
\frac{\lambda}{\sqrt{3}}+\sqrt{3}
\right)^2}
\right]^{1/\lambda},
\end{equation}
and the corresponding Hubble rate is
\begin{equation}
H_{\rm vac}
=
\frac{\Lambda}{\sqrt{3}}
\sqrt{V_0 f_{\rm vac}^{\lambda}}
=
\frac{\Lambda}{3}\,
\frac{C_0}
{\left|
\frac{\lambda}{3}+1
\right|}.
\end{equation}
The tracking case therefore interpolates between an early phase in which the cuscuton's evolution is determined by the dominant background density and a late-time vacuum case with
approximately constant dark energy density. Somewhat counterintuitively, this
vacuum energy is independent of the potential amplitude $V_0$ and is entirely
determined by the kinetic coefficient $C_0$ for a given $\lambda$. The relevant
parameters governing the cosmological behaviour in this regime are therefore
$\lambda$ and $C_0$, with $C_0$ setting the late-time contribution $\Omega_f$.

At early times, the solution lies in the density-dominated regime of \cref{eq:quad_equation}.
This corresponds to the regime in which the
$C_0$ corrections to the quadratic constraint are small,
\begin{equation}
    \frac{C_0^2}{\left|1+\frac{\lambda}{3}\right|}
    \ll
    \frac{\rho}{\Lambda^4}.
\end{equation}
In this limit,
\cref{eq:un3_discuss} gives
\begin{equation}\label{eq:tracker_Scaling_f}
\frac{\Lambda^4 V_0 f^\lambda}{\bar\rho}
\simeq
-\frac{1}{\lambda+3},
\end{equation}
so that $f(\bar\rho)\propto \bar\rho^{1/\lambda}$, showing the potential contribution tracks the ambient cosmological density.

In the matter-dominated epoch, $\bar\rho\simeq\bar\rho_m$, this relation gives
\begin{equation}\label{eq:fscalingdensity}
    \frac{\dot f}{f}
    =\frac{1}{\lambda}
    \frac{\dot{\bar\rho}_m}{\bar\rho_m}.
\end{equation}
Substituting this into the Einstein-frame matter continuity equation
\cref{eq:cont_discuss} to eliminate $f$ gives
\begin{equation}
\bar\rho_m \propto a^{\frac{-3}{1+1/(2\lambda)}}.
\end{equation}
Ensuring the matter density scales approximately as expected for non-relativistic particles therefore requires $|\lambda| \gg 1$ in this area of the parameter space. For $|\lambda|\gg1$, the Einstein-frame matter coupling \cref{eq:einstein_frame_matter_coupling} evolves
only weakly even while the effective potential tracks the background density. This slow
evolution of $f(\phi)$ causes the particle mass drift and the quasi-static fifth force correction
to be small.

Conversely, substituting \cref{eq:fscalingdensity} into
\cref{eq:cont_discuss} to instead eliminate $\rho_m$ gives
\begin{equation}\label{eq:f_H_scaling}
    \frac{\dot f}{f}
    =
    -\frac{3H}{\lambda+\frac{1}{2}},
\end{equation}
for the scaling of $f$ with the Hubble rate. Inserting this scaling into the effective cuscuton energy density \cref{eq:eff_energy}
and using the leading high-density Friedmann relation
$3\Lambda^2H^2\simeq\bar\rho_m$ from \cref{eq:Friedmann_raw} (and using $\dot f / f \ll H$ when $|\lambda|\gg 1$ from \cref{eq:f_H_scaling}), gives
\begin{equation}
    \rho_f
    =
    \Lambda^4 V_0 f^\lambda
    -
    \left[
        \frac{3}{\lambda+\frac{1}{2}}
        +
        \frac{9}{4\left(\lambda+\frac{1}{2}\right)^2}
    \right]
    \bar\rho_m .
\end{equation}
For $|\lambda|\gg1$, using \cref{eq:tracker_Scaling_f}, this becomes
\begin{equation}
    \rho_f
    =-
        \frac{4}{\lambda}
    \bar\rho_m .
\end{equation}
Thus the cuscuton energy density itself inherits the tracking behaviour of
$f$ and makes up an energy contribution suppressed
by powers of $1/\lambda$. The sign of this energy density is determined by the slope of the cuscuton's potential. Here we focus on the $\lambda < 0$ case, corresponding to the potential decreasing with $f(\phi)$ and the energy density remaining positive.

\begin{figure}
    \centering
    \includegraphics[width=\linewidth]{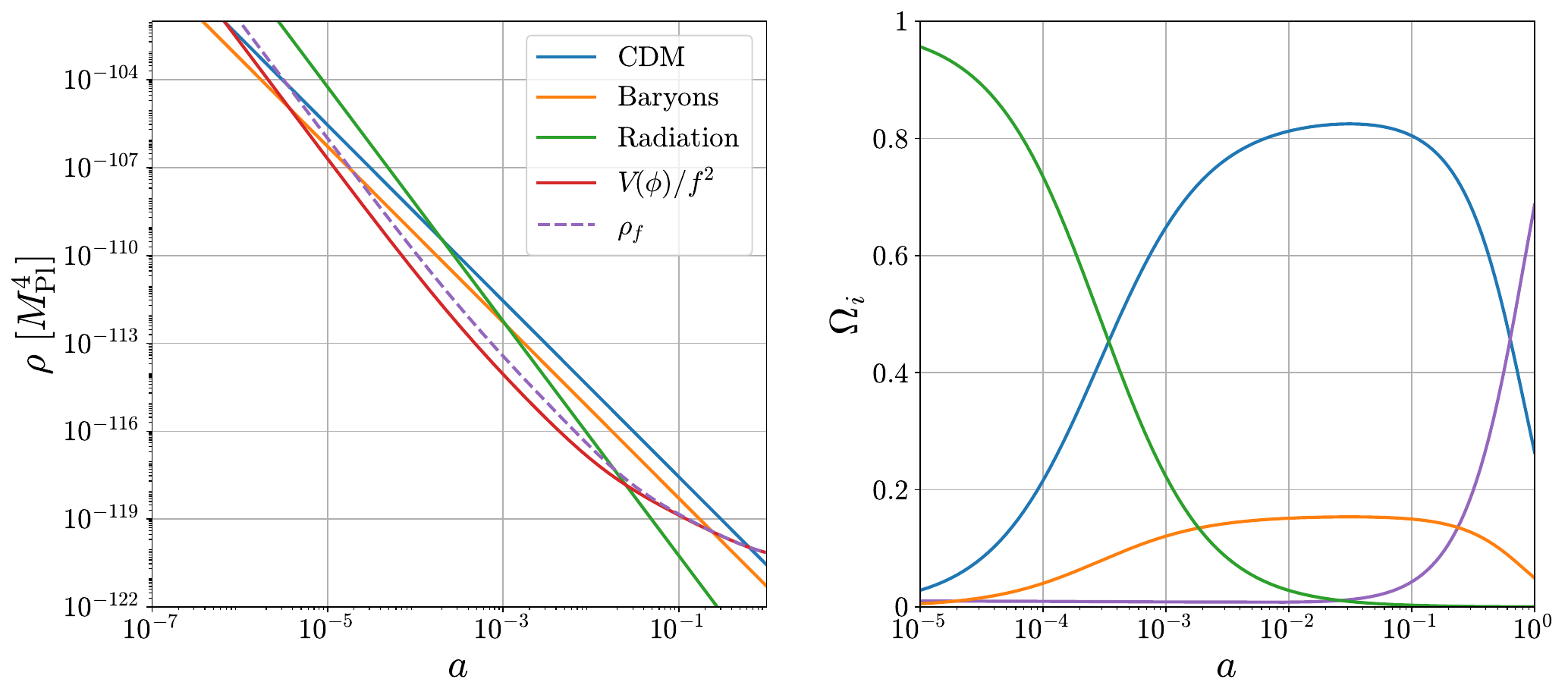}
    \caption{Background cosmology for the tracking case, shown in terms of absolute
    density scales on the left and fractional densities on the right. Here $\zeta\to\infty$ and $\lambda=-5\times10^{2}$, with $\Lambda=M_{\rm Pl}$, $V_0=1$ and
$C_0=2.0\times10^{-58}$.}
    \label{fig:tracking_background}
\end{figure}

This behaviour is shown in \cref{fig:tracking_background}. At early times the cuscuton
contribution follows the dominant cosmological energy scale rather than behaving as a pure
cosmological constant.
At late times the solution approaches the constant density vacuum regime, redshifting slower than the dominant matter species, and begins to act
as a dark energy component.

\begin{figure}
    \centering
    \includegraphics[width=\linewidth]{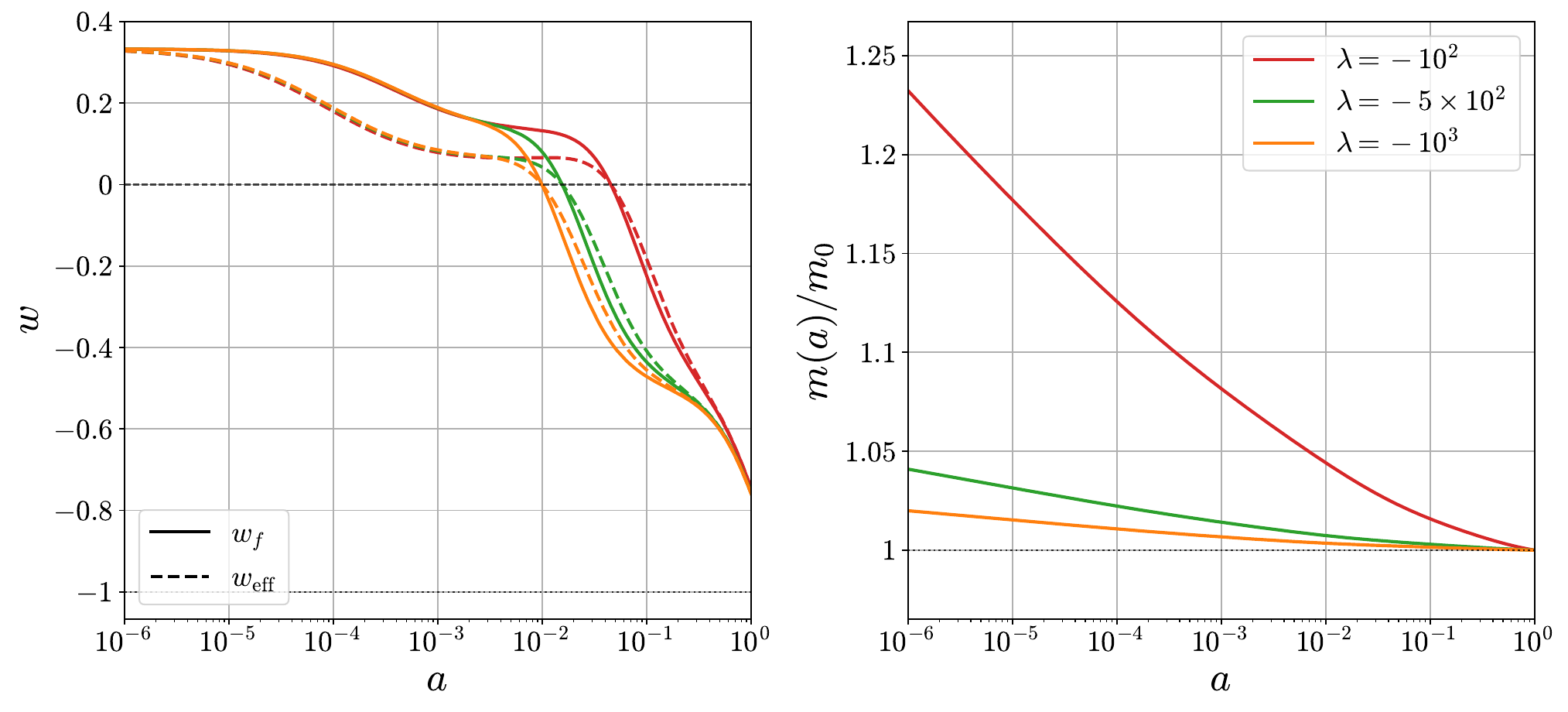}
    \caption{\textbf{Left:} Background equation of state as a function of scale factor for the
    tracking case. Solid lines show the intrinsic cuscuton equation of state
    $w_f=p_f/\rho_f$, while dashed lines show the effective equation of state defined
    in \cref{eq:weff_def_here}. \textbf{Right:} Relative difference in
    Einstein-frame particle masses compared to today's value $m_0$. Here $\zeta\to\infty$ and $\lambda=-10^{2},-5\times10^{2},-10^{3}$ in red, green and
orange respectively, with $\Lambda=M_{\rm Pl}$,
$V_0=1$ and $C_0=3.9\times10^{-59},\,2.0\times10^{-58},\,4.0\times10^{-58}$
respectively.}
    \label{fig:tracking_eos_mass}
\end{figure}

The equation of state and mass evolution of non-relativistic particles are shown in \cref{fig:tracking_eos_mass}. The
intrinsic equation of state $w_f$ and the effective equation of state $w_{\rm eff}$ are close
but not identical, because the latter includes the Einstein-frame reshuffling between the
cuscuton sector and non-relativistic matter as shown in \cref{eq:weff_def_here}. Changing $\lambda$ shifts the time at which the
solution leaves the tracking regime and approaches the negative-pressure vacuum
case, but has little effect on the equation of state reached today in the
displayed models. This is in contrast to a standard quintessence scalar, whose
late-time evolution and equation of state are directly governed by the slope of
its potential. The near-independence of the present-day $w_f$ on $\lambda$ is
an empirical feature of these normalized benchmarks; we do not infer a general
analytic result from it.

Changing $\lambda$ does more than just changing the departure from the tracker. As expected from \cref{eq:f_H_scaling}, it also changes how much $f(\phi)$ evolves and therefore how much coupled particle masses deviate throughout the cosmic history. The right panel of \cref{fig:tracking_eos_mass} shows that the associated mass
drift is controlled by $|\lambda|$, and that it is not small for the benchmarks
plotted here. Near recombination the plotted
models give
$
    \frac{\Delta m}{m_0} \simeq 8\%,\, 1.4\%,\, 0.7\%,
$
for $\lambda=-10^{2},-5\times10^{2},-10^{3}$ respectively, while over the full
scale-factor interval displayed in \cref{fig:tracking_eos_mass} the excursion
reaches $23\%$, $4\%$ and $2\%$. These are at or above the percent-level
changes around recombination that current data already constrain
\cite{Planck:2014ylh,Hart:2019dxi,Schoneberg:2024ynd}. We will see in the next
section that decreasing $|\lambda|$ and increasing the mass variation produces
visibly large departures in the benchmark CMB spectra. Quantifying the allowed
range requires a likelihood analysis.

The tracking case therefore gives a controlled example of a cuscuton sector that follows the cosmological background at early times and becomes dark energy-like at late times. Its observable effects are consequently expected to come from the drift of the particle masses together with the background evolution and the constraint-driven behaviour of the metric perturbations.

\subsection{\texorpdfstring{Sign-switching cuscutons ($\zeta \ll 1$)}{Sign-switching cuscutons}}
\label{sec:sign-switching-case}

We now consider the complementary case in which the effective potential contribution is
constant, while the cuscuton coefficient retains a non-trivial dependence on $f$. This is
obtained from the power-law ansatz by taking
\begin{equation}
    \frac{V(\phi)}{f^2(\phi)}=V_0,
    \qq{and}
    \frac{c_1(\phi)}
    {\sqrt{f(\phi)}\,f_{,\phi}(\phi)}
    =
    C_0 f^{1/\zeta}(\phi).
\end{equation}
In this convention the exponent appearing in the cuscuton coefficient is
large when $\zeta$ is small, so $\zeta$ controls how weakly the background field evolves as the
ambient density redshifts.

Substituting this choice into the algebraic background constraint \cref{eq:un3_discuss} gives
\begin{equation}
\Lambda^4V_0+\bar\rho
=
\frac{\Lambda^4}
{3C_0^{\,2}\bar f^{2/\zeta}}
\left(
\frac{\bar\rho}{\Lambda^4}
+
3V_0
\right)^2 .
\end{equation}
This can be solved directly for the background field,
\begin{equation}
\bar f^{2/\zeta}
=
\frac{
\left(
\frac{\bar\rho}{\Lambda^4}
+
3V_0
\right)^2
}
{
3C_0^{\,2}
\left(
\frac{\bar\rho}{\Lambda^4}
+
V_0
\right)
}.
\end{equation}
At high density, $\bar\rho\gg\Lambda^4V_0$, this reduces to
\begin{equation}
\bar f^{2/\zeta}
\simeq
\frac{\bar\rho}{3C_0^{\,2}\Lambda^4},
\end{equation}
and therefore
\begin{equation}
\frac{d\ln\bar f}{d\ln\bar\rho}
=
\frac{\zeta}{2}.
\end{equation}
Thus, for small positive $\zeta$, the cuscuton tracks the dominant background density only
weakly.

In a matter-dominated epoch, $\bar\rho\simeq\bar\rho_m$, the high-density relation gives
\begin{equation}\label{eq:f_scaling_small_zeta}
\frac{\dot{\bar f}}{\bar f}
=
\frac{\zeta}{2}
\frac{\dot{\bar\rho}_m}{\bar\rho_m}.
\end{equation}
Combining this with the Einstein-frame matter continuity equation (\ref{eq:cont_discuss})
gives
\begin{equation}
\frac{\dot{\bar f}}{\bar f}
=
-\frac{6\zeta}{4+\zeta}H,
\end{equation}
and hence
\begin{equation}
\bar\rho_m
\propto
a^{-12/(4+\zeta)}.
\end{equation}
For $\zeta\ll1$ the standard dust scaling is recovered. The effective cuscuton energy density is given by \cref{eq:eff_energy}. Using the
matter-tracker solution, this becomes
\begin{equation}
\label{eq:lambda0_cusc_energy_density}
\bar{\rho}_f
\simeq
\Lambda^4V_0
-
\frac{3\zeta(8+5\zeta)}
{16(1+\zeta)^2}
\bar\rho_m .
\end{equation}
Therefore, for positive $\zeta$, the high-density tracker contribution is negative,
while the constant term $\Lambda^4V_0$ eventually drives the effective energy density
positive. In the slow-roll limit this reduces to
\begin{equation}
\frac{\bar{\rho}_f}{\bar\rho_m}
\simeq
-\frac{3}{2}\zeta .
\end{equation}
The magnitude of the negative tracker contribution is therefore suppressed by the same small
parameter that controls the rolling of $\bar f$. Its sign is also set by
$\zeta$; for positive $\zeta$ the cuscuton supplies a negative contribution to
the high-density energy budget.

\begin{figure}
    \centering
    \includegraphics[width=\linewidth]{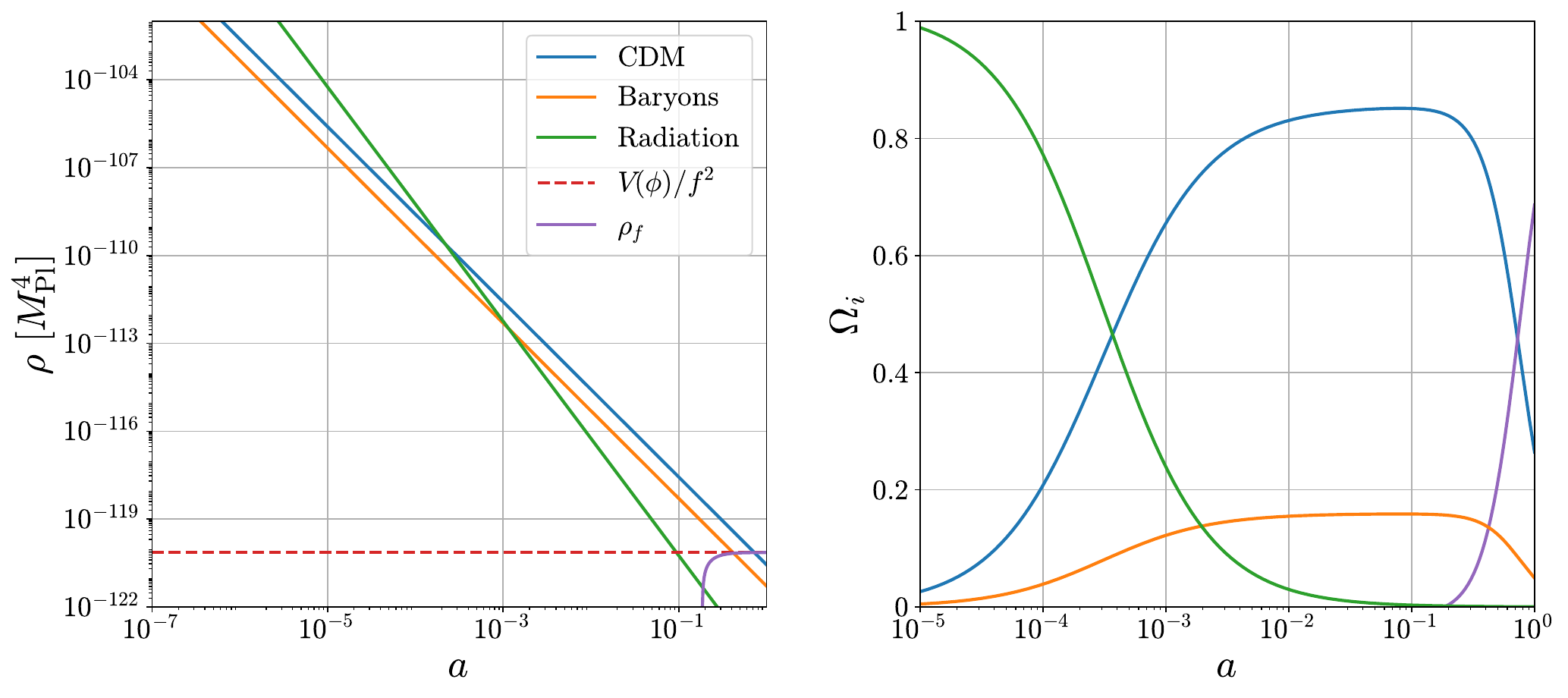}
    \caption{Background cosmology for the sign-switching case in terms of absolute density scales on the left and
    fractional densities on the right. Here $\lambda=0$ and $\zeta=10^{-2}$, with $\Lambda=M_{\rm Pl}$, $C_0=-1$ and
$V_0=7\times10^{-121}$.
}
    \label{fig:lambda_0_background}
\end{figure}

The sign change in the cuscuton energy density occurs when
\begin{equation}
\bar\rho_m^{\,+}
\simeq
\frac{16(1+\zeta)^2}
{3\zeta(8+5\zeta)}
\Lambda^4V_0 ,
\end{equation}
where $\bar\rho_m^{\,+}$ denotes the matter density at which $\rho_f$ crosses zero. 
Thus smaller $\zeta$ suppresses the negative tracker contribution and moves the zero-crossing
to higher density, and hence to higher redshift.

In the vacuum limit, $\bar\rho\ll\Lambda^4V_0$, $\bar{f}$ approaches the constant value
\begin{equation}
\bar f_{\rm vac}
=
\left(
\frac{3V_0}{C_0^{\,2}}
\right)^{\zeta/2},
\end{equation}
and
\begin{equation}
\rho_f
\simeq
\Lambda^4V_0,
\qquad
p_f
\simeq
-\Lambda^4V_0 .
\end{equation}
In the late-time case we have
\begin{equation}
H_{\rm vac}
=
\frac{\Lambda}{\sqrt{3}}\sqrt{V_0},
\end{equation}
therefore the cuscuton transitions from a small negative tracker contribution
at high density to a positive cosmological constant-like component at late
times. Because \cref{fig:lambda_0_background} uses logarithmic, nonnegative
density axes, it displays the onset of the positive part but not the negative
part of $\rho_f$ itself; the sign switch follows from
\cref{eq:lambda0_cusc_energy_density}. Unlike the alternate tracking solutions considered above, where the
late-time vacuum scale is set by $C_0$, in the present sign-switching case the constant
potential contribution $V_0$ directly controls the asymptotic dark energy density.

\begin{figure}
    \centering
    \includegraphics[width=\linewidth]{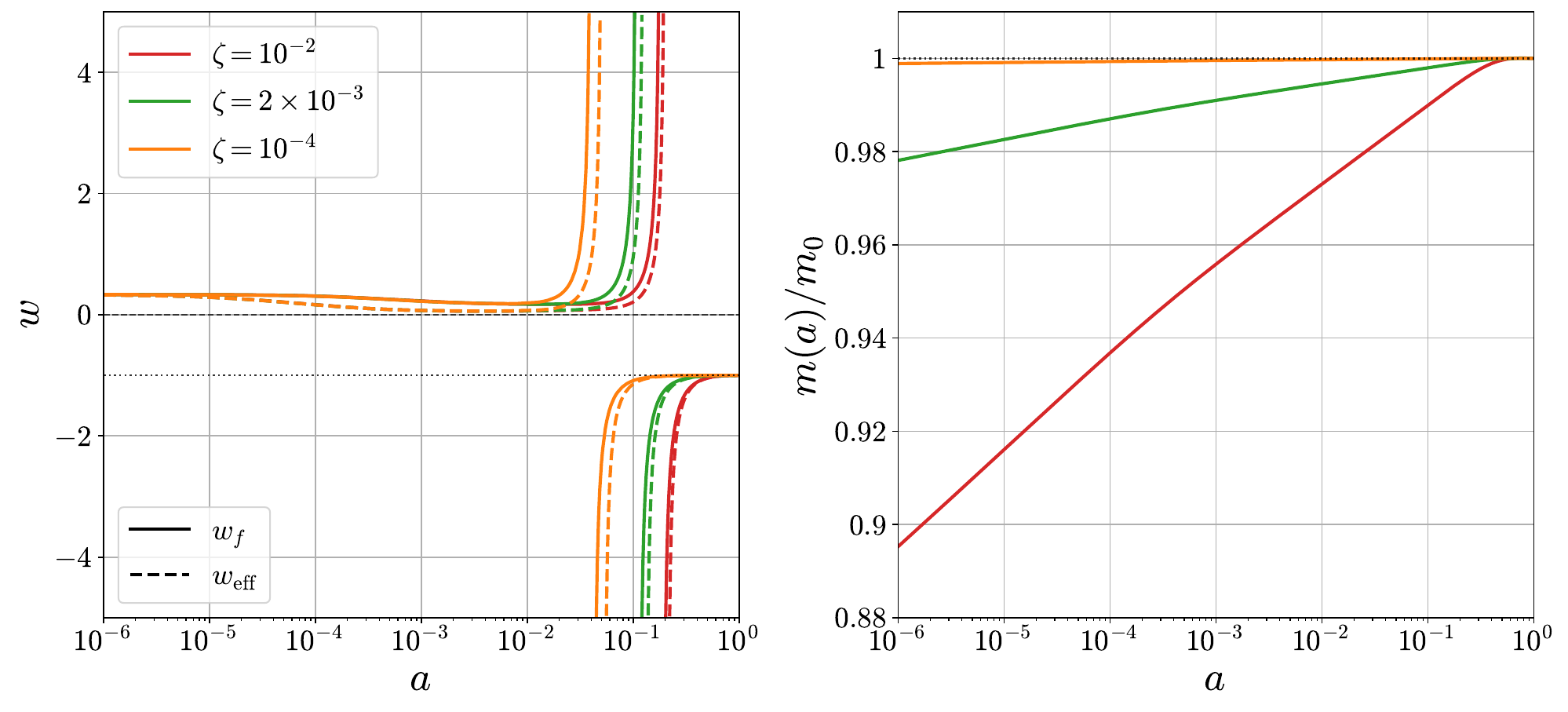}
    \caption{\textbf{Left:} Background equation of state vs scale factor for the cuscuton,
    with dashed and solid lines corresponding to the effective equation of state, defined in
    \cref{eq:weff_def_here}, and intrinsic, $p_f/\rho_f$, equations of state respectively.
    \textbf{Right:} Einstein-frame particle mass evolution relative to today's value $m_0$.
    Here $\lambda=0$ and $\zeta=10^{-2},2\times10^{-3},10^{-4}$ in red, green and orange respectively, with $\Lambda=M_{\rm Pl}$,
$C_0=-1$ and $V_0=7\times10^{-121}$ in all cases.}
    \label{fig:lambda_0_eos_mass}
\end{figure}


This negative-to-positive transition is closely related to phenomenological
sign-switching cosmological-constant models
\cite{Akarsu:2021fol,Akarsu:2025gwi,Bouhmadi-Lopez:2026ckz,Akarsu:2026anp,Gokcen:2026pkq}.
In those models the matter density parameter $\Omega_m$ is tightly constrained, hindering their ability to improve on the Hubble tension, because raising $H_0$ at fixed CMB acoustic scale requires the Hubble rate to be lower at higher redshifts. This in turn requires lowering $\Omega_m$
which BAO and supernova data disfavour \cite{Toda:2024ncp}. The sign-switch in the
present scenario behaves differently because the same rolling of $f$ that
triggers it also makes the Einstein-frame particle masses evolve according to
\cref{eq:mass_scaling_here}. Since $f(\phi)$ decreases with time
(\cref{eq:f_scaling_small_zeta}), particle masses increase with time (\ref{eq:mass_scaling_here}), so
$m(a)/m_0<1$ in the past. The CMB-era matter density is therefore smaller than
would be inferred from the present-day experiments, assuming constant masses, which may allow the sign-switching cusucton to exploit  the degeneracy between $H_0$ and $\Omega_m$ in CMB measurements while still satisfying late-time constraints on the matter density.
Whether this
has observable consequences requires a frame-consistent likelihood analysis and the
analysis presented here does not establish an improved fit to cosmological data.

\begin{figure}
    \centering
    \includegraphics[width=\linewidth]{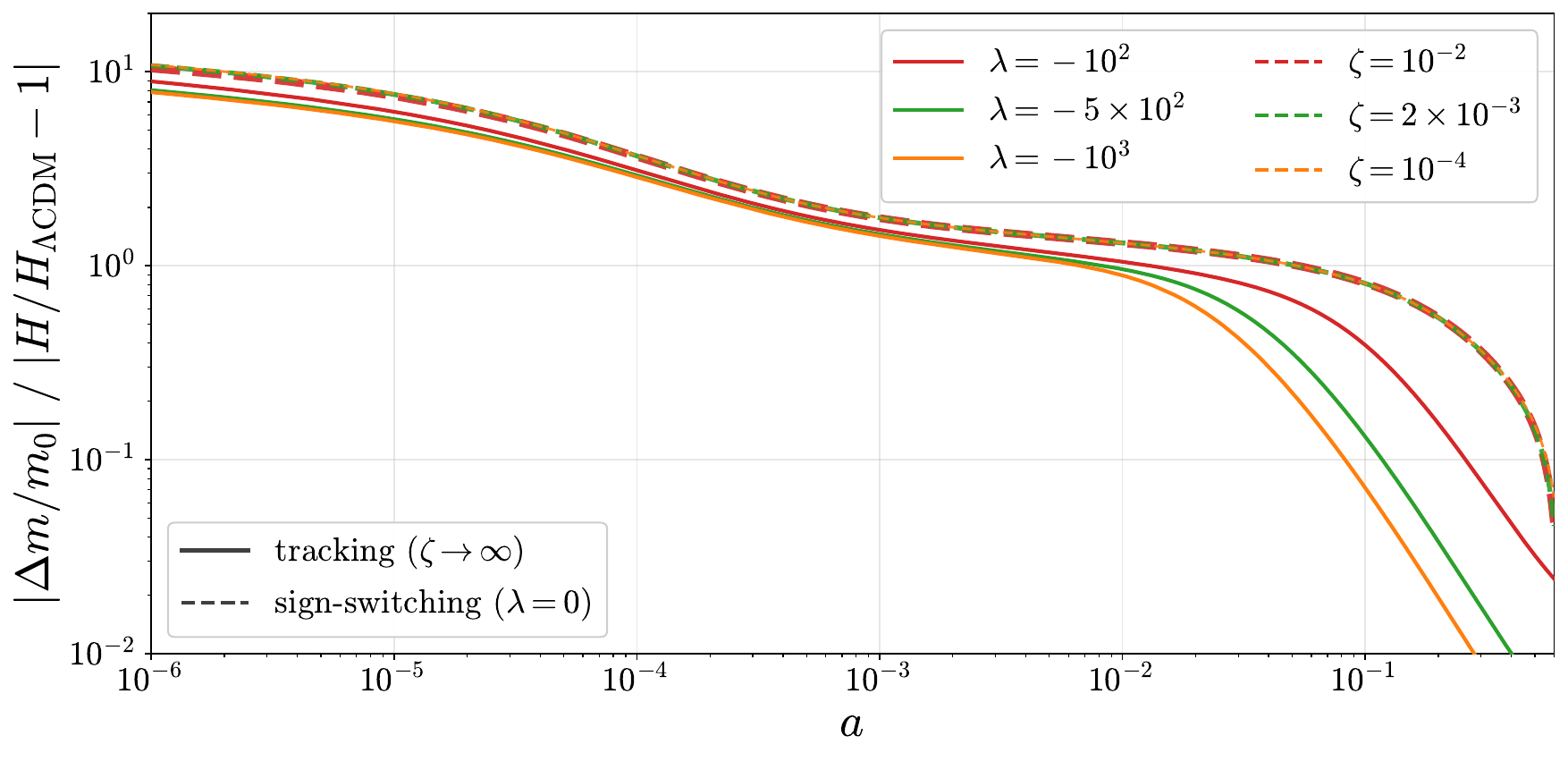}
    \caption{Einstein-frame particle mass drift per unit departure of the expansion
    history from $\Lambda$CDM evaluated at the same scale factor.
    Solid curves show the tracking case at $\zeta\to\infty$ for
    $\lambda=-10^{2},-5\times10^{2},-10^{3}$ in red, green and orange respectively;
    dashed curves show the sign-switching case at $\lambda=0$ for
    $\zeta=10^{-2},2\times10^{-3},10^{-4}$ in the same colours. The plot is
    cut at $a=0.6$, beyond which both quantities approach zero and their ratio is no
    longer informative. Parameters are as in
    \cref{fig:tracking_eos_mass,fig:lambda_0_eos_mass}.}
    \label{fig:drift_per_hubble}
\end{figure}

The equation of state and mass evolution for this case are shown in
\cref{fig:lambda_0_eos_mass}. A sharp feature appears near the zero-crossing of the effective
cuscuton density, where $w_f=p_f/\rho_f$ becomes singular because $\rho_f$ passes through
zero. This feature is pushed to slightly later times for the effective equation of state defined in \cref{eq:weff_def_here}
when the corresponding matter-sector reshuffling is included. 
Relative to the departure of the expansion history from $\Lambda$CDM, particle
masses drift further in this case than in any of the tracking benchmarks.
\Cref{fig:drift_per_hubble} shows the ratio of $|\Delta m/m_0|$ to
$|H/H_{\Lambda\rm CDM}-1|$, where the two cases are comparable at early times, but after $a\simeq10^{-3}$ the tracking ratio falls away, and by
$a=0.1$ the sign-switching case delivers between two and twelve times more mass
drift per unit deviation, depending on $\lambda$. This is because the late-time
dark energy evolution is less pronounced than in the tracking examples above, as
can be seen from $w_f$ settling near $-1$ in \cref{fig:lambda_0_eos_mass} over a
broader range of the expansion history, which permits stronger mass evolution
before it departs comparably far from $\Lambda$CDM. The sign-switching ratio is
moreover independent of $\zeta$, since both the mass drift and the change in the
expansion rate are linear in $\zeta$.

Although this case can produce a larger mass drift than the large-$|\lambda|$ tracker, the
direction of the drift in the examples shown here is the wrong one for the usual sound-horizon Hubble-tension alleviation
mechanism based on varying the electron mass
\cite{Hart:2019dxi,Sekiguchi:2020teg,Schoneberg:2024ynd,Smith:2025icl,Smith:2025uaq}, which is one of the
few proposals that eases the tension between the CMB-inferred and locally
measured expansion rates \cite{Riess:2021jrx,DiValentino:2021izs}. Since $m(a)/m_0<1$ in the past, particle
masses were smaller at recombination than they are today. The mass drift therefore has
the wrong sign for playing a role in alleviating the Hubble tension.

\subsection{Fifth forces and linear observables}
\label{sec:fifth forces-linear-observables}

\begin{figure}
    \centering

    \begin{subfigure}{\linewidth}
        \centering
        \includegraphics[width=0.92\linewidth]{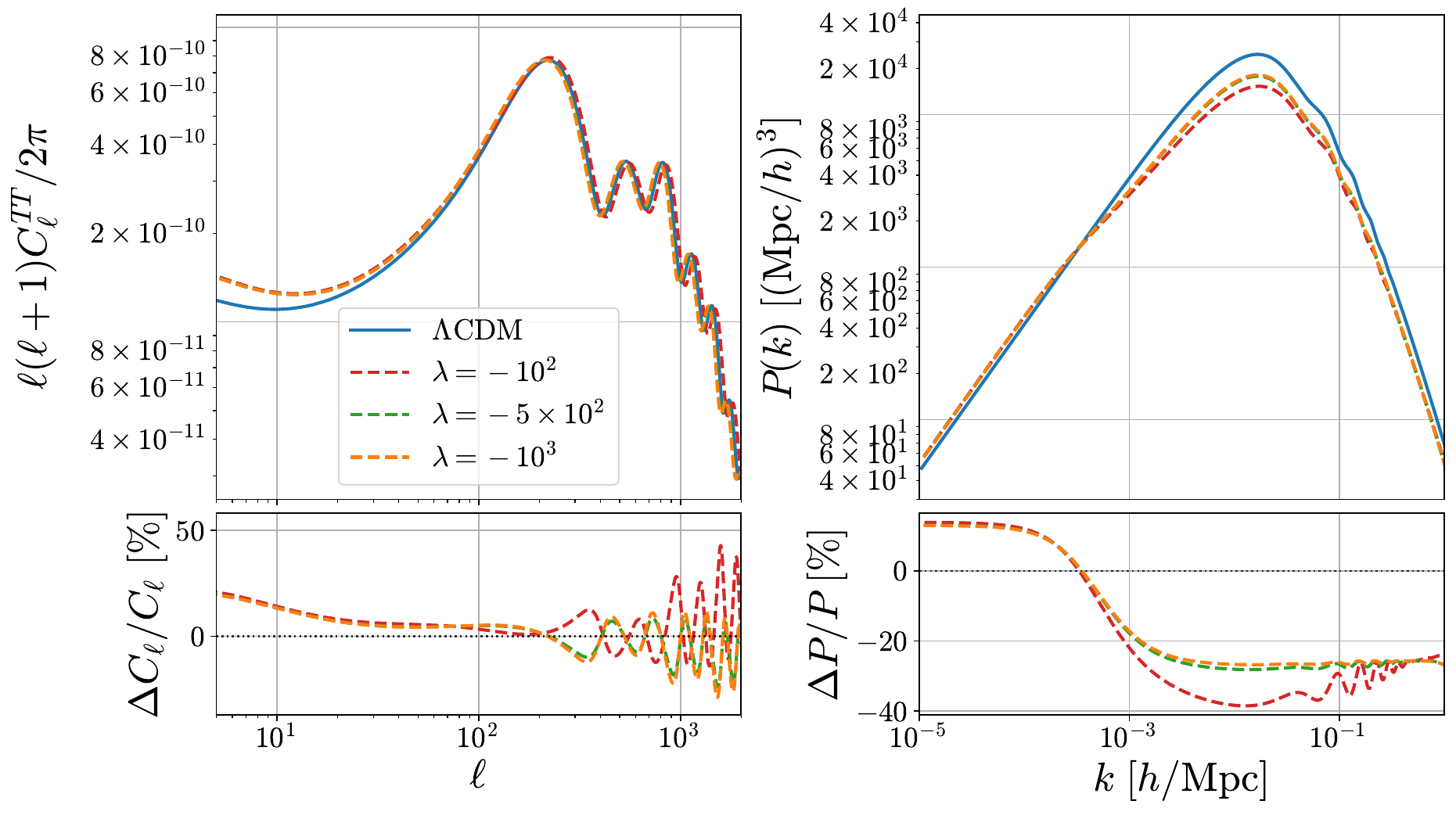}
        \label{fig:tracking_pk_cl}
    \end{subfigure}

    \vspace{0.5cm}

    \begin{subfigure}{\linewidth}
        \centering
        \includegraphics[width=0.92\linewidth]{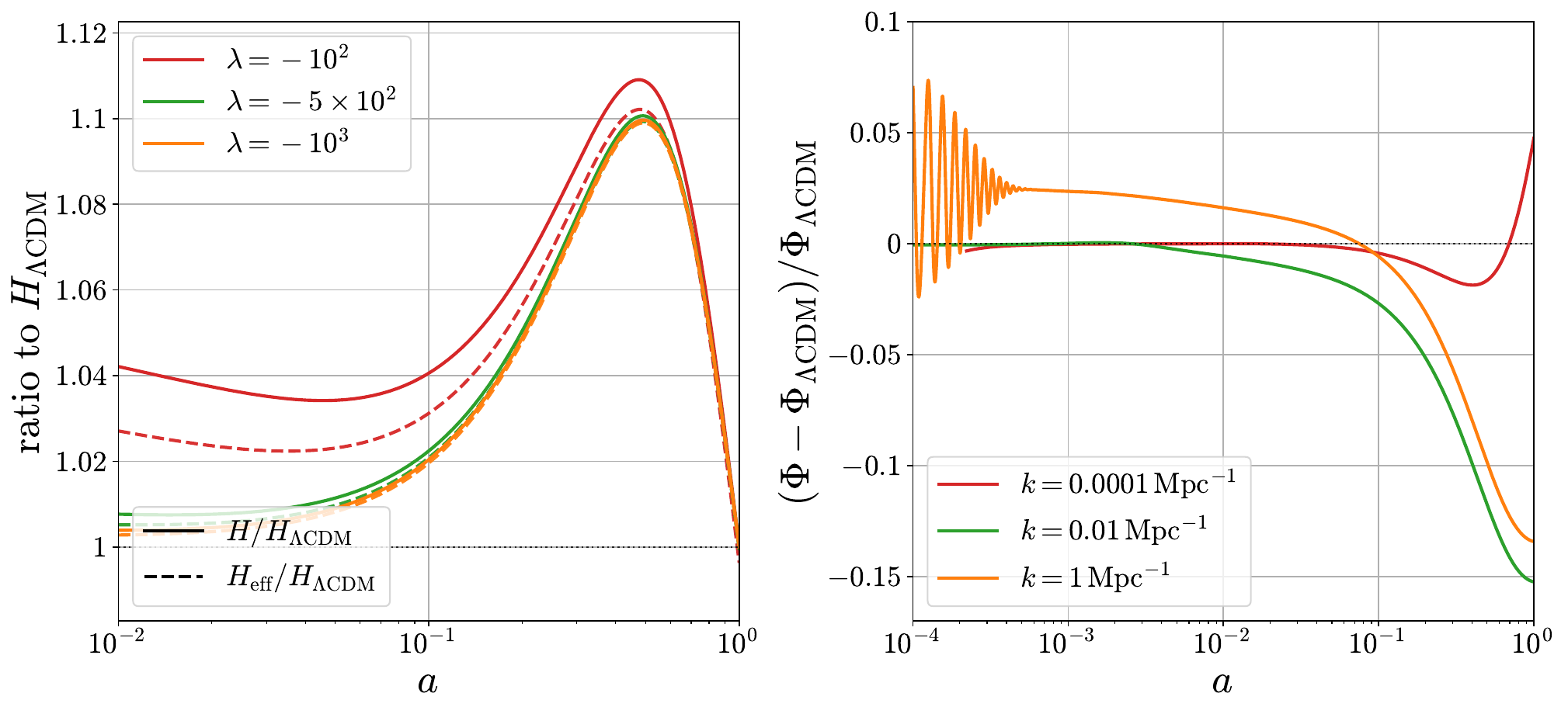}
        \label{fig:tracking_phi}
    \end{subfigure}

\caption{Linear observables in the tracking regime. \textbf{Top:} Dimensionless CMB
temperature angular power spectrum on the left and matter power spectrum on the right,
with $\Lambda$CDM in blue and $\lambda=-10^2,-5\times10^2,-10^3$ in red, green and orange
respectively. The panels underneath show the fractional differences from
$\Lambda$CDM. \textbf{Bottom:} The left panel shows the ratio $H/H_{\Lambda{\rm CDM}}$
(solid) and the effective friction coefficient $H_{\rm eff}/H_{\Lambda{\rm CDM}}$ (dashed)
for $\lambda=-10^2,-5\times10^2,-10^3$ in the same colours, while the right
panel shows the fractional departure of the Newtonian potential,
$(\Phi-\Phi_{\Lambda{\rm CDM}})/\Phi_{\Lambda{\rm CDM}}$, for the
$\lambda=-5\times10^2$ benchmark at $k=0.0001,0.01,1\,{\rm Mpc}^{-1}$ in red,
green and orange respectively.
Here $\zeta\to\infty$ in every panel, with
$\Lambda=M_{\rm Pl}$ and $V_0=1$ in all cases, giving
$C_0\simeq3.9\times10^{-59},2.0\times10^{-58},4.0\times10^{-58}$ for
$\lambda=-10^2,-5\times10^2,-10^3$ respectively.}
    \label{fig:tracking_linear_observables}
\end{figure}

Having painted the different cosmological scenarios in broad brushstrokes, we now turn to
the cupboard where many modified gravity models keep their skeletons: the CMB and the
growth of structure. The
background analysis above already shows that the different tracking and sign-switching cases
realise qualitatively different dark energy histories. However, the cuscuton also modifies gravity, and therefore affects the relation between matter perturbations,
metric potentials and the growth of structure. These observables expose effects
that are not visible in a background-only comparison.

The linear angular and matter power spectra for the tracking case are shown in the top row of
\cref{fig:tracking_linear_observables}. The dominant effect in the angular power
spectrum is a displacement of the acoustic peaks. The peaks
move because the conformal coupling makes the electron mass drift. This shifts
recombination\footnote{The varying electron mass shifts the binding energy of hydrogen which in return changes the time recombination happens. This cna be understood because a larger electron mass, compare with today, will lead to a larger binding energy and shift recombination to earlier times.} and hence the sound horizon, while the enhanced late-time expansion shown in the bottom left panel
shortens the distance to last scattering. Now, it is easy to understand, the two effects oppose one another and
their relative size sets the direction of the shift of the peaks. If we aberage over the first six
peaks,  the shift is $+4\%$ in $\ell$ for $\lambda=-10^{2}$, and by $-3\%$
and $-4\%$ for $\lambda=-5\times10^{2}$ and $-10^{3}$, which is the oscillating
residual in the $\Delta C_\ell/C_\ell$ panel. In addition to the shift in the peaks, the residual deviation in the angular power spectrum is an enhancement at
low $\ell$ of about $20\%$ at $\ell=2$, almost the same for all three benchmarks. This indicates that the remaining
CMB sensitivity comes from the late-time evolution of the metric potentials and
the associated integrated Sachs Wolfe (ISW)  contribution, and this sensitivity is well within the cosmic variance at small multipoles.

The matter power spectrum in the top right panel of \cref{fig:tracking_linear_observables} shows a more pronounced response.
Relative to $\Lambda$CDM, the tracking models enhance power on the largest
scales, by about $13\%$, while suppressing power on all smaller linear scales,
from $k\simeq3.5\times10^{-4}\,h\,{\rm Mpc}^{-1}$ onwards. The suppression is
largest near $k=10^{-2}\,h\,{\rm Mpc}^{-1}$, where it reaches $38\%$ for
$\lambda=-10^{2}$ and $28\%$ and $27\%$ for $\lambda=-5\times10^{2}$ and $-10^{3}$,
and it is between $25\%$ and $27\%$ for all three at $k=1\,h\,{\rm Mpc}^{-1}$. The suppression is
a background effect. The cuscuton density is taken to match the dark energy
density today, but since $w_f>-1$ in the past it decreases with time and $\rho_f$ was therefore
 larger than $\rho_{f}(0)$ at earlier times. This gives the enhanced
expansion rate shown in the bottom-left panel of
\cref{fig:tracking_linear_observables}, which holds back the growth of matter
perturbations.

The bottom-right panel of \cref{fig:tracking_linear_observables} shows the same
effect in the metric rather than in the matter, and explains why that
suppression applies to many different scales. On sub-horizon scales the Poisson
equation ties $\Phi$ to $\delta_m$, so slower growth appears directly as a lower
potential. The $k=10^{-2}$ and $k=1\,{\rm Mpc}^{-1}$ (green and orange respectively) modes enter the
horizon decades apart, the first near matter-radiation equality and the second
deep in radiation domination where its early oscillations are the usual decay of
the potential after entry, yet both arrive at almost the same fractional
departure of about $-0.14$ today. The enhanced expansion rate does not depend on
when a mode entered, so at $\lambda=-5\times10^{2}$ every sub-horizon mode is
held back by nearly the same amount. At $\lambda=-10^{2}$ the suppression instead
weakens towards small scales, so a further scale-dependent effect operates there.

The largest-scale mode (red) is the exception, since it never enters the horizon and
reaches only $k/aH\simeq0.5$ by the present day. On these scales the potential
decays only once the expansion accelerates, and the same $w_f>-1$ of
\cref{fig:tracking_eos_mass} that raised the expansion rate at earlier times
leaves the cuscuton a weaker accelerator than a cosmological constant today,
with $w_{\rm tot}=-0.52$ against $-0.69$ in $\Lambda$CDM in all three benchmark $\lambda$ values. The potential
therefore decays less and ends $5\%$ above its $\Lambda$CDM value, producing
both the low-$\ell$ ISW excess and the enhancement of matter power on the
largest scales.

\begin{figure}
    \centering
    \includegraphics[width=\linewidth]{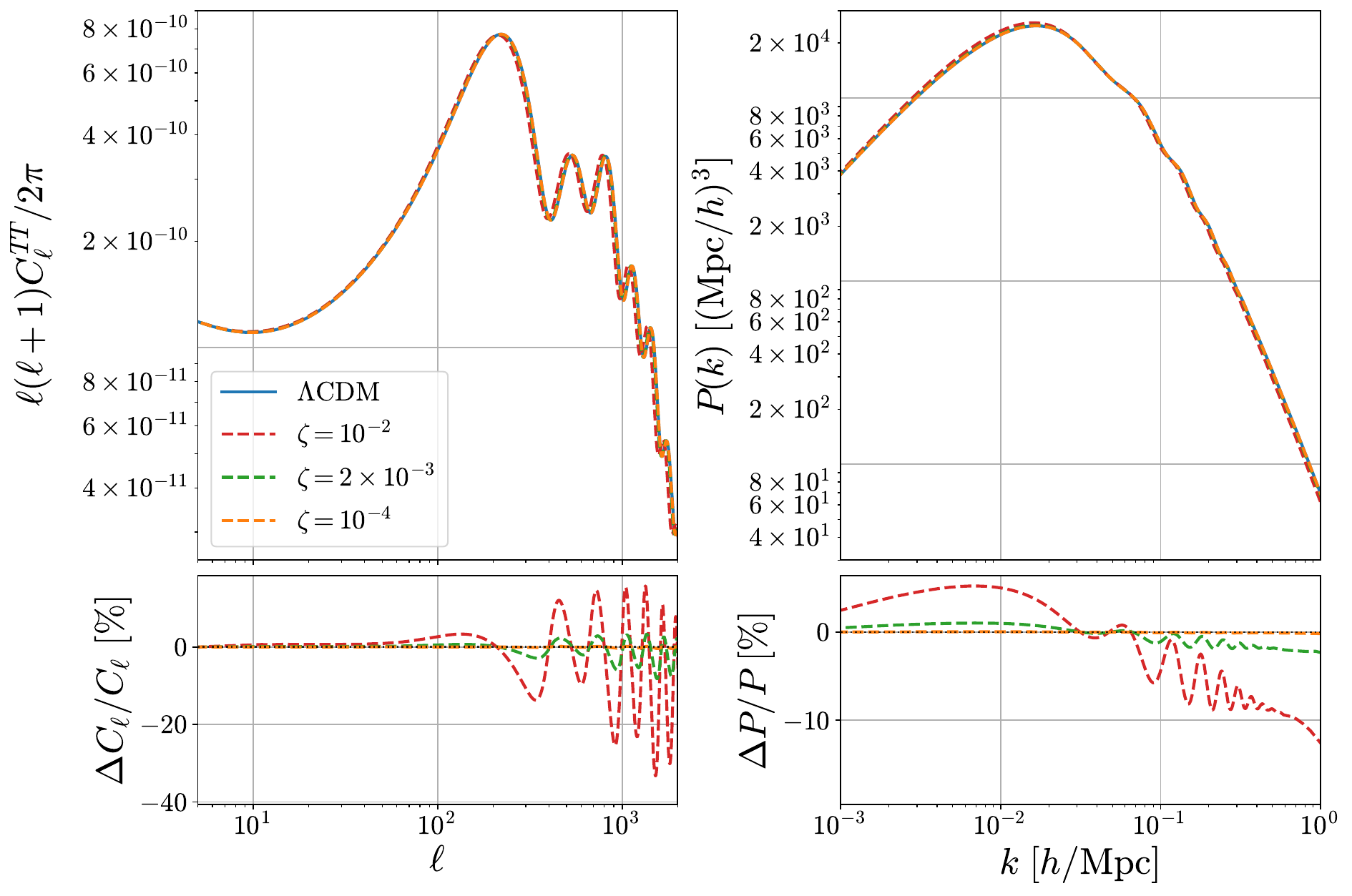}
\caption{
    Linear observables in the sign-switching case, $\lambda = 0$, with
    $\zeta = 10^{-2}, 2\times10^{-3}, 10^{-4}$. \textbf{Top:} dimensionless CMB
    temperature angular power spectrum on the left and linear matter power
    spectrum on the right, with the $\Lambda$CDM benchmark in solid blue.
    \textbf{Bottom:} the same quantities as fractional differences from
    $\Lambda$CDM.
    Here $\Lambda = M_{\rm Pl}$, $C_0 = -1$ and $V_0 = 7\times10^{-121}$.
    }
    \label{fig:lambda_0_observables}
\end{figure}

\begin{figure}
    \centering
    \includegraphics[width=\linewidth]{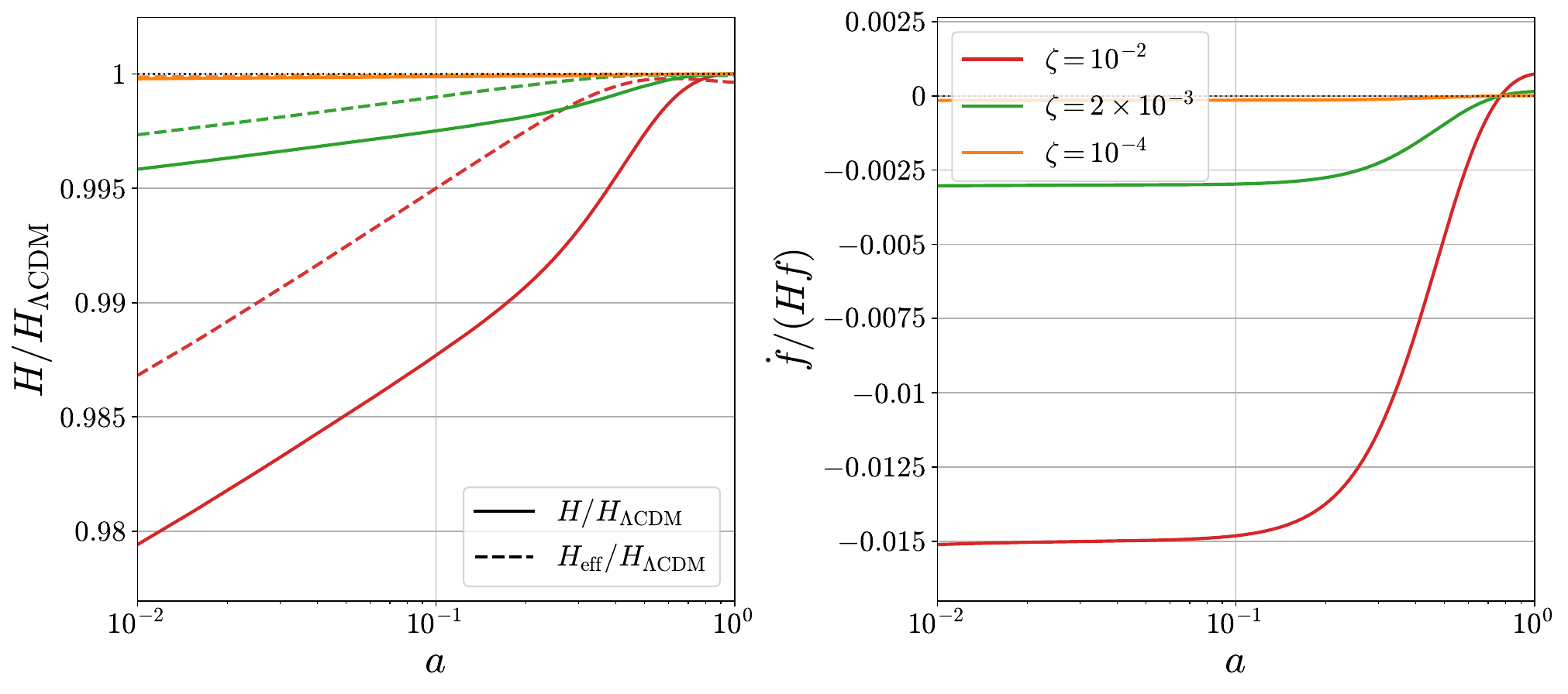}
\caption{
    Background expansion and effective friction in the sign-switching case,
    $\lambda = 0$, for $\zeta = 10^{-2}, 2\times10^{-3}, 10^{-4}$. The solid
    curves in the left panel show the ratio $H/H_{\Lambda\mathrm{CDM}}$, while
    the dashed curves show the effective friction coefficient
    \cref{eq:eff_hubble_friction}. The right panel shows the rate of evolution
    of $f$ relative to the Hubble rate. Here $\Lambda = M_{\rm Pl}$, $C_0 = -1$
    and $V_0 = 7\times10^{-121}$.
    }
    \label{fig:lambda_0_friction}
\end{figure}

The corresponding spectra for the sign-switching case are shown in
\cref{fig:lambda_0_observables}. The acoustic peaks are displaced towards lower
$\ell$, by $-4\%$, $-1\%$ and $0\%$ for
$\zeta=10^{-2},2\times10^{-3},10^{-4}$, and the mass drift established above is
what moves them. Recombination happens later due to having a smaller electron mass compared to today and this raises the comoving sound horizon by
$6\%$ at $\zeta=10^{-2}$, while the negative-to-positive transition in the
cuscuton density increases the distance to last scattering by only $1\%$. Since
the angular sound horizon, $\theta_s$, is their ratio the change in distance partially cancels the
sound-horizon shift rather than reinforcing it, unlike phenomenological
sign-switching dark energy models where the distance carries the whole effect
\cite{Akarsu:2021fol,Bouhmadi-Lopez:2026ckz}. The two combine to raise
$\theta_s$ by $5\%$, which predicts that the peak multipoles move by
$-4.7\%$. The first 6 peaks in \cref{fig:lambda_0_observables} move on average by $-4.2\%$, so the
displacement is accounted for by the change in the background acoustic scale
alone. There are subdominant effecs in the acoustic structure which we do not consider as they have low impact in the dynamics. 

\begin{figure}
    \centering
    \includegraphics[width=\linewidth]{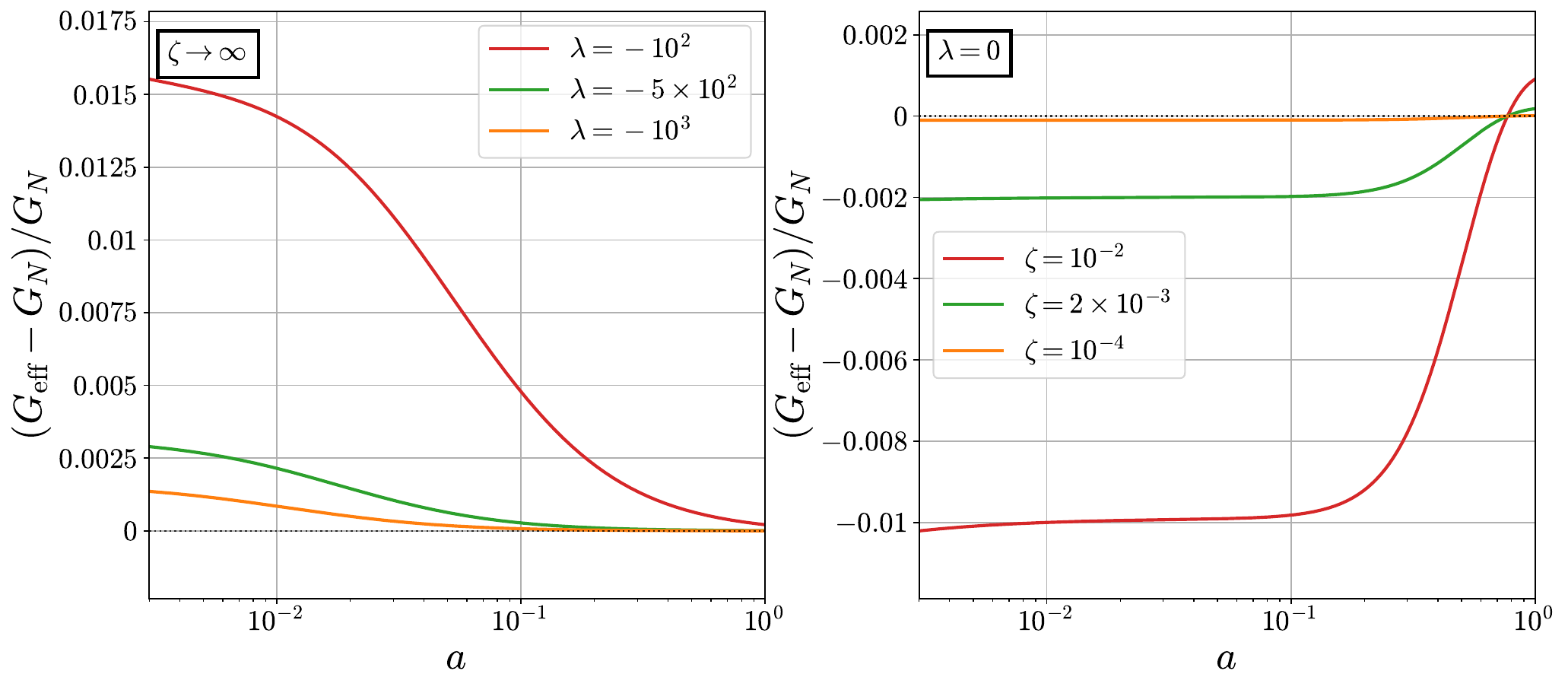}
\caption{
    \textbf{Left:} Quasi-static estimate of the fifth force correction
    $(G_{\rm eff}-G_N)/G_N$ in the tracking case, $\zeta\to\infty$, for
    $\lambda=-10^2,-5\times10^2,-10^3$ in red, green and orange respectively.
    \textbf{Right:} The same quantity in the sign-switching case, $\lambda=0$,
    for $\zeta=10^{-2},2\times10^{-3},10^{-4}$ in the same colours.
    The expression is evaluated along the numerical background solutions and is
    used only as a diagnostic of the characteristic magnitude and sign of the
    additional force.
    Here $\Lambda=M_{\rm Pl}$ throughout, with $V_0=1$ and
    $C_0\simeq3.9\times10^{-59},2.0\times10^{-58},4.0\times10^{-58}$ for
    $\lambda=-10^2,-5\times10^2,-10^3$ in the left panel, and $C_0=-1$ and
    $V_0=7\times10^{-121}$ in the right panel.
    }
    \label{fig:fifth_force_comparison}
\end{figure}

The matter power spectrum in the upper panel of \cref{fig:lambda_0_observables}
shows an enhancement of $5\%$ near $k\simeq10^{-2}\,h\,{\rm Mpc}^{-1}$ that
crosses to suppression above $k\simeq3\times10^{-2}\,h\,{\rm Mpc}^{-1}$ and
reaches $-13\%$ at $k=1\,h\,{\rm Mpc}^{-1}$, with the amplitude of matter
fluctuations in $8\,h^{-1}{\rm Mpc}$ spheres, $\sigma_8$, lower by $3\%$. Two
effects of opposite sign produce this shape, on top of a nearly scale-independent suppression of sub-horizon growth by the fifth force, which is repulsive in this case (\cref{fig:fifth_force_comparison,fig:residuals_fifth_force}).

The first raises growth on all scales. Since particle masses are growing in
\cref{fig:lambda_0_eos_mass}, implying $\dot{\bar f}/\bar f<0$ from
\cref{eq:mass_scaling_here}, the conformal coupling raises the effective
friction of \cref{eq:eff_hubble_friction} above the intrinsic Hubble rate.
This is more than offset by the reduction in the expansion rate itself, so
that $H_{\rm eff}/H_{\Lambda{\rm CDM}}$ stays below unity throughout, as shown
in \cref{fig:lambda_0_friction}. The net friction is therefore slightly
reduced, and matter perturbations grow more efficiently than in $\Lambda$CDM on
every sub-horizon scale.

The second suppresses small scales through the transfer function. Because the
Einstein-frame matter density was smaller in the past, matter-radiation
equality is delayed from $z_{\rm eq}=3400$ to $3300$ at $\zeta=10^{-2}$, and the
turnover of $P(k)$ moves by $-4\%$ compared to $\Lambda$CDM. Modes entering the
horizon during radiation domination therefore spend longer in the regime where
matter perturbations grow only inefficiently and acquire a smaller
transfer-function amplitude. This mainly affects modes that entered well before
equality, so on large scales the reduced friction dominates and gives the
enhancement, while above $k\simeq3\times10^{-2}\,h\,{\rm Mpc}^{-1}$ the
transfer-function suppression takes over.

\subsubsection{Fifth forces}

The quasi-static fifth force estimates for the same benchmark regimes are
shown in \cref{fig:fifth_force_comparison}. In the tracking case, smaller values
of $|\lambda|$ produce a stronger evolution of $f$ and hence
larger changes in the background, the particle masses, and the recombination
history. The matter-era scaling of $f$ in
\cref{eq:f_H_scaling} suppresses the rolling of the field as
$\dot{\bar f}/\bar f\propto 1/\lambda$, and consequently the
correction becomes smaller.  Since the same slow evolution
controls the Einstein-frame mass drift and the quasi-static fifth force
correction, the two shrink together. The largest tracking fifth force
estimate shown in \cref{fig:fifth_force_comparison} is approximately
$10^{-2}$, at $\lambda=-10^{2}$, falling to $10^{-3}$ at $\lambda=-10^{3}$.


The fifth force correction in the sign-switching case is comparable in magnitude
to the strongest tracking benchmark but opposite in sign, and is negative until $a\simeq0.8$ for
the examples shown. This
sign is noteworthy because a conventional propagating scalar with a healthy
kinetic term and a universal conformal coupling ordinarily mediates an
attractive interaction \cite{Damour:1990tw,Amendola:1999er,Joyce:2014kja},
whereas the non-dynamical cuscuton response gives
$G_{\rm eff}<G_N$ and therefore weakens the effective gravitational attraction
between matter perturbations. The magnitude of the correction reaches
approximately $10^{-2}$ in the largest displayed example, but is
nevertheless far too small to account for
the shift of the CMB peaks, although it contributes a few percent to the change in the matter power spectrum (\cref{sec:consistency}). The remaining
effects are instead driven by the modified expansion history and the altered
early universe dynamics caused by
the evolving particle masses.

Whether a correction of comparable magnitude persists around localised sources, for example in the solar system where tests on fifth forces are most-sensitive \cite{Bertotti:2003rm,Will:2014kxa},
cannot be inferred from the cosmological quasi-static analysis and requires a
separate treatment of the configuration of the cuscuton around a  static, spherically symmetric compact object.

\section{Conclusions}\label{sec:conclusions}

We have shown that the absence of a propagating scalar does not preclude a
fifth force. In effective cuscuton theory, the scalar perturbation is fixed by a
constraint rather than by independent initial conditions, but its algebraic response
to matter still enters the Euler equation. For matter minimally coupled in the
Jordan frame, in the Einstein-frame description we  explicitly see three important effects; evolving particle masses, modified friction in the growth
equation, and a direct constrained force. In the quasi-static limit these combine
into the effective Hubble rate \cref{eq:eff_hubble_friction} and effective
gravitational coupling \cref{eq:eff_newt}.

The direct force cannot be adjusted independently of the background. In the
power-law family studied here, the rate at which the conformal factor evolves is
set by $\lambda$ in the tracking case and by $\zeta$ in the sign-switching
case. The displayed tracking benchmarks therefore
have fifth force corrections reaching approximately $10^{-2}$ at $a=10^{-2}$ for
$\lambda=-10^{2}$ and falling by an order of magnitude for
$\lambda=-10^{3}$. The sign-switching case also reaches approximately
$10^{-2}$ in the largest displayed example. Its correction is
negative over much of the plotted history,
so the constrained cuscuton response weakens rather than strengthens the
effective attraction between matter perturbations, behaving as a repulsive fifth force.

The most visible cosmological effects are nevertheless not caused by this
direct force. In the tracking case the modified late-time expansion and metric
constraints raise the low-$\ell$ ISW contribution and suppress the matter power
spectrum for $k\gtrsim 3.5\times10^{-4}\,h\,{\rm Mpc}^{-1}$. The suppression reaches around $25\%$ at $k=1\,h\,{\rm Mpc}^{-1}$ for all benchmarks. The acoustic peaks are displaced by a few percent, towards
higher multipoles at $\lambda=-10^{2}$ and towards lower multipoles at the two
larger values of $|\lambda|$. In the sign-switching case, a negative high-density cuscuton
contribution evolves into a positive late-time vacuum, while the same conformal
factor changes the particle masses and recombination. The resulting shifts in the CMB peaks
and the suppression of the  matter-transfer function  are much larger than the
quasi-static correction to $G_N$. The cuscuton fifth force is therefore inseparable from the background and
mass-evolution history that generates it.

That conclusion rests on a small set of benchmarks rather than on a fit, so the
present results are illustrative predictions rather than parameter constraints.
What they establish is that fifth force in cuscuton gravity can have notable observational consequences. Whether the data allow
them to be large is a separate question, and answering it requires a likelihood
analysis over $\lambda$, $\zeta$ and the remaining cosmological parameters,
which we defer to a follow-up study. 

There is still space for refinements within the linear cosmological regime studied
here, such as a complete treatment of massive neutrinos in
the conformally coupled sector. In this analysis neutrinos were taken to be massless. The exact cuscuton source from a species is
proportional to the trace of its stress tensor, $\rho_I-3p_I$, so it vanishes in
the massless limit and turns on continuously as a
massive species becomes non-relativistic.

 Two things follow, and both matter for
the analysis proposed above:

\begin{enumerate}
 \item Neutrinos are the only species whose non-relativistic transition occurs late
enough that the cuscuton is still evolving appreciably while it happens, and in
the conformally coupled sector their masses drift with $\bar f$ exactly as the
electron mass does. The redshift of that transition is therefore predicted by
the background solution and the cuscuton source
will be displaced from the solutions shown in this paper by their non-relativistic transition and acquire a feature at a definite epoch instead of the smooth power-law evolution
of the tracking family. This is the one place where the mass drift leaves a
localised imprint rather than a slowly accumulating one.
\item Suppression of the linear matter power spectrum is also the observable that
currently bounds the neutrino mass sum
\cite{Lesgourgues:2006nd,DESI:2025zgx}, and the tracking benchmarks suppress it
by a comparable amount through an enhanced intermediate expansion rate. The two are separable in principle, since the neutrino
suppression deepens towards small scales while the tracking suppression at
$\lambda=-10^{2}$ does the opposite, weakening from $38\%$ at
$k=10^{-2}\,h\,{\rm Mpc}^{-1}$ to $25\%$ at $k=1\,h\,{\rm Mpc}^{-1}$. That
separation can only be demonstrated in an analysis that varies both, so a bound
on $\lambda$ or $\zeta$ obtained with massless neutrinos should not be taken at
face value. 
\end{enumerate}

The question that lies outside the linear regime is whether the cosmological
fifth forces found here can be tested by Cassini, lunar laser ranging or
planetary ephemerides
\cite{Bertotti:2003rm,Williams:2004qba,Fienga:2023ocw,Will:2014kxa}. The quasi-static
calculation presented here is not sufficient for such a comparison, since these
experiments probe the field of a localized source rather than the linear
response about a homogeneous FRW background.

For the present theory this is more than a modelling detail, since it calls
into question whether the effective description can be extended consistently
to strongly inhomogeneous configurations. The cuscuton action is only real
when the field gradient is timelike,
\begin{equation}
    X\equiv-g^{\mu\nu}\partial_\mu\phi\,\partial_\nu\phi>0,
\end{equation}
for which the constant-$\phi$ hypersurfaces define a regular foliation. In an
inhomogeneous configuration this condition is schematically
\begin{equation}
    X\simeq\dot{\phi}^{\,2}
    -|{\nabla}\phi|^2>0.
\end{equation}
Although it is satisfied by the homogeneous cosmological background and its
linear perturbations, a localized source can generate spatial gradients that
become comparable to the cosmological time-derivative. At
$|{\nabla}\phi|=|\dot{\phi}|$ one reaches $X=0$, where the unit
normal $n_\mu=\partial_\mu\phi/\sqrt{X}$ becomes singular. If the spatial
gradient grows further, then $X<0$ and the cuscuton kinetic terms become
imaginary, so the configuration has left the regime on which the
effective theory is defined.

A local-gravity prediction therefore cannot be obtained by simply
extrapolating the cosmological value of $G_{\rm eff}$ into the nonlinear
regime, and we make no solar system bound in this work. Such a bound requires a
static or quasi-static spherical solution that
approaches the cosmological background at large radius, together with an
explicit demonstration that $X>0$ everywhere within the region probed by the
experiment. If the timelike-gradient condition fails there, the
effective theory in its present form does not provide a real prediction to
bound.

The cuscuton fifth force is thus fully determined by the expansion and
mass-evolution history that produces it. That makes it sharply predictive in
linear cosmology, where precisely that history is what the data constrain, and
leaves it undetermined on solar system scales until the theory is shown to admit
a regular configuration there.

\section*{Acknowledgements}
We thank Carsten van de Bruck and Eleonora Di Valentino for valuable input and
discussions. AS is supported by the W.D. Collins Scholarship and is grateful
for the hospitality of Perimeter Institute and Kavli IPMU, where part of this
work was carried out. IA is supported by the Estonian Research Council through
the Centre of Excellence ``Foundations of the Universe'' (TK202U4) and grant
PRG2608. MM is supported by the Institute for Basic Science (IBS) as a Senior Researcher at CTPU-CGA under the project code, IBS-R018-D3by, and is grateful
for the hospitality of Perimeter Institute. MM's visit to Perimeter Institute
was supported by a grant from the Simons Foundation (1034867, Dittrich). IA's visit to Perimeter Institute
was supported by the Erasmus+ grant.
NA is
supported by the University of Waterloo and the Natural Sciences and
Engineering Research Council of Canada. Research at Perimeter Institute is
supported in part by the Government of Canada through NSERC and by the Province
of Ontario through the Ministry of Colleges and Universities.

\appendix

\section{Perturbations in Unitary \& Synchronous gauges}\label{sec:perts_in_other_gauges}
The perturbation analysis in the main text is presented in Newtonian gauge,
which is the frame in which the fifth force is easiest to interpret. The underlying theory, however, is constructed in
unitary gauge, where the constant-$\phi$ hypersurfaces are the coordinate slices
and $\delta\phi=0$ identically, as in the ADM formulation of the effective field
theory of dark energy \cite{Gubitosi:2012hu}. Synchronous gauge is the language Boltzmann codes
speak most naturally, being numerically stable and fixing the initial conditions
unambiguously \cite{Ma:1995ey}. This appendix records the linear system in those two gauges that were ultimately used for these respective roles.

For compactness, in this appendix we use a field variable $\chi$ in which the quadratic derivative term in the Einstein-frame action \eqref{eq:einstein_frame_action} is canonically normalized by defining
\begin{equation}
\chi\equiv\sqrt{\frac{3}{2}}\ln f(\tphi),
\qquad
f=e^{\sqrt{2/3}\,\chi}.
\label{eq:chi_definition}
\end{equation}
To avoid a collision with the Jordan-frame function $V(\tphi)$ in the main
text, we further define the effective Einstein-frame potential
\begin{equation}
\mathcal V(\chi)\equiv\frac{V(\tphi)}{f^2(\tphi)},
\label{eq:chi_W_definitions}
\end{equation}
used
throughout the appendices.
The action \eqref{eq:einstein_frame_action} thus becomes
\begin{equation}
S=\Lambda^2\int d^4x\,\sqrt{-g}\left[
\frac12R+\frac12\partial_\mu\chi\,\partial^\mu\chi
-\sqrt{\frac23}\sqrt{X_\chi}\,K
-\Lambda^2\mathcal V(\chi)
+\sqrt{\frac23}\Lambda W(\chi)\sqrt{X_\chi}
\right]+S_{\rm sm}
\label{eq:chi_action}
\end{equation}
where $X_\chi \equiv - \partial_\mu \chi\, \partial^\mu \chi$,
and 
\begin{equation}
    W(\chi)\equiv
\frac{c_1(\tphi)}{\sqrt{f(\tphi)}f_{,\tphi}(\tphi)}.
\end{equation}
We denote unitary gauge variables with u subscript/superscript, Newtonian gauge quantities remain with no subscripts as in the main text and Synchronous gauge quantities appear with S subscripts.

\subsection{Unitary gauge}\label{sec:unitary_gauge}

Fixing the time slicing by $\phi=\text{const}$ exhausts the temporal gauge
freedom, so the shift cannot also be set to zero and survives as an independent
scalar potential $B_u$. Only the spatial freedom remains, and it is used to keep
the scalar part of the spatial metric diagonal. With $N = 1 + \Phi_u$ the metric
is
\begin{equation}
    ds^2 = -\left(1+2\Phi_u\right)dt^2
    +2a\partial_i B_u\,dt dx^i
    +\left(1-2\Psi_u\right)a^2(t)d\mathbf{x}^2 ,
\end{equation}
so that $d\mathbf{x}^2=\delta_{ij}dx^i dx^j$, and thus in ADM language
$\gamma_{ij}=a^2(1-2\Psi_u)\delta_{ij}$,
$N_i=a\partial_iB_u$, and $N=1+\Phi_u$. The three potentials $\Phi_u$, $\Psi_u$
and $B_u$ carry between them everything that $\Phi$, $\Psi$ and $\delta\chi$
carry in Newtonian gauge.

We take the perturbed fluid four--velocity to be
\begin{equation}
u_{(I)}^{\mu} = \left( \frac{1}{N},\; \frac{v_I^{\,i}}{a} \right) = \left( 1 - \Phi_u,\; \frac{v_I^{\,i}}{a} \right),
\qquad
v_I^{\,i} = \partial^i v_I .
\end{equation}
From
\begin{equation}
T^{\mu\nu}_{(I)}
= (\rho_I + p_I)\, u_{(I)}^{\mu} u_{(I)}^{\nu}
+ p_I\, g^{\mu\nu},
\end{equation}
the momentum--density component is
\begin{equation}
T_{\mu\nu}n^\mu \gamma^\nu_{\;i}
= -a(\rho_I + p_I)\, \left(\partial_i v_{I} + \partial_i B_u \right)
=a^2(\rho_I + p_I)\,\frac{ik_i}{k^2}\left(\theta_I^{(u)}-\frac{k^2}{a}B_u\right).
\end{equation}
which defines the velocity divergence $\theta^{(u)}_I\equiv-k^2v_I/a$. The shift
appears alongside $\theta^{(u)}_I$ in this combination throughout, because it is
the velocity relative to the normal to the slices, not the coordinate velocity,
that is physical here.

Everything specific to the cuscuton enters through $\sqrt{X}$ and the extrinsic
curvature of the constant-$\phi$ slices, since those are the only objects the
cuscuton terms in \cref{eq:einstein_frame_action} depend on. In unitary gauge
$\sqrt{X}=\dot{\bar\phi}/N$, so the first is fixed by the lapse. Perturbed, the
second is
\begin{align}
    K &= \frac{\gamma^{ij}}{2N}\left(-\dot{\gamma}_{ij} + D_i N_j + D_j N_i\right) \\
    &= -3H + 3\left(\dot\Psi_u +H\Phi_u\right) -\frac{1}{a}k^2 B_u.
\end{align}

Because $\delta\chi$ vanishes identically in this gauge, the perturbed scalar
field equation is not an evolution equation for anything. It is a relation among
the metric potentials and the matter sources, and it is the unitary-gauge form
of the algebraic constraint that replaces a propagating scalar mode. Using the
background equation \eqref{eq:cuscback_discuss} it reads
\begin{align}
-\sqrt{6}\Bigg[
\Lambda W(\chi)\Big(\dot\Psi_u-H\Phi_u &- \frac{k^2}{3a}B_u\Big)
+ \ddot\Psi_u + 6H\dot\Psi_u+H\dot{\Phi}_u -\frac{k^2}{3a^2}\,\Phi_u
- \frac{k^2}{3a}\dot B_u
- 5\frac{Hk^2}{3a}B_u
\Bigg]
\nn\\
&+2\EqCorrection{\Phi_u}\Lambda^2 \mathcal V_{,\chi}
+\dot\chi\left(\dot\Phi_u + 3\dot\Psi_u-\frac{k^2}{a}B_u\right)
= \frac{1}{\Lambda^2\sqrt{6}}\,
\left(\EqCorrection{\delta_m^{(u)}}+2\Phi_u\right)\bar{\rho}_m.
\end{align}
The matter density contrast appears on the right-hand side, so the metric
potentials respond algebraically to matter through this equation. That response
is the origin of the fifth force discussed in \cref{sec:quasistatic}.


The Einstein equations are, respectively, the Hamiltonian constraint, the
momentum constraint and the metric equation. The first two fix $\Psi_u$ and
$B_u$ in terms of the matter sources at each instant, 
\begin{equation}
\left[
\frac{2k^2}{a^2}\Psi_u
+6 H\left(\dot\Psi_u+H\Phi_u\right)
-\frac{2 Hk^2}{a}\,B_u
\right]
=
-\dot\chi^{\,2}\,\Phi_u
+\sqrt{\frac{2}{3}}\dot\chi\!\left(3\dot\Psi_u+6H\Phi_u-\frac{k^2}{a}B_u\right)
-\frac{\delta\rho}{\Lambda^2}\;,
\end{equation}
\begin{equation}
\frac{k^{2}}{a^2}\bigg(\dot\Psi_u+H\Phi_u - \frac{\dot\chi}{\sqrt{6}}\Phi_u\bigg)
= \,\frac{1}{2\Lambda^{2}}\EqCorrection{\sum_{I}}(\bar\rho_I+\bar p_I)\,
\left(\theta^{(u)}_I-k^2\frac{B_u}{a}\right),
\end{equation}
while the third evolves
$\Psi_u$
\begin{align}
\,\ddot{\Psi}_u
&+3H\,\dot{\Psi}_u
+\Big(H-\frac{\dot{\chi}}{\sqrt{6}}\,\Big)\dot{\Phi}_u
+\Bigg(3H^{2}+2\dot{H}-\sqrt{\frac{2}{3}}\,\ddot{\chi}-\frac{1}{2}\dot{\chi}^{2}+\frac{\Lambda}{\sqrt6}W(\chi)\dot{\chi}\Bigg)\,(\Phi_u-\Psi_u)
\nn\\
&+\frac{k^2}{3a^{2}}(\Psi_u-\Phi_u)
-\frac{k^2}{3a}\!\left(\dot B_u+2H B_u\right)
=
\Psi_u\!\left(\frac{\Lambda}{\sqrt{6}}W(\chi)\dot{\chi}
-\Lambda^{2}\mathcal V(\chi)
\right)
+\frac{\delta p_u}{2\Lambda^2}+\frac{\bar{p}}{\Lambda^2}\,\Psi_u \; .
\end{align}

The traceless part of the spatial equation gives the slip between the two
potentials. Unlike the Newtonian-gauge case, the slip does not vanish with the
anisotropic stress, because the shift contributes to it,
\begin{equation}
    \Psi_u -\Phi_u= a\left(2H B_u + \dot B_u\right) + \frac{1}{2}\pi_{u} .
\end{equation}


Here
$\delta\rho\equiv\delta\rho_m+\delta\rho_r
=\delta_m\bar\rho_m+\delta_r\bar\rho_r$.

The matter equations are the continuity and Euler equations for the
conformally coupled species. The friction coefficient in the second is $2H$
rather than the usual value, shifted by the conformal factor through
$\dot{\bar f}/\bar f$, and the same combination reappears alongside the shift.
This is the unitary-gauge form of the modified friction of
\cref{eq:eff_hubble_friction},
\begin{align}
&\dot{\EqCorrection{\delta_m^{(u)}}} + \theta^{(u)}_m -3\dot\Psi_u =0\\
&\dot\theta^{(u)}_m
+ \left(2H - \frac{1}{2}\frac{\dot{\bar f}}{\bar f}\right)\theta^{(u)}_m
- \frac{k^2}{a^2}\,\Phi_u
- \frac{k^2}{a}\left[\dot B_u + \left(H - \frac{1}{2}\frac{\dot{\bar f}}{\bar f}\right) B_u\right]
= 0 .
\end{align}

\subsubsection*{Gauge transformation}

The shift generates the transformation back to the Newtonian variables of the
main text. Note that $\delta\chi$ is fixed entirely by $B_u$, so the two
descriptions carry the same information despite the apparent extra variable,
\begin{align}
\Phi &= \Phi_u + \partial_t(aB_u), \;\;\;
\Psi = \Psi_u - aH\,B_u, \;\;\;
\delta\chi = \dot{\chi}\,aB_u, 
\end{align}
and for the matter components 
\begin{align}
    \delta_m = \delta_m^{(u)} +\frac{\dot{\bar{\rho}}_m}{\bar\rho_m}aB_u,
    \qq{and}
    \delta_r = \delta_r^{(u)} +\frac{\dot{\bar{\rho}}_r}{\bar\rho_r}aB_u.
\end{align}
The velocity divergences differ by the same shift, written either way,
\begin{equation}
    \theta_I^{(u)}
    = \theta_I + \EqCorrection{\frac{k^2}{a}B_u}
    = \theta_I + k^2\frac{\delta\chi}{a^2\dot\chi} .
\end{equation}

\subsection{Synchronous gauge}\label{sec:synchronous_gauge}

Synchronous gauge sets the lapse to unity and the shift to zero, so the
coordinate observers are freely falling and the residual freedom is fixed by
taking the cold dark matter to be at rest in the initial conditions. The
constant-$\phi$ slices are no
longer the coordinate slices, so $\delta\chi_S$ reappears as an explicit
variable. It is still not an independent degree of freedom, and demonstrating
that is the object of the constraint derived at the end of this subsection.

We use the standard scalar decomposition
\begin{equation}
 ds^2=-dt^2+a^2(t)\left(\delta_{ij}+h_{ij}\right)dx^i dx^j,
 \qquad
 h_{ij}(\mathbf k,t)=\hat k_i\hat k_j h
 +6\left(\hat k_i\hat k_j-\frac13\delta_{ij}\right)\eta .
 \label{eq:synchronous_metric_convention}
\end{equation}
A dot denotes a cosmic-time derivative, while a prime below denotes a
conformal-time derivative. The combination of metric derivatives that plays the
role of the shear, and which appears throughout the system below, is
\begin{equation}
\alpha
=
\frac{a^2(t)}{2k^2}
\left(
\dot{h}+6\dot{\eta}
\right).
\end{equation}

This differs from the conformal-time Ma \& Bertschinger definition
\cite{Ma:1995ey} by $\alpha_{\rm MB}=\alpha/a$.

The perturbed cuscuton equation is written below as it comes out of the
variation, before any use of the Einstein equations. In this form it appears to
be second order in $\delta\chi_S$, which would suggest a propagating mode. The
second derivative is removed below and the appearance is spurious.
\begin{align}
0={}&
\delta\ddot\chi_S
-\frac{1}{\sqrt{6}}\,\ddot h
-\left(
\sqrt{6}\,H
+\frac{\Lambda W(\bar\chi)}{\sqrt{6}}
-\frac{\dot{\bar\chi}}{2}
\right)\dot h
+\frac{\bar\rho_m}{\sqrt{6}\Lambda^2}\,\delta_m^{(S)}
\EqCorrection{+}
3H\delta\dot\chi_S
\nonumber\\[0.5ex]
&-
\delta\chi_S
\Bigg[\sqrt{\frac{2}{3}}
\frac{k^2}{a^2\dot{\bar\chi}}
\left(
4\,H
+\Lambda W - \sqrt{\frac{3}{2}}\dot{\bar \chi}
\right)
+\Lambda^2 \mathcal V_{,\chi\chi}(\bar\chi)
+\sqrt{6}\,\Lambda H\,W_{,\chi}(\bar\chi)
\Bigg].
\end{align}


The time-time, momentum and spatial-trace Einstein equations supply the
relations needed to eliminate it
\begin{align}
0={}&
\frac{2k^2}{a^2}\eta
+\frac{\delta\rho^{(S)}}{\Lambda^2}
+\left(\sqrt{6}H-\dot{\bar\chi}\right)\delta\dot\chi_S
+\left(\frac{\dot{\bar\chi}}{\sqrt{6}}-H\right)\dot h
+
\delta\chi_S
\left(
\sqrt{\frac{2}{3}}\frac{k^2}{a^2}
+\Lambda^2 \mathcal V_{,\chi}(\bar\chi)
\right),
\label{eq:Einstein00_synchronous}
\end{align}

\begin{align}
\frac{k^2}{a^2}
\left[
\dot\eta
+
\left(
\frac{\dot{\bar\chi}}{2}
-\frac{\Lambda W(\bar\chi)}{\sqrt{6}}-\sqrt\frac{3}{2}H
\right)
\delta\chi_S + \frac{\delta\dot\chi_S}{\sqrt6}
\right]
=
\frac{1}{2\Lambda^2}
\EqCorrection{\sum_I}(\bar\rho_I+\bar p_I)\theta_I^{(S)} ,
\label{eq:Einstein0i_synchronous_general}
\end{align}

\begin{align}
0={}&
-\frac{\bar\rho_r}{\Lambda^2}\,\delta_r^{(S)}
-\ddot h
-3H\dot h
+\frac{2k^2}{a^2}\eta+
\left(3\dot{\bar\chi}
-\sqrt{6}\,\Lambda W(\bar\chi)
\right)\delta\dot\chi_S\nonumber\\&\qquad\qquad\qquad\qquad
+\sqrt{6}\,\delta\ddot\chi_S
+
\delta\chi_S
\left(
3\Lambda^2 \mathcal V_{,\chi}(\bar\chi)
-\sqrt{6}\,\Lambda W_{,\chi}(\bar\chi)\dot{\bar\chi}
\right) .
\label{eq:ij_synchronous_corrected}
\end{align}
Note that the second derivative
$\delta\ddot\chi_S$ occurs in the spatial-trace equation as well, with a
coefficient that differs from the one above, which is what makes the
elimination possible.

The traceless part of the spatial equation carries the anisotropic stress and
closes the metric sector,
\begin{equation}
\ddot h + 6\ddot\eta
+ 3H\left(\dot h + 6\dot\eta\right)
- \frac{2k^2}{a^2}\eta
=
-\frac{k^2}{a^2}\pi .
\end{equation}

Eliminating $\delta\ddot\chi_S$ between the cuscuton equation and
\cref{eq:ij_synchronous_corrected} removes every second derivative of
$\delta\chi_S$. What survives is algebraic in $\delta\chi_S$, and is solved at
each step in place of an evolution equation. This is the technical content of
the statement that the cuscuton carries no independent degree of freedom, and
the cancellation is the check reported in \cref{sec:consistency}. The result is
\begin{align}
&\delta\chi_S\Bigg[\frac{k^2}{a^2\dot{\bar\chi}}\Big(8H+2\Lambda W(\bar\chi)-\sqrt6\,\dot{\bar\chi}\Big)
+6\Lambda H\,W_{,\chi}(\bar\chi)-\sqrt6\,\Lambda W_{,\chi}(\bar\chi)\,\dot{\bar\chi}
+3\Lambda^2\mathcal V_{,\chi}(\bar\chi)
\nonumber\\&\qquad+\sqrt6\,\Lambda^2\mathcal V_{,\chi\chi}(\bar\chi)\Bigg]+\Bigg(3H+\Lambda W(\bar\chi)-\sqrt{\frac{3}{2}}\,\dot{\bar\chi}\Bigg)\big(\dot h-\sqrt6\,\delta\dot\chi_S\big)
+\frac{2k^2}{a^2}\,\eta-\frac{\delta\rho^{(S)}}{\Lambda^2} = 0.
\end{align}

Using the remaining Einstein equations to remove $\dot h$ and $\delta\dot\chi_S$
as well leaves $\delta\chi_S$ expressed purely in terms of $\eta$ and the matter
density perturbation, with no derivative of the field appearing anywhere,
\begin{align}
0={}&\frac{k^2}{a^2}\Big(4H+\Lambda W(\bar\chi)-2\sqrt{\tfrac23}\,\dot{\bar\chi}\Big)\Big(2\eta+\frac{2H}{\dot{\bar\chi}}\,\delta\chi_S\Big)
+\Big(2H+\Lambda W(\bar\chi)-\sqrt{\tfrac23}\,\dot{\bar\chi}\Big)\frac{\delta\rho^{(S)}}{\Lambda^2}
\nonumber\\
&+\delta\chi_S\Big[\Lambda W_{,\chi}(\bar\chi)\big(\sqrt6H-\dot{\bar\chi}\big)^2
+\Lambda^2\mathcal V_{,\chi}(\bar\chi)\big(6H+\Lambda W(\bar\chi)-\sqrt6\,\dot{\bar\chi}\big)
\nonumber\\
&\qquad\qquad
+\Lambda^2\mathcal V_{,\chi\chi}(\bar\chi)\big(\sqrt6H-\dot{\bar\chi}\big)\Big],
\end{align}
This is the equation solved numerically for $\delta\chi_S$ at each step.

\section{Numerical Methodology \& Consistency Checks}\label{app:numerics}

A model is only as novel as it is reproducible, and reproducibility demands
being explicit about the choices made in solving it. To this end, we collect
here the implementation choices behind the runs shown in this paper, the
numerical form of the equations actually evaluated where it differs from the
presentation in the main text, and the residuals we use to check that the whole
thing holds together.

\subsection{Implementation choices and assumptions}

\paragraph{Neutrinos.}
All runs use three massless species, $N_{\rm ur}=3.044$ and $\sum m_\nu = 0$.
The continuity equations \cref{eq:cont_discuss} split the total density into a
pressureless part, which exchanges energy with the scalar, and a relativistic
part, which remains uncoupled. A massive neutrino belongs to both of these
groups at different times, so its contribution to the two sources changes
continuously through its non-relativistic transition and cannot be assigned to
either. Including it requires the constraint to be rewritten with
$\rho_{\rm ncdm}$ and $p_{\rm ncdm}$ as a third source with its own
pressure \cite{Lesgourgues:2006nd,Lesgourgues:2011re}, which we have not done. The comparison $\Lambda$CDM runs use the same
neutrino content, so the differences shown compare like with like, although
their size may change once massive neutrinos are included.

\begin{figure}
    \centering
    \includegraphics[width=\linewidth]{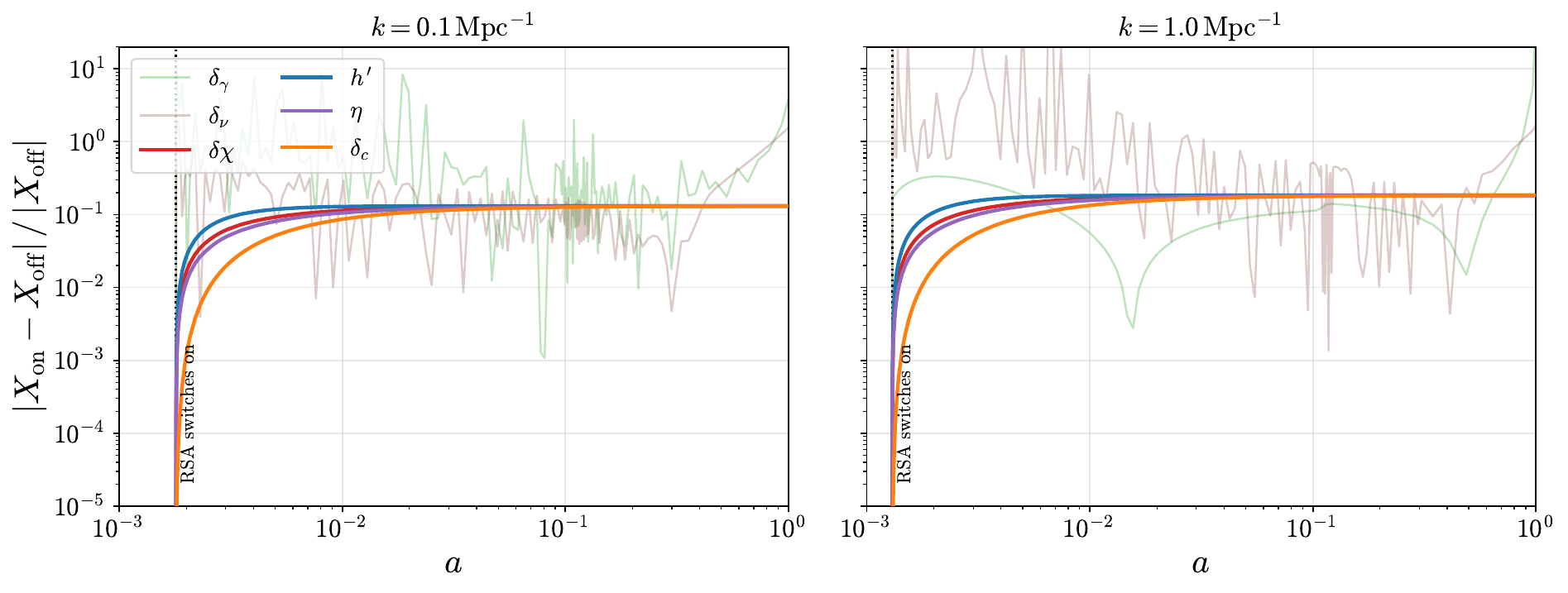}
    \caption{Relative difference between two otherwise identical tracking runs at
    $\lambda=-5\times10^{2}$, one with the radiation streaming approximation turned on and one with it disabled, at $k=0.1\,{\rm Mpc}^{-1}$ (left)
    and $k=1\,{\rm Mpc}^{-1}$ (right). Faded curves show the photon and neutrino
    density contrasts, solid curves denote the cuscuton perturbation $\delta\chi$, the synchronous-gauge
    metric perturbations $\dot h$ and $\eta$, and the cold dark matter contrast
    $\delta_c$. The dotted vertical line marks the scale factor at which the two
    runs first differ. Here $\Lambda=M_{\rm Pl}$, $V_0=1$ and $\zeta\to\infty$.}
    \label{fig:rsa_error_propagation}
\end{figure}

\paragraph{Radiation streaming approximation.}
This approximation is disabled. It replaces the photon and neutrino hierarchies
by a truncated expression once a mode is deep inside the horizon and photons
have decoupled \cite{Diego_Blas_2011}. That expression is derived using the
general-relativistic Einstein equations, which are modified here by the
cuscuton terms in \cref{eq:ij_synchronous_corrected}, so it does not solve the
system being integrated. \Cref{fig:rsa_error_propagation} compares runs with
the approximation turned on and off. The two runs agree exactly until the
approximation switches on, after which $\delta\chi$, $\dot h$, $\eta$ and
$\delta_c$ rise smoothly to a fixed offset and hold it to the present day,
reaching $13\%$ in $\delta_c$ at $k=0.1\,{\rm Mpc}^{-1}$ and $18\%$ at
$k=1\,{\rm Mpc}^{-1}$. The tight-coupling and ultra-relativistic fluid
approximations \cite{Ma:1995ey,Diego_Blas_2011} are unaffected by this argument
and are left at their defaults.

\subsection{Closed background system used numerically}

The background equations of \cref{sec:background_cosmology} never determine $\dot f$ and
$H$ on their own. As noted below \cref{eq:un2_discuss}, the combination
$\sqrt{6}H-\sqrt{3/2}\,\dot f/f$ is the only place the time derivative of the
scalar appears, and the Friedmann equation \cref{eq:Friedmann_raw} fixes that
same combination.

Obtaining $\dot f$ to break this degeneracy requires one further relation, and the only one available is
the statement that the cuscuton satisfies its constraint as the densities
evolve. Differentiating \cref{eq:un3_discuss} along the solution supplies $\dot f$,
but does so circularly at first sight, since it gives $\dot f$ in terms of
$\dot{\bar\rho}$, and $\dot{\bar\rho}$ contains $H$ through the continuity
equations (\ref{eq:cont_discuss}) while $H$ depends in turn on $f$ and $\dot f$.
Here we eliminate $H$ between the two and obtain the evolution of the field in
closed form from the background densities alone, which is what the code
evaluates. Besides being cleaner, this removes an inner solve from every one of
the many background evaluations the root finder performs.

It is convenient to trade $f$ for the variable $\chi$ and the functions
$\mathcal V(\chi)$ and $W(\chi)$ defined in \cref{eq:chi_W_definitions}, in
which the background constraint becomes
\begin{equation}
F(\chi,\bar\rho)\;\equiv\;
\sqrt{\,\mathcal V(\chi)+\dfrac{\bar\rho}{\Lambda^{4}}\,}
\;+\;
\sqrt{\tfrac{1}{3}}\,
\dfrac{\bar\rho+3\Lambda^{4}\mathcal V(\chi)}{W(\chi)\,\Lambda^{4}}
\;+\;
\dfrac{\mathcal V_{,\chi}(\chi)}{W(\chi)\,\sqrt{2}}
\;=\;0,
\qquad
\bar\rho=\bar\rho_m+\bar\rho_r ,
\label{eq:F_constraint_new}
\end{equation}
which we keep unsquared so that the physical solution satisfying \cref{eq:un2_discuss} is
selected automatically.

Using \cref{eq:cont_discuss} for $\dot{\bar\rho}$ together with the exact
identity
\begin{equation}
\sqrt{6}\,H
\;=\;
\dot\chi
+
\frac{\sqrt{2}}{\Lambda}\,
\sqrt{\Lambda^{4}\mathcal V(\chi)+\bar\rho} \, ,
\label{eq:H_identity_new}
\end{equation}
which follows from \cref{eq:un1_discuss}, the $H$ dependence cancels and
\begin{equation}
\dot\chi
=
\frac{\sqrt{2}\,\Lambda\,\sqrt{\mathcal V(\chi)+\bar\rho/\Lambda^{4}}\;
F_{,\rho}\,(3\bar\rho_m+4\bar\rho_r)}
{\sqrt{6}\,F_{,\chi}-4\bar\rho\,F_{,\rho}} \, ,
\label{eq:chi_prime_solution_new}
\end{equation}
with
\begin{align}
F_{,\rho} \;&=\;
\frac{1}{2\Lambda^{4}}\,\frac{1}{\sqrt{\,\mathcal V(\chi)+\bar\rho/\Lambda^{4}\,}}
\;+\;
\frac{1}{\sqrt{3}}\;\frac{1}{W(\chi)\,\Lambda^{4}} \, ,
\label{eq:F_rho_new}
\\[6pt]
F_{,\chi} \;&=\;
\frac{ \mathcal V_{,\chi}(\chi) }{ 2\,\sqrt{\,\mathcal V(\chi)+\bar\rho/\Lambda^{4}\,} }
\;+\;
\frac{1}{\sqrt{3}}\,\frac{1}{\Lambda^{4}}
\left(
\frac{3\Lambda^{4}\,\mathcal V_{,\chi}(\chi)}{W(\chi)}
\;-\;
\frac{\bar\rho+3\Lambda^{4}\mathcal V(\chi)}{W(\chi)^{2}}\,W_{,\chi}(\chi)
\right)
\notag\\[4pt]
&\hspace{1.7cm}
+\;\frac{1}{\sqrt{2}}
\left(
\frac{ \mathcal V_{,\chi\chi}(\chi) }{ W(\chi) }
\;-\;
\frac{ \mathcal V_{,\chi}(\chi)\,W_{,\chi}(\chi) }{ W(\chi)^{2} }
\right) .
\label{eq:F_chi_new}
\end{align}
Given $\bar\rho_m$ and $\bar\rho_r$, \cref{eq:F_constraint_new} is solved for
$\chi$ at each step and \cref{eq:chi_prime_solution_new} then supplies
$\dot\chi$ and hence $\dot f$ directly, after which $H$ follows from \cref{eq:H_identity_new}.

\subsection{Consistency Checks}
\label{sec:consistency}

This model requires heavy modification of \texttt{CLASS}. The scalar driving the
late-time expansion cannot be evolved as a separate degree of freedom, because it satisfies the algebraic constraint
\cref{eq:F_constraint_new} as a function of the rest of the background, so it
has to be obtained by solving the background and the constraint simultaneously
with a root finder at every step, rather than by integrating an equation of
motion. The same is true at linear order, where $\delta\chi$ is fixed
algebraically and $\delta\dot\chi$ follows from the time derivative of that
relation. Given how far this departs from the structure the public code is built
around, extensive consistency checks are shown below to validate the numerics used in the main text.

In each case below the residual is normalised by the largest individual term in
the equation, so that unity means complete failure and $10^{-16}$ machine
precision.

\begin{figure}
    \centering
    \includegraphics[width=0.9\linewidth]{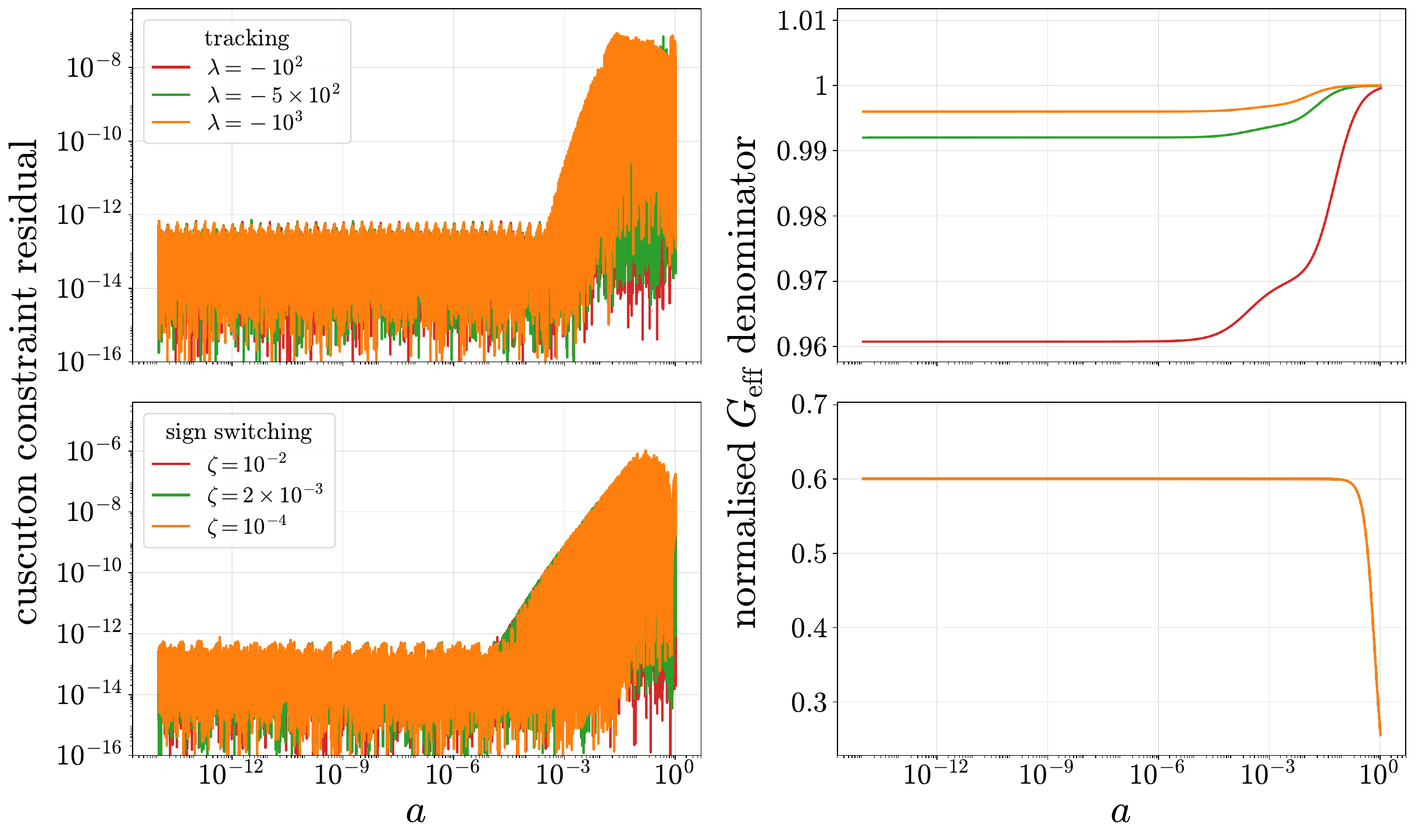}
    \caption{\textbf{Left:} residual of the algebraic cuscuton constraint \cref{eq:F_constraint_new}, normalised to the largest individual term. \textbf{Right:} denominator of \cref{eq:eff_newt}, normalised by the sum of the magnitudes of its terms. \textbf{Top:} tracking case, $\zeta\to\infty$, with $\lambda = -10^2, -5\times10^2, -10^3$. \textbf{Bottom:} sign-switching case, $\lambda = 0$, with $\zeta = 10^{-2}, 2\times10^{-3}, 10^{-4}$. Here $\Lambda = M_{\rm Pl}$.}
    \label{fig:residuals_background}
\end{figure}

\subsubsection*{Closure of the background constraint}
The first check is whether the stored background still satisfies the relation
that generated it. The left column of \cref{fig:residuals_background} shows the
residual of \cref{eq:F_constraint_new}. We use this unsquared form, so the check
also confirms that the retained root is the one satisfying
\cref{eq:un2_discuss} rather than the spurious solution introduced by squaring.
For every benchmark the residual sits at a median of $1.3\times10^{-13}$ across
the integration, rising to at most $9\times 10^{-8}$ in the tracking case and
$1\times 10^{-6}$ in the sign-switching case. The rise tracks the epoch at which
the field becomes dynamically relevant. Its size varies smoothly across the
benchmarks, growing with $|\lambda|$ in the tracking case and with decreasing
$\zeta$ in the sign-switching case, and no benchmark departs from this pattern.

\subsubsection*{Regularity of the fifth force}
The second background check concerns the analytical fifth force expression \cref{eq:eff_newt} itself.
Its denominator can in principle vanish, which would make the derived fifth force
response singular, so we verify that this does not occur in any of the displayed cases. The right column of \cref{fig:residuals_background} shows the
denominator normalised by the sum of the magnitudes of its three terms. The
three values of $\zeta$ are indistinguishable because the $\dot{\bar f}/\bar f$
term is of order $\zeta H$, while at high density the constraint fixes the
middle term to $-H$, so the normalised denominator is $3H/5H=0.6$. The
displayed solutions therefore remain regular throughout.

\begin{figure}
    \centering
    \includegraphics[width=0.9\linewidth]{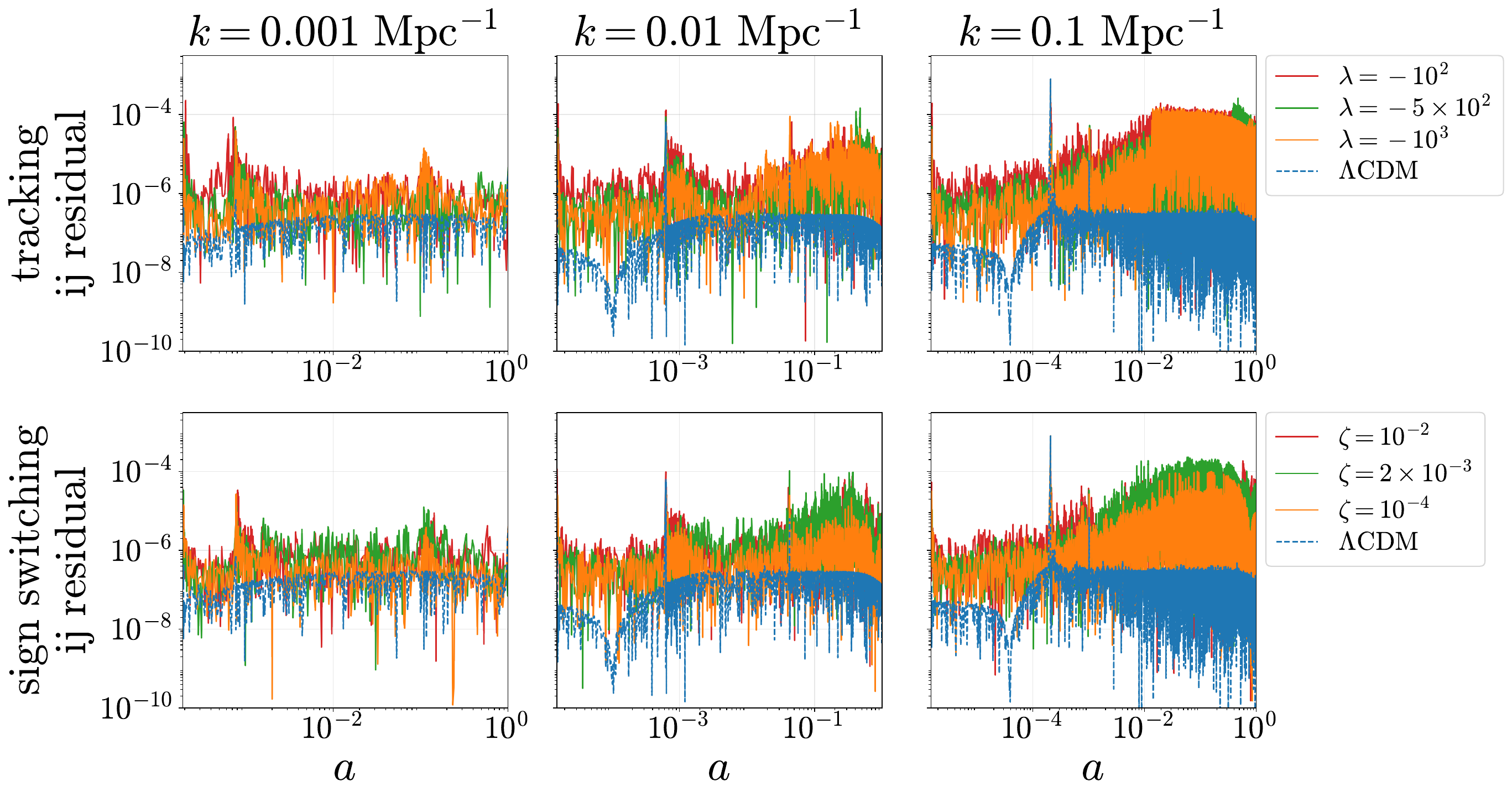}
    \caption{Residual of the trace of the $ij$ Einstein equation \cref{eq:ij_synchronous_corrected}, normalised to the largest individual term, at $k = 10^{-3}, 10^{-2}, 10^{-1}\,{\rm Mpc}^{-1}$. \textbf{Top:} tracking case with $\lambda = -10^2, -5\times10^2, -10^3$. \textbf{Bottom:} sign-switching case with $\zeta = 10^{-2}, 2\times10^{-3}, 10^{-4}$. The dashed blue curve is $\Lambda$CDM through the identical pipeline. All runs use $\Lambda = M_{\rm Pl}$ with the radiation streaming approximation disabled.}
    \label{fig:residuals_perturbations}
\end{figure}

\subsubsection*{Closure of the perturbed Einstein system}

Not all of the Einstein equations are independent. Once the matter and
cuscuton equations hold, the Bianchi identity makes the $ij$ equations follow
from the 00 and 0i equations. The solver therefore integrates the matter and
radiation hierarchies, obtains $\dot h$ from \cref{eq:Einstein00_synchronous} and
$\dot\eta$ from \cref{eq:Einstein0i_synchronous_general}, and fixes $\delta\chi$
algebraically from the linearised cuscuton constraint. Nothing then enforces the
$ij$ equations, so their
residuals measure whether the pieces are mutually consistent.
Of the two $ij$ equations, we use the trace,
\cref{eq:ij_synchronous_corrected}. The traceless part carries no cuscuton terms
at all, since the cuscuton contributes no anisotropic stress at linear order, so
its residual would test only the radiation hierarchy. We evaluate it
by differentiating the $\dot{h}$ obtained from \cref{eq:Einstein00_synchronous}. The Bianchi identity guarantees this agrees with the corresponding solution obtained driectly from the $ij$ equation only if all of them are
correct.

\Cref{fig:residuals_perturbations} shows the result at
$k = 10^{-3}, 10^{-2}$ and $10^{-1}\,{\rm Mpc}^{-1}$. A $\Lambda$CDM run through
the identical pipeline measures the error floor of the diagnostic itself, sitting at a
median of $2\times 10^{-7}$. The cuscuton runs lie above this floor but remain
small. At $k = 10^{-1}\,{\rm Mpc}^{-1}$ the median residual lies between
$1\times 10^{-5}$ and $2\times 10^{-5}$ across the tracking benchmarks and
between $4\times 10^{-6}$ and $2\times 10^{-5}$ across the sign-switching
ones. In both
cases it falls by about an order of magnitude as $k$ decreases from $k = 10^{-1}$ to $10^{-3}\,{\rm Mpc}^{-1}$. The largest
individual excursions reach $8\times 10^{-4}$, which is the same size as the
largest excursion in the $\Lambda$CDM baseline. The perturbation equations are therefore
consistent to better than one part in $10^{3}$ throughout the evolution at each
wavenumber sampled.

\subsubsection*{The quasi-static reduction against the full evolution}
\begin{figure}
    \centering
    \includegraphics[width=0.9\linewidth]{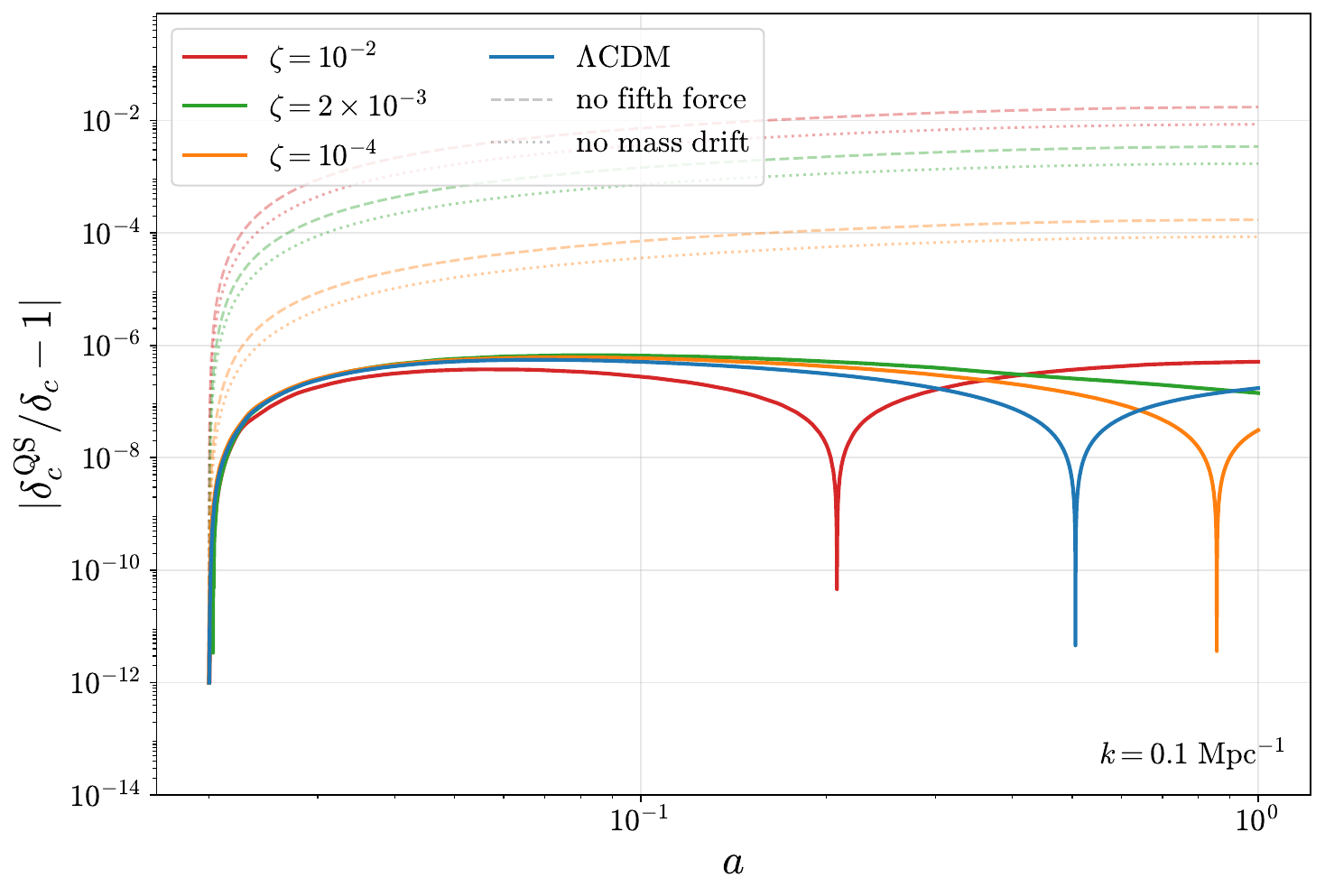}
    \caption{Fractional difference between the cold-dark-matter density contrast of a full \texttt{CLASS} solve and the contrast obtained by integrating the quasi-static growth equation \cref{eq:growth_geff} on top of that same run, with $G_{\rm eff}$ from \cref{eq:eff_newt}, at $k = 0.1\,{\rm Mpc}^{-1}$. The integration starts at $a = 0.02$ with initial conditions $\delta_c$ and $\dot\delta_c$ taken from the run at that time. \textbf{Solid:} the sign-switching case with $\zeta = 10^{-2}, 2\times10^{-3}, 10^{-4}$, and $\Lambda$CDM through the identical pipeline in blue. \textbf{Faded dashed:} the same integration with the fifth force term of \cref{eq:eff_newt} set to zero. \textbf{Faded dotted:} the same integration with the mass-drift contribution to the friction \cref{eq:eff_hubble_friction} set to zero, retaining the fifth force. Here $\Lambda = M_{\rm Pl}$ and $C_0 = -1$.}
    \label{fig:residuals_fifth_force}
\end{figure}

The final check tests the analytic quasi-static result obtained in \cref{sec:quasistatic}.
\Cref{eq:growth_geff}, with $G_{\rm eff}$ given by \cref{eq:eff_newt}, is a
closed second-order equation for the pressureless matter contrast once the
background is known, so we can ask whether the clustering of dark matter,  $\delta_c$, produced by the full
system satisfies it. We do this by taking initial conditions from the run at
$a = 0.02$ and integrating the quasi-static result \cref{eq:growth_geff} forward to the present,
comparing the result with the full CLASS run at every step.
\Cref{fig:residuals_fifth_force} shows this at $k = 0.1\,{\rm Mpc}^{-1}$.

This should be seen strictly as a residual test. Integrating forward
and comparing is a slightly stronger form of substituting $\delta_c$ into
\cref{eq:growth_geff} and asking whether the residual vanishes, since an error
in the equation accumulates over the whole interval rather than being read off
at an instant.


For all three sign-switching benchmarks the quasi-static prediction reproduces
the evolved $\delta_c$ to within $6\times10^{-7}$ by the present day. A
$\Lambda$CDM run through the identical procedure returns $2\times10^{-7}$, which
is the floor of the comparison. The integrated contrast therefore never leaves
the trajectory it was compared against, and the quasi-static description of the growth is therefore
accurate to the precision the test can resolve.

To show that this agreement is not an insensitivity of the test, repeating the integration
with the fifth force term of \cref{eq:eff_newt} set to zero, shown dashed,
displaces the prediction by $2\times10^{-2}$ at $\zeta = 10^{-2}$,
$3\times10^{-3}$ at $\zeta = 2\times10^{-3}$ and $2\times10^{-4}$ at
$\zeta = 10^{-4}$, between three and five orders of magnitude above the floor.
Discarding instead the mass-drift contribution to the friction,
\cref{eq:eff_hubble_friction}, while retaining the fifth force displaces it by
$9\times10^{-3}$, $2\times10^{-3}$ and $9\times10^{-5}$ respectively.
Both friction and fifth force terms are therefore required, and at comparable strength, to recover the
growth of the full solution.

\section{Weyl-transformation identities}\label{app:useful_formulae}

This appendix collects the transformation rules used in \cref{sec:cuscuton} to
carry the Jordan-frame action \cref{eq:action} into the Einstein-frame, and
records the intermediate steps of that calculation. The step worth isolating is
the coefficient of the term quadratic in $\partial_\mu f$, which picks up
contributions from three separate places. They do not cancel, and the residue
is what appears in \cref{eq:einstein_frame_action}. From $g_{\mu\nu}^J=f^{-1} g_{\mu\nu}$ it follows that $X_J=fX$, so that $n^J_\mu=f^{-1/2}n_\mu$ and $n_J^\mu=f^{1/2}n^\mu$. Under the Weyl rescaling \cref{eq:conf_trans}, the trace of the extrinsic
curvature of the constant-$\tphi$ hypersurfaces and the Ricci scalar transform
as
\begin{align}
K_J&=\nabla^{J}_\mu n_J^\mu
=\sqrt{f}\left(\,K-\frac{3}{2f}\,n^\mu\partial_\mu f\right),
\label{eq:weyl_K}\\[2pt]
R_J&=f\left(\,R+\frac{3}{f}\,\Box f-\frac{9}{2f^2}\,\partial_\mu f \partial^\mu f\right),
\label{eq:weyl_R}
\end{align}
with the Einstein-frame normal and extrinsic curvature built from $g_{\mu\nu}$
as in \cref{eq:normal_def}.

Using these transformation rules, the intermediate Einstein-frame action is
\begin{align}\label{ef_action}
S = \int d^4x&\,\sqrt{-g}\,\bigg\{
\frac{\Lambda^2}{2} R
+3\Lambda^2\bigg( \frac{\Box f}{2f} - \frac{3}{4}\frac{\partial_\mu f\partial^\mu f}{f^2}\bigg)
\nn\\&-\Lambda^4 \frac{ V(\tphi)}{f^2}
+\Lambda^3\, \frac{c_1(\tphi)}{f^{3/2}}\, \sqrt{X}
\nn\\&+\,\Lambda^2 \frac{c_2(\tphi)}{f}\,
\Big( \sqrt{X}\,K - \frac{3}{2}\frac{\sqrt{X}}{f}\, n^\mu\partial_\mu(f) \Big)
\bigg\} + S_{\rm sm}( \tphi, g_{\mu\nu}, \Psi_i).
\end{align}
The $\Box f$ term and the derivative piece of the transformed $c_2$ operator are not independent of one another. For the $\Box f$ term we use the identity
\begin{equation}
\frac{\Box f}{f}=\nabla_\mu\left(\frac{\nabla^\mu f}{f}\right)+\frac{\nabla_\mu f\,\nabla^\mu f}{f^2}.\label{eq:box_f_identity}
\end{equation}
Integrating the $\Box f$ term of \cref{ef_action} by parts, and dropping the
boundary term, gives
\begin{equation}
S_{\Box}
=\frac{3}{2}\Lambda^2\int d^4x\,\sqrt{-g}\,\frac{\Box f}{f}
=\frac{3}{2}\Lambda^2\int d^4x\,\sqrt{-g}\,
\frac{\partial_\mu f\,\partial^\mu f}{f^2}.
\label{eq:S_box_ibp}
\end{equation}
For the derivative piece of the transformed $c_2$ operator, the foliation defined by $\phi$ gives
\begin{equation}
\partial^\mu f=\sqrt{X}\,f_{,\phi}\,n^\mu,\qquad n^\mu\partial_\mu f=-\sqrt{X}\,f_{,\phi},\qquad \partial_\mu f\,\partial^\mu f=-Xf_{,\phi}^{\,2}.
\label{eq:f_normal_identities}
\end{equation}
Using $c_2=-f_{,\phi}$ together with \cref{eq:f_normal_identities}, the derivative part of the transformed $c_2$ operator becomes
\begin{equation}
-\frac{3}{2}\Lambda^2\frac{c_2}{f^2}\sqrt{X}\,n^\mu\partial_\mu f
=-\frac{3}{2}\Lambda^2\frac{Xf_{,\phi}^2}{f^2}
=\frac{3}{2}\Lambda^2\frac{\partial_\mu f\,\partial^\mu f}{f^2}.
\end{equation}
Together with the $+3/2$ contribution from the integration by parts of the $\Box f$ term and the $-9/4$ contribution inherited from the transformed Ricci scalar, the Einstein-frame action reduces to
\begin{align}
S = \int d^4x\,\sqrt{-g}\,\bigg\{
\frac{\Lambda^2}{2} R
+\Lambda^2 \frac{3}{4}\frac{\partial_\mu f\partial^\mu f}{f^2} &-\,\Lambda^2 \frac{\EqCorrection{f_{,\tphi}}}{f}\,
\sqrt{X}\,K
\nn\\&-\Lambda^4 \frac{ V(\tphi)}{f^2}
+\Lambda^3\, \frac{c_1(\tphi)}{f^{3/2}}\, \sqrt{X}
 \bigg\}+ S_{\rm sm}( \tphi, g_{\mu\nu}, \Psi_i),
\label{eq:einstein_frame_action_appendix}
\end{align}
which is \cref{eq:einstein_frame_action} of the main text. The term quadratic in
$\partial_\mu f$ is the residue of the mixing between the cuscuton operators and gravity under
the Weyl transformation, and the field remains non-dynamical, as the background
and linear analyses of \cref{sec:background_cosmology,sec:pert_newtonian} show
explicitly.

\bibliographystyle{JHEP}
\bibliography{bibliography}

\end{document}